\documentclass{aa}

\usepackage{lscape}
\usepackage{graphicx}
\usepackage{adjustbox}
\usepackage{tabularx}
\usepackage{xtab,booktabs}
\usepackage{txfonts}
\usepackage{natbib}
\usepackage{siunitx}
\usepackage{url}
\usepackage{float}
\usepackage{caption}
\usepackage{ulem}
\usepackage{xcolor}
\usepackage{soulutf8}
\usepackage{textcomp}
\usepackage{subfigure}
\usepackage{longtable}
\usepackage{starfont}
\usepackage[colorlinks=true,linkcolor=black,urlcolor=blue,citecolor=blue]{hyperref} 

\begin{document}

\title{The compact multi-planet system GJ~9827 revisited with ESPRESSO\thanks{\'Echelle SPectrograph for Rocky Exoplanets and Stable Spectroscopic Observations. Based on guaranteed time observations collected at the European Southern Observatory under ESO programme(s) 1102.C-0744, 1104.C-0350 by the ESPRESSO Consortium} \thanks{Extracted radial velocities from ESPRESSO and HARPS used in this work are only available in electronic form at the CDS via anonymous ftp to cdsarc.u-strasbg.fr (130.79.128.5) or via http://cdsweb.u-strasbg.fr/cgi-bin/qcat?J/A+A/.}}

\titlerunning{GJ~9827 revisited with ESPRESSO}

\author{V.\,M.~Passegger\inst{\ref{iac},\ref{uiac},\ref{hs}}, 
        A.~Su{\'a}rez Mascare{\~n}o\inst{\ref{iac},\ref{uiac}},
        R.~Allart\thanks{Trottier Postdoctoral Fellow}\inst{\ref{montreal},\ref{unige1}},
        J.\,I.~Gonz{\'a}lez Hern{\'a}ndez\inst{\ref{iac},\ref{uiac}},
        C.~Lovis\inst{\ref{unige1}}, 
        B.~Lavie\inst{\ref{unige1}},
        A.\,M.~Silva\inst{\ref{porto1},\ref{porto2}},
        H.\,M.~M{\"u}ller\inst{\ref{hs}},
        H.\,M.~Tabernero\inst{\ref{ucm}},
        S.~Cristiani\inst{\ref{INAF-Trieste}},
        F.~Pepe\inst{\ref{unige2}},
        R.~Rebolo\inst{\ref{iac},\ref{uiac},\ref{csic}},
        N.\,C.~Santos\inst{\ref{porto1},\ref{porto2}},
        V.~Adibekyan\inst{\ref{porto1},\ref{porto2}},
        Y.~Alibert\inst{\ref{unibe}},
        C.~Allende~Prieto\inst{\ref{iac},\ref{uiac}},
        S.\,C.\,C.~Barros\inst{\ref{porto1},\ref{porto2}},
        F.~Bouchy\inst{\ref{unige1}},
        A.~Castro-Gonz\'{a}lez\inst{\ref{villanueva}},
        V.~D'Odorico\inst{\ref{INAF-Trieste}},
        X.~Dumusque\inst{\ref{unige1}},
        P.~ Di~Marcantonio\inst{\ref{INAF-Trieste}},
        D.~Ehrenreich\inst{\ref{unige1}},
        P.~Figueira\inst{\ref{unige1},\ref{porto1}},
        R.~G\'enova Santos\inst{\ref{iac},\ref{uiac}},
        G.~Lo~Curto\inst{\ref{ESO}},
        C.~J.~A.~P.~Martins\inst{\ref{porto1},\ref{porto3}},
        A.~Mehner\inst{\ref{ESO}},
        G.~Micela \inst{\ref{INAF-Palermo}},
        P.~Molaro\inst{\ref{INAF-Trieste}},
        N.~Nari\inst{\ref{iac},\ref{uiac},\ref{light}},
        N.\,J.~Nunes\inst{\ref{lisbon}},
        E.~Pall\'e\inst{\ref{iac},\ref{uiac}},
        E.~Poretti\inst{\ref{INAF-TNG}},
        J.~Rodrigues\inst{\ref{porto1},\ref{porto2},\ref{OFXB}},
        S.\,G.~Sousa\inst{\ref{porto1}},
        A.~Sozzetti\inst{\ref{INAF-Torino}},
        S.~Udry\inst{\ref{unige1}},
        M.\,R.~Zapatero Osorio\inst{\ref{villanueva}}
    }

\authorrunning{V.\,M.~Passegger et al.}
 
\institute{
        Instituto de Astrof\'{\i}sica de Canarias, c/ V\'ia L\'actea s/n, 38205 La Laguna, Tenerife, Spain \label{iac}
        \and
        Departamento de Astrof\'{\i}sica, Universidad de La Laguna, 38206 La Laguna, Tenerife, Spain \label{uiac}
        \and
        Hamburger Sternwarte, Gojenbergsweg 112, 21029 Hamburg, Germany \label{hs}
        \and
        Department of Physics, and Trottier Institute for Research on Exoplanets, Universit\'e de Montr\'eal, Montr\'eal H3T 1J4, Canada \label{montreal}
        \and
        Observatoire astronomique de l’Universit\'e de Gen\`{e}ve, Chemin Pegasi 51b, 1290 Versoix, Switzerland \label{unige1}
        \and
       Instituto de Astrof\'isica e Ci\^{e}ncias do Espa\c{c}o, CAUP, Universidade do Porto, Rua das Estrelas, 4150-762 Porto, Portugal \label{porto1}
        \and
        Departamento de F\'isica e Astronomia, Faculdade de Ci\^{e}ncias, Universidade do Porto, Rua do Campo Alegre, 4169-007 Porto, Portugal \label{porto2}
        \and
        Departamento de F{\'i}sica de la Tierra y Astrof{\'i}sica \& IPARCOS-UCM (Instituto de F\'{i}sica de Part\'{i}culas y del Cosmos de la UCM), Facultad de Ciencias F{\'i}sicas, Universidad Complutense de Madrid, E-28040 Madrid, Spain \label{ucm}
        \and
        INAF – Osservatorio Astronomico di Trieste, Via Tiepolo 11, I-34143 Trieste, Italy \label{INAF-Trieste}
        \and
         D\'epartement d'astronomie de l’Universit\'e de Gen\`{e}ve, Chemin Pegasi 51, CH-1290 Versoix Switzerland \label{unige2}
        \and
        Consejo Superior de Investigaciones Cient{\'\i}ficas (CSIC), E-28006 Madrid, Spain \label{csic}
        \and
        Physikalisches Institut \& Center for Space and Habitability, Universit\"{a}t Bern, Gesellschaftsstrasse 6, 3012 Bern, Switzerland \label{unibe}
        \and
        Centro de Astrobiolog\'ia, CSIC-INTA, Camino Bajo del Castillo s/n, E-28692 Villanueva de la Ca\~{n}ada, Madrid, Spain \label{villanueva}
        \and
        European Southern Observatory, Av. Alonso de Cordova 3107, Vitacura, Santiago de Chile, Chile \label{ESO}
        \and
        Centro de Astrof\'{\i}sica da Universidade do Porto, Rua das Estrelas, 4150-762 Porto, Portugal \label{porto3}
        \and
        INAF – Osservatorio Astronomico di Palermo, Piazza del Parlamento 1, 90134 Palermo, Italy \label{INAF-Palermo}
        \and
        Light Bridges S.L., Avda. Alcalde Ram\'{\i}rez Bethencourt, 17, 35004 Las Palmas de Gran Canaria, Canarias, Spain \label{light}
        \and
        Instituto de Astrof\'{\i}sica e Ci\^{e}ncias do Espa\c{c}o, Faculdade de Ci\^{e}ncias da Universidade de Lisboa, Campo Grande, PT1749-016 Lisboa, Portugal \label{lisbon}
        \and
        INAF – Osservatorio Astronomico di Brera, Via Bianchi 46, 23807 Merate, Italy \label{INAF-TNG}
        \and
        Observatoire Fran\c{c}ois-Xavier Bagnoud -- OFXB, 3961 Saint-Luc, Switzerland\label{OFXB}
        \and
        INAF – Osservatorio Astrofisico di Torino, Strada Osservatorio, 20 10025 Pino Torinese (TO), Italy \label{INAF-Torino}
        }

\date{Received 13 November 2023 / Accepted 04 January 2024}

\abstract
{GJ~9827 is a bright, nearby K7V star orbited by two super-Earths and one mini-Neptune on close-in orbits. The system was first discovered using K2 data and then further characterized by other spectroscopic and photometric instruments. Previous literature studies provide several mass measurements for the three planets, however, with large variations and uncertainties. 
To better constrain the planetary masses, we added high-precision radial velocity measurements from ESPRESSO to published datasets from HARPS, HARPS-N, and HIRES and we performed a Gaussian process analysis combining radial velocity and photometric datasets from K2 and TESS. This method allowed us to model the stellar activity signal and derive precise planetary parameters. We determined planetary masses of $M_b = 4.28_{-0.33}^{+0.35}$~M${_\oplus}$, $M_c = 1.86_{-0.39}^{+0.37}$~M${_\oplus}$, and $M_d = 3.02_{-0.57}^{+0.58}$~M${_\oplus}$, and orbital periods of $1.208974 \pm 0.000001$ days for planet b, $3.648103_{-0.000010}^{+0.000013}$ days for planet c, and $6.201812 \pm 0.000009$ days for planet d. We compared our results to literature values and found that our derived uncertainties for the planetary mass, period, and radial velocity amplitude are smaller than the previously determined uncertainties. We modeled the interior composition of the three planets using the machine-learning-based tool ExoMDN and conclude that GJ~9827~b and c have an Earth-like composition, whereas GJ~9827~d has an hydrogen envelope, which, together with its density, places it in the mini-Neptune regime. }
%
\maketitle
\authorrunning{}
%

\section{Introduction}
\label{introduction}

GJ~9827 is a K7V star orbited by two super-Earth planets and one mini-Neptune at periods of 1.209, 3.648, and 6.201 days corresponding to planets b, c, and d, respectively. The orbital periods of planet b and c are near the 3:1 mean motion resonance, and b and d are near a 5:1 resonance. The planets were first discovered by \cite{Niraula2017} and \cite{Rodriguez2018} analysing K2 Campaign 12 data \citep{Howell2014}. The three planets span the radius gap for small close-in planets described by \cite{Fulton2017}. They occupy the transition region from rocky to gaseous planets, which makes characterizing their interior structures and atmospheres particularly interesting. 

At the time of discovery, the system was the closest planetary system detected by K2, located at a distance of 29.652~pc from the Sun \citep{Gaia2020}. The star is relatively bright with a V magnitude of 10.25~mag \citep{Henden2016}, which makes it an excellent target for atmospheric composition studies through transmission spectroscopy with the James Webb Space Telescope \citep{Gardner2006} and the future ANDES (ArmazoNes high Dispersion Echelle Spectrograph) at the ELT \citep[Extremely Large Telescope,][]{Marconi2022}. However, a robust mass measurement is crucial to retrieve accurate molecular abundances in the planet's atmosphere. The strength of atmospheric features depends on the scale height of the atmosphere which, in turn, depends on the surface gravity of the planet and, therefore, its mass and radius \citep{Kreidberg2018}.
Precise masses and radii are also important for estimations on the interior composition of the planets and the evolution of the planetary system. 
So far, literature determinations of the planetary masses have shown large variations, ranging from 3.42~M$_{\oplus}$ \citep{Rodriguez2018} to 8.16~M$_{\oplus}$ \citep{Teske2018} for planet b, 0.67~M$_{\oplus}$ \citep{Rice2019} to 2.56 \citep{Teske2018} for planet c, and 2.35~M$_{\oplus}$ \citep{PrietoArranz2018} to 5.58~M$_{\oplus}$ \citep{Teske2018} for planet d, with uncertainties as large as 100\%. 
Based on their results, \cite{PrietoArranz2018} and \cite{Rice2019} argue that the two inner planets have high densities, whereas the outer planet has a low density, which suggests that during the evolution of the system migration or photoevaporation could have been important factors. 

In an attempt to better constrain the masses and other planetary parameters, we used 54 observations obtained by the \'Echelle SPectrograph for Rocky Exoplanets and Stable Spectroscopic Observations (ESPRESSO) through the Guaranteed Time Observation (GTO) consortium (Program ID 1102.C-744, 1104.C-0350, PI: F. Pepe) between October 2018 and December 2019. These observations are part of the GTO subprogram focusing on the radial velocity follow-up of K2 and TESS (Transiting Exoplanet Survey Satellite) transiting candidates \citep{Damasso2020,Toledo2020,Sozzetti2021,Mortier2020,Demangeon2021,Lavie2023,Damasso2023,Castro-Gonzalez2023}.
We complement them with published measurements from HARPS (High Accuracy Radial velocity Planet Searcher), HARPS-N, and HIRES (High Resolution Echelle Spectrometer) and carried out a combined analysis with K2 and TESS photometry using Gaussian process (GP) regression to model stellar activity and planetary orbits simultaneously. In the following we provide a description of the data in Section~\ref{sec:observations} and a discussion on previous analyses of the planetary system in the literature in Section~\ref{sec:star}. We also explain our GP method in Section~\ref{sec:analysis} and present our results together with a literature comparison in Section~\ref{sec:results}.  

\section{Observations}
\label{sec:observations}

\subsection{Spectroscopic observations}

\begin{figure}[ht]
\centering
\includegraphics[width=0.5\textwidth]{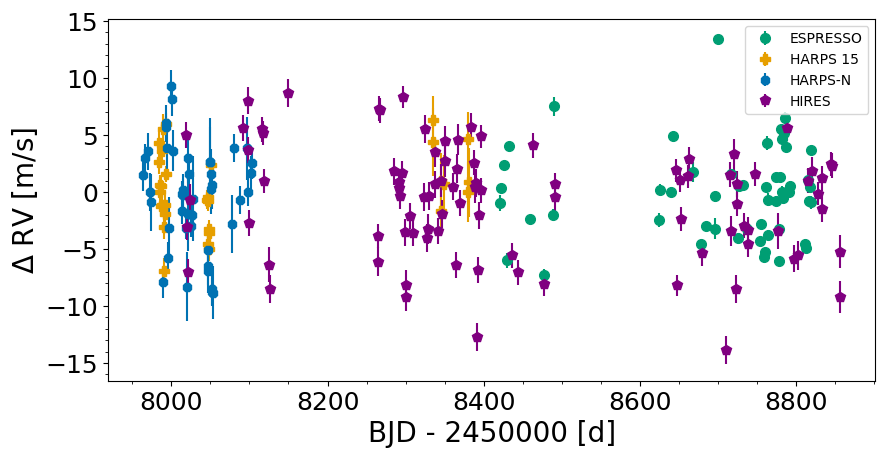}
\caption{Time-series observations of all RV measurements used in this work.}
\label{fig:RV-data}
\end{figure}

\begin{figure}[ht]
\centering
\includegraphics[width=0.5\textwidth]{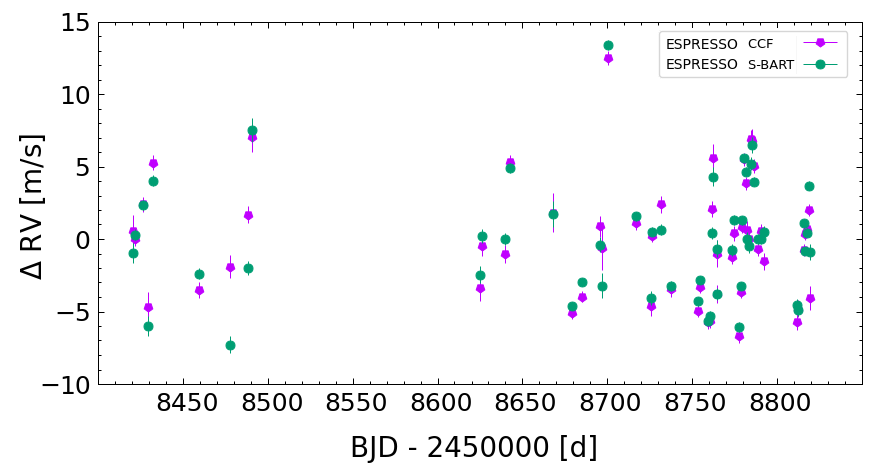}
\caption{Comparison between S-BART- and CCF-extracted RVs for ESPRESSO DRS.}
\label{fig:ESP_sbart-drs}
\end{figure}

We use radial velocity (RV) measurements from ESPRESSO, HARPS, HARPS-N, and HIRES and present all measurements in Fig.~\ref{fig:RV-data}.
ESPRESSO is a high-resolution fibre-fed echelle spectrograph mounted at the ESO Very Large Telescope (VLT) array on Paranal, Chile \citep{Pepe2013,Pepe2021}. The instrument operates in a spectral range from 380 to 788~nm with a resolution of R$\approx$140,000. A Fabry P\'erot Etalon is used for simultaneous calibrations, allowing an RV precision of down to 10~cm~s$^{-1}$ to be reached \citep{Wildi2010}. The observations of GJ~9827 were collected between October 29, 2018, and December 2, 2019, during the GTO using HR11 observing mode. The data were reduced using the ESPRESSO data reduction pipeline (DRS) version 3.0.0.
We compared the RVs extracted by the ESPRESSO DRS, which uses a cross-correlation function (CCF), to RVs extracted with the S-BART (Semi-Bayesian Approach for RVs with Template matching) pipeline \citep{Silva2022}, using the template matching technique within a semi-Bayesian framework. 
Figure~\ref{fig:ESP_sbart-drs} shows a comparison between S-BART RVs and RVs from the ESPRESSO DRS. Although the root mean square (rms) for both sets is very similar (4.042~m~s$^{-1}$ for S-BART and 4.003~m~s$^{-1}$ for CCF), the errors from S-BART are significantly smaller. On average, S-BART provides errors of about 0.426~m~s$^{-1}$, whereas the CCF gives errors of 0.617~m~s$^{-1}$. Therefore, we decided to use the S-BART RVs. After removing three measurements with a low exposure time ($< 900$~s), 54 measurements remained for our analysis. 

HARPS \citep{Mayor2003} is mounted at the ESO 3.6~m telescope at La Silla Observatory, Chile. This high-resolution fiber-fed echelle spectrograph observes in a wavelength range between 380 and 690 nm with a resolution of R$\approx$115,000. The instrument has demonstrated a long-term precision of about 0.8~ms$^{-1}$ \citep[e.g.,][]{Dumusque2012}, and after the installation of a laser-frequency comb in 2018 the precision is expected to improve down to 0.5~ms$^{-1}$ \citep{LoCurto2015,Coffinet2019}. HARPS was involved in the discovery of numerous exoplanets in the southern hemisphere \citep[e.g.,][]{Mayor2011,Astudillo2017,Unger2021}. 
We used data obtained under programs 099.C-0491(A), 0100.C-0808(A) between August 19 and October 24, 2017. We also included out-of-transit measurements from program 0101.C-0788, which were obtained between August 4 and December 19, 2018. These observations were taken with a shorter exposure time than the rest of the HARPS measurements, leading to higher photon noise and larger uncertainties. As for ESPRESSO, we extracted the RVs with S-BART, which leads to an average error of 1.34~m~s$^{-1}$ compared to 2.14~m~s$^{-1}$ from the CCF. 
To reduce the effect of short-term stellar variability, we binned data points obtained within a time span of three hours using a weighted average. For the ESPRESSO data this step was not necessary, since all measurements lie more than three hours apart. We ended up with 31 measurements from HARPS.

The HARPS-N spectrograph \citep{Cosentino2012} is located in the northern hemisphere at the 3.6~m Telescopio Nazionale Galileo (TNG) at the Observatorio del Roque de los Muchachos in La Palma, Spain. Having the same instrumental specifications as HARPS, it can also reach an RV stability of better than 1~ms$^{-1}$. 
We used data published in literature from \cite{PrietoArranz2018} and \cite{Rice2019}, which have average errors of 1.63 and 1.87~ms$^{-1}$, respectively. Both works extracted the RVs using a cross-correlation function with a K5-dwarf mask. Applying the same binning criterion as for HARPS, we had 44 measurements from HARPS-N observed from July 29 to December 15, 2017. 

Lastly, we added 92 measurements from \cite{Kosiarek2021} obtained with HIRES \citep{Vogt1994} mounted at the Keck I telescope on Maunakea, Hawai'i. HIRES is a grating cross-dispersed echelle spectrograph operating between 300 and 1000~nm with an resolution of up to 85,000. 
The observations were taken between September 22, 2017, and January 8, 2020. The average error of the measurements is 1.17~m~s$^{-1}$. 

\subsection{Photometric observations}
\label{sec:photometry}

\begin{figure}[!ht]
\centering
\includegraphics[width=0.49\textwidth]{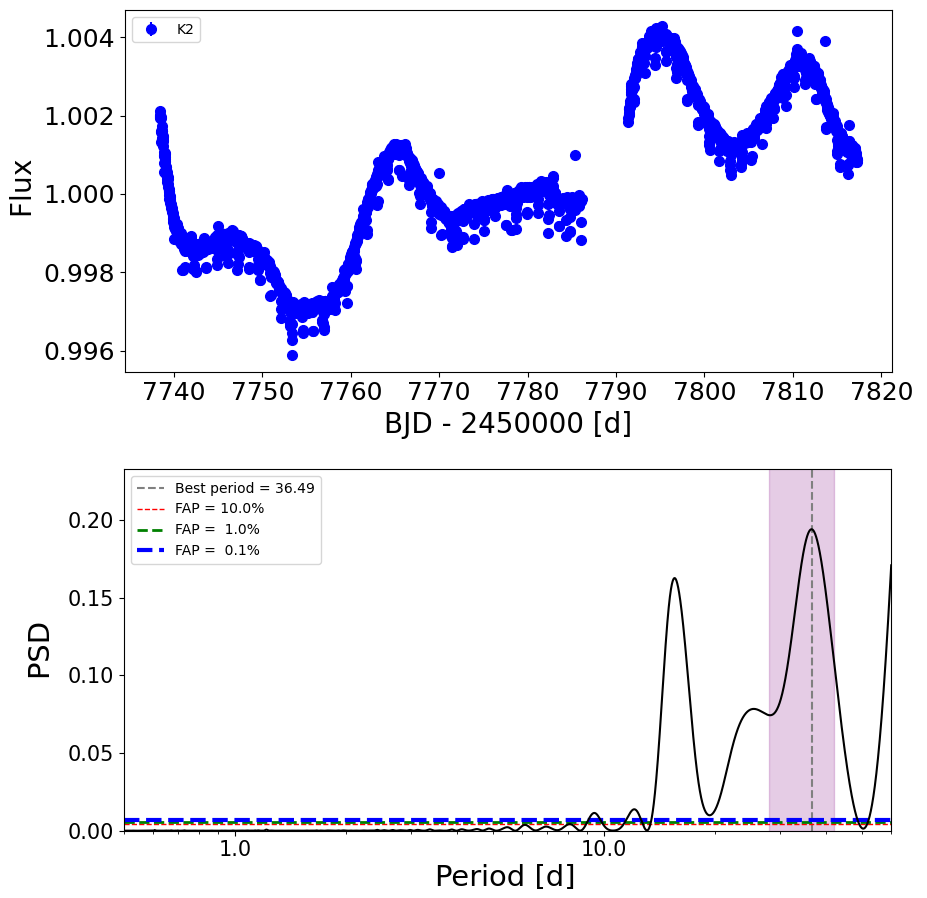}
\caption{K2 light curve (top) and corresponding GLS periodogram (bottom). The purple area indicates the approximate rotation period of the star, with the gray line presenting the highest period peak found by GLS, which corresponds to the rotation period of the star. }
\label{fig:K2}
\end{figure}

\begin{figure*}[!ht]
\centering
\includegraphics[width=0.9\textwidth]{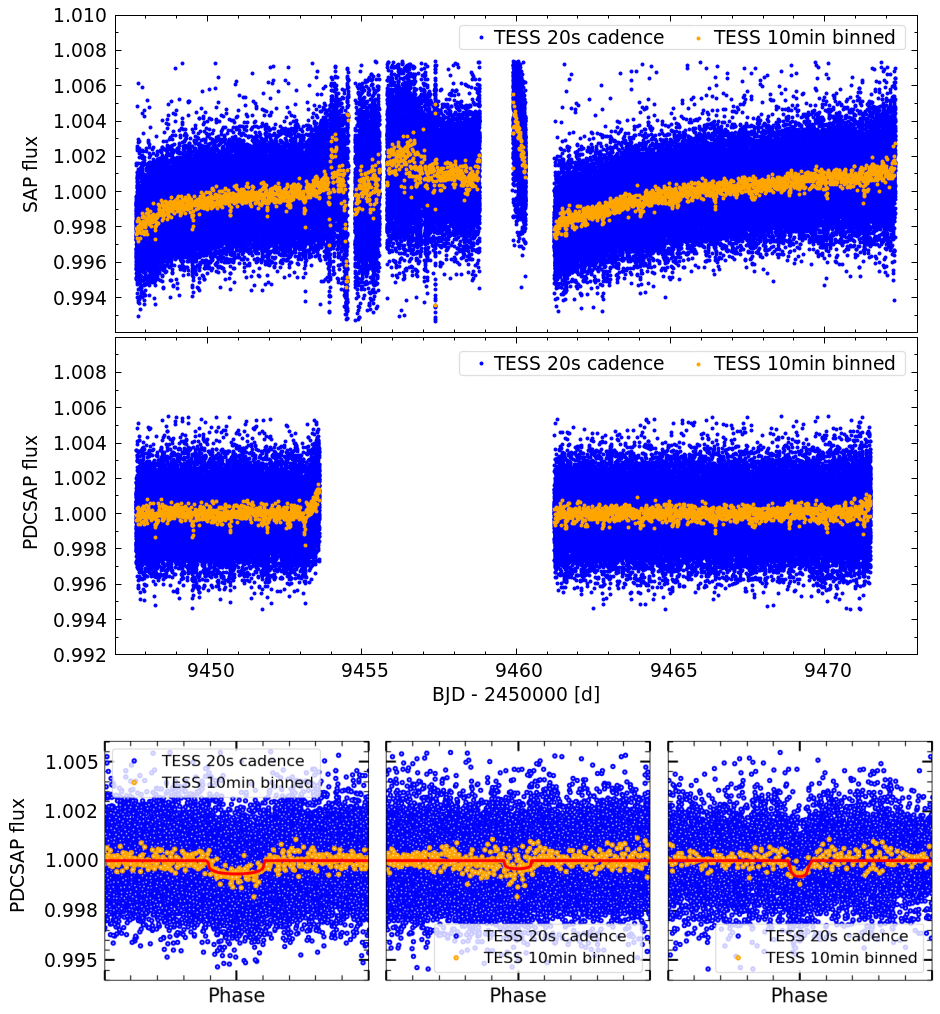}
\caption{TESS light curve SAP (top panel) and PDCSAP flux (middle panel) with original cadence (blue) and binned to 10 minutes (orange). The bottom panel shows phase folds according to planetary periods and mid-transit times derived by \cite{Rodriguez2018} for planet b (left), planet c (middle), and planet d (right). The red line presents a transit model based on parameters from \cite{Rodriguez2018} for illustrative purpose.}
\label{fig:TESS}
\end{figure*}

After the failure of two of its four reaction wheels in May 2013, the Kepler spacecraft \citep{Borucki2010} started its extended K2 mission and observed several fields near the ecliptic to obtain high-precision photometry, each field being observed for about 80~days at a time \citep{Howell2014}. 
GJ~9827 was monitored during K2 Campaign 12 from December 16, 2016, to March 4, 2017, under the identifier EPIC 246389858. K2 collected 3379 data points with a cadence of about 30 minutes. We used the light curve produced by the EVEREST (EPIC Variability Extraction and Removal for Exoplanet Science Targets) K2 pipeline \citep{Luger2016,Luger2018}, more specifically, the PDCSAP flux (Presearch Data Conditioning Simple Aperture Photometry). Each data point has a precision of about 20~ppm. The light curve and corresponding GLS periodogram are presented in Fig.~\ref{fig:K2}. The rotation period of the star clearly shows as highest peak in the periodogram with the second highest peak corresponding to half the rotation period. 
After subtracting the offset, linear trend, and stellar activity contribution to the light curve using a Gaussian Process with a simple harmonic oscillator (SHO) kernel (see Sect.~\ref{sec:activity}), we measured a standard deviation of 15~ppm over the whole light curve. We adopted this value as jitter term in the combined model. 


GJ~9827 (TIC 301289516) was observed by TESS \citep{Ricker2014} in sector 42 from August 21 to September 13, 2021. The sector was processed by the Science Processing Operations Center pipeline
\citep[SPOC,][]{Jenkins2016}, which searches for transit-like signatures using an adaptive, wavelet-based transit detection algorithm \citep{Jenkins2002,Jenkins2010}. In the process, planet b (TOI 4517.01) was identified with an period of 1.2089 $\pm$ 0.0001~days, T0=2459448.316 $\pm$ 0.001~days, and a radius of 1.62 $\pm$ 1.39~R$_{\oplus}$. 
The observations were performed in high-cadence with a data point every 20 seconds. To reduce high-frequency noise and computing time, which scales linearly to the number of points, we binned the data to 10 minutes. We measured a standard deviation of 32~ppm in the binned PDCSAP flux, and as for K2, we adopted this value as jitter term for this dataset. 
Figure~\ref{fig:TESS} presents the TESS light curve, showing SAP and PDCSAP flux, both in original cadence and binned to 10 minutes, as well as the phasefolds for all three planets according to periods derived from K2 \citep{Rodriguez2018} for illustrative purposes. 

\section{Current and previous analyses of system properties}
\label{sec:star}

\subsection{Stellar parameters from this work}
\label{sec:param_this_work}

\begin{table}[]
\caption{Summary of stellar properties for GJ~9827.}
\label{tab:parameters}
\centering 
\renewcommand{\arraystretch}{1.4}
\begin{tabular}{l cc}
    \hline 
    \hline 
    \noalign{\smallskip}
    Parameter & GJ~9827 & Ref. \\
\noalign{\smallskip}
\hline
\noalign{\smallskip}
RA (J2000) &  23:27:04.84 & 1 \\
DE (J2000) & -01:17:10.58 & 1 \\
$\mu_{RA}$ [mas yr$^{-1}$] & 375.977 $\pm$ 0.018 & 1 \\
$\mu_{DE}$ [mas yr$^{-1}$] & 215.870 $\pm$ 0.012 & 1 \\
$\pi$ [mas] & 33.7247 $\pm$ 0.0169 & 1 \\
$d$ [pc] & 29.6519 $\pm$ 0.0148 & 1 \\
B [mag] & $11.569 \pm 0.010$ & 2 \\
V [mag] & $10.250 \pm 0.138$ & 3 \\
J [mag] & $7.984 \pm 0.020$ & 4 \\
Spectral type & K7V & 5 \\
$T_{\rm eff}$ [K] & $4236 \pm 12$ & 0 \\
$[$Fe/H$]$ [dex] & $-0.29 \pm 0.03$ & 0 \\
$\log g_{\rm spec}$ [dex] & $4.70 \pm 0.05$ & 0 \\
$\log g_{\rm para}$ [dex] & $4.719 \pm 0.013$ & 0 \\
Age [Gyr] & $ 5.465 \pm 4.058$ & 0 \\
$\log{R'_\textrm{HK}}$ & $-5.28 \pm 0.05$ & 0 \\
$v \sin{i}$ [km~s$^{-1}$] & $< 1.75$ & 0 \\
$M$ [M$_{\odot}$] & $0.62 \pm 0.04$ &  0 \\
$R$ [R$_{\odot}$] & $0.58 \pm 0.03$ &  0 \\
$P_\textrm{rot}$ [d] & $28.16_{-2.66}^{+3.38}$ & 0 \\
 
\noalign{\smallskip}
\hline

\end{tabular}
\tablebib{
0 - This work, 1 - \cite[][, DR3]{Gaia2022}, 2 - \cite{Zacharias2012}, 3 - \cite{Henden2016}, 4 - \cite{Cutri2003}, 5 - \cite{Dressing2019}, 6 - \cite{Houdebine2017}
   }
\end{table}

Table~\ref{tab:parameters} summarizes the stellar properties of GJ~9827. 
We derived stellar parameters $T_{\rm eff}$, $\log g$, $[$Fe/H$]$, and the total line-broadening velocity, $v_{\rm broad}$, using a high-S/N template generated from the ESPRESSO spectra and the {\tt SteParSyn} code\footnote{https://github.com/hmtabernero/SteParSyn/} described in \cite{Tabernero2022}. The instrumental broadening $v_{\rm ins} \sim 2.1$~km~s$^{-1}$ 
($R\sim 140,000$) was also taken into account as a fixed parameter. From this, we determined $T_{\rm eff} = 4236 \pm 12$~K, $\log g_{\rm spec} = 4.70 \pm 0.10$~dex and $[$Fe/H$] = -0.29 \pm 0.03$~dex, which are in good agreement with stellar parameters determined in previous studies, for example, \cite{Niraula2017} ($T_{\rm eff} = 4255 \pm 110$~K, $[$Fe/H$]$ = $-0.28 \pm 0.12$~dex), \cite{Rodriguez2018} ($T_{\rm eff} = 4269^{+98}_{-99}$~K), \cite{PrietoArranz2018} ($T_{\rm eff} = 4219 \pm 70$~K, $[$Fe/H$]$ = $-0.29 \pm 0.12$~dex), \cite{Rice2019} ($T_{\rm eff} = 4340^{+48}_{-53}$~K, $[$Fe/H$]$ = $-0.26 \pm 0.09$~dex), and \cite{Kosiarek2021} ($T_{\rm eff} = 4294 \pm 52$~K, $[$Fe/H$]$ = $-0.26 \pm 0.08$~dex). 
We derived a $v_{\rm broad} = 1.75 \pm 0.03$~km~s$^{-1}$. All uncertainties reported here are statistical errors coming only from the {\tt SteParSyn} computation and do not account for any systematic errors. We disentangled the broadening velocity into macroturbulent velocity $v_{\rm mac} = 1.418 \pm 0.003$~km~s$^{-1}$ and projected rotational velocity $v \sin{i} = 1.02 \pm 0.05$~km~s$^{-1}$ using Equation 2 from \cite{Brewer2016} and the assumption that $v_{\rm broad}^2 = (v \sin{i})^2 + v_{\rm mac}^2$. The stated uncertainties are probably underestimated due to the fact that they involve only statistical errors. Additionally, the temperature of GJ~9827 lies outside the calibration range used by \cite{Brewer2016} to derive Equation 2. However, since $v_{\rm broad}$ is rather low, we expected $v \sin{i}$ to be low as well and assumed an upper limit of $v \sin{i} < 1.75$~km~s$^{-1}$.

We put our derived $T_{\rm eff}$ and $[$Fe/H$]$ into PARAM1.3\footnote{http://stev.oapd.inaf.it/cgi-bin/param{\_}1.3} \citep{daSilva2006} together with the parallax from $Gaia$ DR3 \cite{Gaia2022} and $V_{mag}$ from \cite{Henden2016} to compute the stellar age, mass, and radius using PARSEC isochrones version 1.1 \citep{Bressan2012}. \cite{Marfil2021} present a systematic offset in $T_{\rm eff}$ compared to interferometric measurements of 72~K. As done by \cite{Murgas2023}, we added this systematic error to our temperature uncertainty of 12~K to obtain a more conservative error estimation, therefore using an uncertainty of 84~K in PARAM1.3. For [Fe/H], we doubled the uncertainty provided by {\tt SteParSyn}. 
As a result we got an age of $5.465 \pm 4.058$~Gyr, a mass of $0.592 \pm 0.012$~M$_{\odot}$, and a radius of $0.542 \pm 0.012$~R$_{\odot}$. We note that the stellar age is not very well defined. A recent work by \cite{Engle2023} presents an age of $2.92 \pm 0.60$~Gyr using an age-rotation relation, however, \cite{Rice2019} claim 5--10~Gyr from an isochrone analysis, and \cite{Kosiarek2021} derive $6.05^{+2.50}_{-2.89}$~Gyr from an MCMC analysis. 
We used these values of stellar mass and radius together with other literature values summarized in Table~\ref{tab:all_literature} to calculate a weighted average that served as a prior in our GP analysis. The width of the priors is defined by the uncertainties added in quadrature. The posterior values for mass and radius from this analysis are presented in Table~\ref{tab:parameters}.

We also derived the stellar age, mass, and radius using the 2-step method of PARAM1.5\footnote{http://stev.oapd.inaf.it/cgi-bin/param}. Therefore, we input the same values as above for $T_{\rm eff}$, $[$Fe/H$]$, and parallax, as well as our spectroscopically derived $\log g$ with twice the uncertainty determined by {\tt SteParSyn}, and the 2MASS magnitudes J = $7.984 \pm 0.02$ mag, H = $7.379 \pm 0.042$ mag, and K$_{\rm s} = 7.193 \pm 0.024$. The result is very similar to the output of PARAM1.3, with an age of $6.524^{+4.760}_{-4.485}$~Gyr, $M = 0.586 \pm 0.02$~M$_{\odot}$, $R = 0.582^{+0.018}_{-0.019}$~R$_{\odot}$, and $\log g = 4.675^{+0.019}_{-0.018}$. 

We would like to point out here that stellar mass and radius derived with PARAM1.3 significantly depends on the value of V$_{\rm mag}$ used. On the other hand, these parameters are less sensitive to changes in the parallax. This is demonstrated in Table~\ref{tab:PARAM_test}, where we tested two different V$_{\rm mag}$ (V$_{\rm mag} = 10.101 \pm 0.001$~mag from UCAC4 \citep{Zacharias2012} and V$_{\rm mag} = 10.250 \pm 0.138$~mag from APASS DR9 \citep{Henden2016}) and two different parallaxes ($\pi = 33.7247 \pm 0.00169$~mas from $Gaia$ DR3 \citep{Gaia2022} and the zero-point corrected value of $\pi = 33.762 \pm 0.019$~mas from $Gaia$ eDR3 \citep{Kervella2022}) with PARAM1.3. 
We decided to use the magnitude from \cite{Henden2016} and the corresponding stellar parameters because of the more conservative error compared to \cite{Zacharias2012}.

\begin{table}[]
\caption{Stellar parameters derived with PARAM1.3 using different input parallaxes and magnitudes.}
\label{tab:PARAM_test}
\centering 
\renewcommand{\arraystretch}{1.4}
\begin{tabular}{ll}
    \hline 
    \hline 
    \noalign{\smallskip}
    Input & Output \\
\noalign{\smallskip}
\hline
\noalign{\smallskip}
$^1\pi = 33.7247 \pm 0.00169$~mas & age = $5.165 \pm 3.920$~Gyr \\
$^3$V$_{\rm mag}$ = $10.101 \pm 0.001$~mag & M = $0.610 \pm 0.012$~M$_{\odot}$ \\
                            & R = $0.558 \pm 0.012$~R$_{\odot}$ \\
                            & $\log g = 4.701 \pm 0.013$~dex \\
\hline
$^1\pi = 33.7247 \pm 0.00169$~mas & age = $5.465 \pm 4.058$~Gyr \\
$^4$V$_{\rm mag}$ = $10.250 \pm 0.138$~mag & M = $0.592 \pm 0.012$~M$_{\odot}$ \\
                            & R = $0.542 \pm 0.012$~R$_{\odot}$ \\
                            & $\log g = 4.719 \pm 0.013$~dex \\
\hline
$^2\pi = 33.762 \pm 0.019$~mas & age = $5.152 \pm 4.095$~Gyr \\
$^3$V$_{\rm mag}$ = $10.101 \pm 0.001$~mag & M = $0.609 \pm 0.013$~M$_{\odot}$ \\
                            & R = $0.558 \pm 0.012$~R$_{\odot}$ \\
                            & $\log g = 4.701 \pm 0.014$~dex \\
\hline
$^2\pi = 33.762 \pm 0.019$~mas & age = $5.247 \pm 4.083$~Gyr \\
$^4$V$_{\rm mag}$ = $10.250 \pm 0.138$~mag & M = $0.593 \pm 0.012$~M$_{\odot}$ \\
                            & R = $0.542 \pm 0.012$~R$_{\odot}$ \\
                            & $\log g = 4.719 \pm 0.013$~dex \\
\noalign{\smallskip}
\hline

\end{tabular}
\tablefoot{Input: $T_{\rm eff} = 4236 \pm 70$~K and $[$Fe/H$] = -0.29 \pm 0.06$~dex for all cases. 1 - \cite[][, DR3]{Gaia2022}, 2 - \cite[][eDR3-ZPcorr]{Kervella2022}, 3 - \cite{Zacharias2012}, 4 - \cite{Henden2016}
   }
\end{table}

\subsection{Planetary parameters from the literature}

\cite{Niraula2017} first detected the transits using a Box Least-Squared \citep[BLS,][]{Kovacs2002} search. Then, they fitted all transits simultaneously with the {\tt batman} model \citep{Kreidberg2015} and derived uncertainties on the fitted parameters with the Markov-Chain-Monte-Carlo (MCMC) method implemented in emcee \citep{ForemanMackey2013}. This results in radii for all three planets of 1.75 $\pm$ 0.18~R$_{\oplus}$, 1.36 $\pm$ 0.14~R$_{\oplus}$, and 2.11$_{–0.21}^{+0.22}$~R$_{\oplus}$, respectively. 
They also collected 7 high-resolution spectra with the FIbre-fed \'Echelle Spectrograph \citep[FIES,][]{Frandsen1999,Telting2014} at the Nordic Optical Telescope on La Palma, Spain. No significant RV variation is found, but they used the coadded spectrum to derive stellar parameters of the host star. 

Almost at the same time, \cite{Rodriguez2018} announced their discovery of the GJ~9827 system, also from K2 data. They modeled the system using EXOFASTv2 \citep{Eastman2017,Eastman2019}, an MCMC algorithm to simultaneously fit multiple planets. With this, they derived planetary radii of 1.62 $\pm$ 0.11~R$_{\oplus}$, 1.269$_{-0.089}^{+0.087}$~R$_{\oplus}$, and 2.07 $\pm$ 0.14~R$_{\oplus}$, for planets b, c, and d, respectively. Within the global model, EXOFASTv2 can also estimate the planetary masses using the mass-radius relation from \cite{ChenKipping2017}, resulting in values of 3.42$_{-0.76}^{+1.2}$~M$_{\oplus}$, 2.42$_{-0.49}^{+0.75}$~M$_{\oplus}$, and 5.2$_{-1.2}^{+1.8}$~M$_{\oplus}$, respectively. 

The first mass measurement from RV data was done by \cite{Teske2018}, using 36 high-resolution spectra from the Carnegie Planet Finder Spectrograph \citep[PFS,][]{Crane2006,Crane2008,Crane2010} on Magellan II at Las Campanas Observatory, Chile, observed between January 2010 and August 2016. They derived planetary masses using two methods, SYSTEMIC \citep{Meschiari2009} and {\tt radvel} \citep{Fulton2018}. Both methods model Keplarian orbits using a maximum likelihood function and incorporate an MCMC method to determine median values and errors. For SYSTEMIC, \cite{Teske2018} report a mass of 7.50 $\pm$ 1.52~M$_{\oplus}$ for planet b, and upper limits of 2.56~M$_{\oplus}$ and 5.58~M$_{\oplus}$ for planets c and d, respectively. The {\tt radvel} analysis gives masses of 8.16$_{-1.54}^{+1.56}$~M$_{\oplus}$ for planet b, 2.45$_{-2.24}^{+2.20}$~M$_{\oplus}$ for planet c, and 3.93$_{-2.76}^{+2.65}$~M$_{\oplus}$ for planet d. For both methods the resulting masses are not well constrained with uncertainties up to 100\%. 

Additionally to the seven FIES spectra presented by \cite{Niraula2017}, \cite{PrietoArranz2018} obtained 35 high-resolution spectra with HARPS and 23 spectra with HARPS-N. They performed a joint analysis of the RV and photometric K2 data with the pyaneti code \citep{Barragan2017}, which incorporates an MCMC algorithm. With this, they determined for planet b $R_b =$ 1.58$_{-0.13}^{+0.14}$~R$_{\oplus}$ and $M_b =$ 3.69$_{-0.46}^{+0.48}$~M$_{\oplus}$, for planet c $R_c =$ 1.24 $\pm$ 0.11~R$_{\oplus}$ and $M_c =$ 1.45$_{-0.57}^{+0.58}$~M$_{\oplus}$, and for planet d $R_d =$ 2.04 $\pm$ 0.18~R$_{\oplus}$ and $M_d =$ 2.35$_{-0.68}^{+0.70}$~M$_{\oplus}$. Compared to the previous first mass measurements by \cite{Teske2018}, these results are more precise, however, the uncertainties still reach 40\% for planet c and about 30\% for planet d. Furthermore, the mass of planet b is significantly lower compared to \cite{Teske2018}. \cite{PrietoArranz2018} also modeled the stellar activity in the RVs with a Gaussian process and obtained similar planetary parameters with this technique. 

\cite{Rice2019} added another 43 HARPS-N measurements to the previously published HARPS and HARPS-N data from \cite{PrietoArranz2018}, PFS data \citep{Teske2018}, and FIES data \citep{Niraula2017}. 
They simultaneously modeled stellar activity and RVs with the package PyORBIT \citep{Malavolta2016}, which implements a GP quasi-periodic kernel through the GEORGE package \citep{Ambikasaran2016}. As a result, they derived $M_b =$ 4.91 $\pm$ 0.49~M$_{\oplus}$, $M_c =$ 0.84~M$_{\oplus}$ with an upper limit of 1.50~M$_{\oplus}$ and $M_b =$ 4.04$_{-0.84}^{+0.82}$~M$_{\oplus}$. Comparing these results to \cite{PrietoArranz2018} shows that the uncertainties do not improve and the mass of planet d is significantly larger. 

\cite{Kosiarek2021} collected 92 RV measurements with HIRES with an average uncertainty of 1.17~ms$^{-1}$. Together with already published RV data, this makes a total of 234 measurements for GJ~9827. They analyzed the RV data using {\tt radvel} and a Gaussian Process and determined masses of $M_b =$ 4.87 $\pm$ 0.37~M$_{\oplus}$, $M_c =$ 1.92 $\pm$ 0.49~M$_{\oplus}$, and $M_d =$ 3.42 $\pm$ 0.62~M$_{\oplus}$. So far, these are the best constrained values with uncertainties better than 25\%. A comprehensive collection of all planetary parameters determined in the literature is given in Table~\ref{tab:all_literature}.

\section{Analysis}
\label{sec:analysis}

\begin{figure*}[!ht]
\centering
\includegraphics[width=0.9\textwidth]{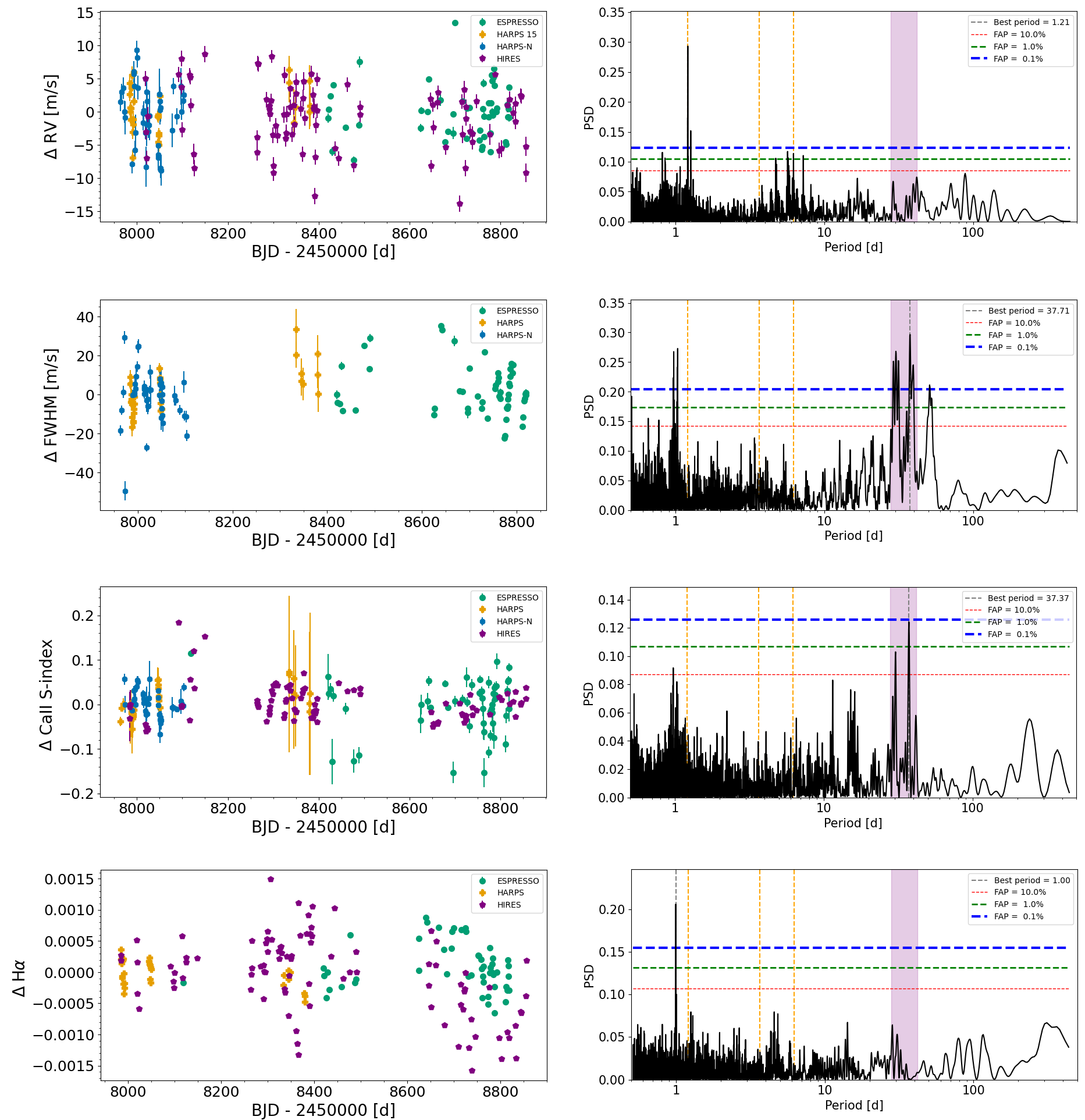}
\caption{$Left$: ESPRESSO, HARPS, HARPS-N, and HIRES spectroscopic data used in this study. We present plots for RV, FWHM, CaII S-index, and H$\alpha$. $Right$: Corresponding GLS periodograms. The orange dashed lines mark the periods of the three planets. The purple area indicates the approximate rotation period of the star. The gray line presents the highest period peak found by GLS for each panel.}
\label{fig:GLS_RV}
\end{figure*}

In this work, we performed a combined analysis of spectroscopy and photometry within the Gaussian Process regression framework \citep[GP,][]{RasmussenWilliams2006,Roberts2012}, following the approach described in \cite{Suarez2020,Suarez2023}.
Within the last decade, GPs have become very successful in characterizing and modeling stellar activity in RV time series \citep[e.g.,][]{Haywood2014}. GPs are flexible in modeling quasi-periodic signals, such as stellar rotation signals induces by active regions on the stellar surface. They are able to account for changes in the amplitude, phase, and period of the signal, however, they can also easily overfit the data and erase the planetary signal. One way to overcome this is to simultaneously model activity indicators, which should be independent of the planet-induced periodic signal, and radial velocity with shared hyper parameters \citep[e.g.,][]{Suarez2020,Faria2022}. Another possibility is to use multi-dimensional GPs as done by, for example, \cite{Rajpaul2015}, \cite{Barragan2022}, and \cite{Delisle2022}, where the whole time series is joined under a single covariance matrix. This approach is especially useful in cases where there is a good correlation between the activity indicator and the activity-induced RV signal, like in the case of Proxima Centauri \citep{Suarez2020} or GJ~1002 \citep{Suarez2023}. We did not see any correlation between RV and full-width at half maximum (FWHM) in our data (see Fig.~\ref{fig:RV-FWHM}), however, we followed the approach as described in \cite{Suarez2023}, because the correlation might still exist with the gradient as in the case of Proxima, and as described in the FF' formalism \citep{Aigrain2012}. We used the S+LEAF code \citep{Delisle2022} together with Dynesty \citep{Speagle2020}. 

S+LEAF is an open-source software, which is able to simultaneously model several time series with the same Gaussian processes and their derivatives. It provides a flexible noise model to account for instrument calibration errors \citep{Delisle2020}, and offers a number of different GP kernels with different properties.

Dynesty uses nested sampling \citep{Skilling2004,Skilling2006} to infer the Bayesian evidence of the model. It performs a random walk or random slice sampling strategy \citep{Handley2015a,Handley2015b} to explore large parameter spaces. Due to the large dataset, the number of live points was set to 10 x $n_{\rm par}$ and the number of slices was 2 x $n_{\rm par}$, where $n_{\rm par}$ is the number of free model parameters.

\subsection{Stellar activity}
\label{sec:activity}
We analyzed different activity indicators with the Generalised Lomb-Scargle periodograms \citep[GLS,][]{Zechmeister2009}. Figure~\ref{fig:GLS_RV} presents a plot of the spectroscopic data used in this study and their corresponding GLS periodograms. The RV data (top row) shows a significant peak at the period of planet b. We marked the periods of the three planets with orange dashed lines. 
The following rows present the activity indicators, FWHM, CaII S-index \citep{Vaughan1978}, and H$\alpha$. Not all indicators are available for every dataset. The FWHM shows a significant signal around the rotation period of the star (indicated by the purple region), which is determined by \cite{Rodriguez2018} with 31 $\pm$ 1~d, \cite{PrietoArranz2018} with 30.7 $\pm$ 1.4~d, \cite{Rice2019} with 28.72$^{+0.18}_{-0.22}$~d, and \cite{Kosiarek2021} with 28.62$^{+0.48}_{-0.38}$~d. \cite{Niraula2017} also find a signal around 30 days, however, they favor a period at 16.9$^{+2.14}_{-1.51}$~d. 
The CaII S-index also shows some signal around 30~days, however, less significant than in FWHM. Therefore, we used FWHM in our following analysis. 

\begin{figure}[ht]
\centering
\includegraphics[width=0.48\textwidth]{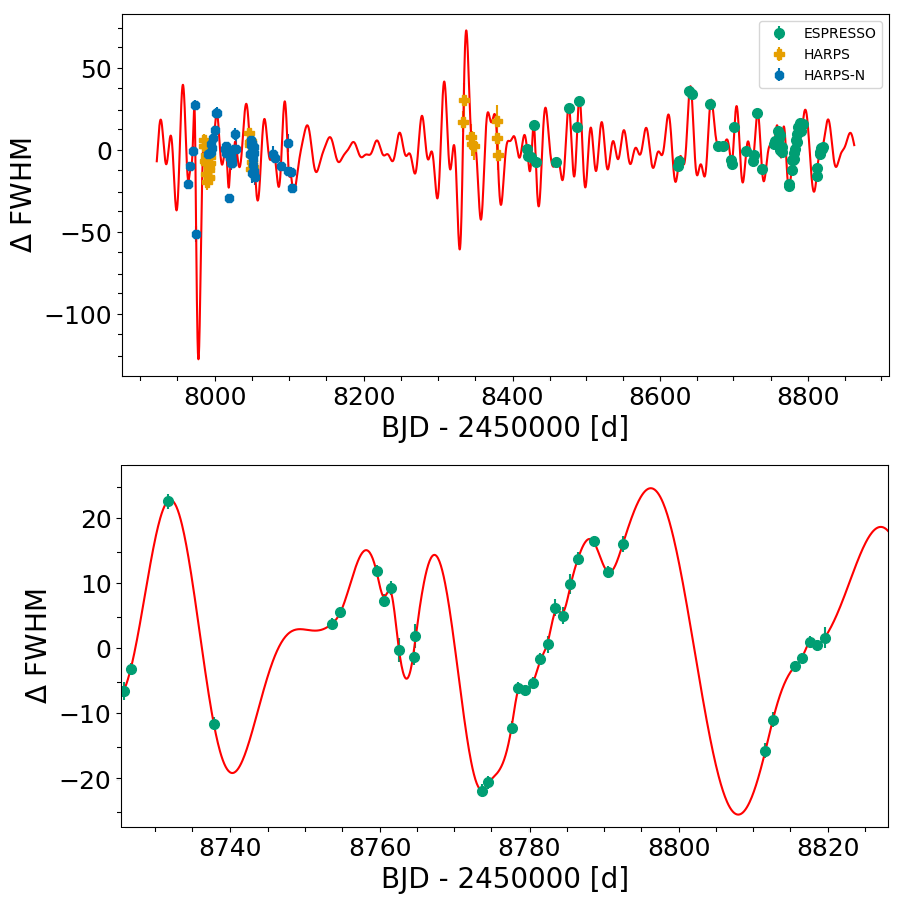}
\caption{FWHM from ESPRESSO, HARPS, and HARPS-N with best GP model. The bottom panel shows a zoom-in on ESPRESSO data. }
\label{fig:GP_FWHM}
\end{figure}

We analyzed the stellar activity by fitting a GP to the FWHM and RV data without a planetary model to determine the rotational period of the star. We used a GP with two SHO kernels at $P_{\rm rot}$ and $P_{\rm rot}/2$ with a normal distributed prior centered around the expected rotation period. The GP amplitudes were the same for all datasets and we used a log uniform jitter term. 
We found a rotation period of $29.98_{-1.06}^{+1.07}$~days. Figure~\ref{fig:GP_FWHM} shows the best model fit to the FWHM. 
We confirmed the rotation period by using the median value of the ESPRESSO S-index ($0.66 \pm 0.06$) to compute $\log{R'_\textrm{HK}}$, following \cite{Suarez2016}. We obtained a measurement of $\log{R'_\textrm{HK}} = -5.28 \pm 0.05$ and from the rotation-activity relationship of \cite{Suarez2016} we estimated a rotation period of $32.9 \pm 1.4$~d. These values are consistent with the great majority of the stellar rotation period measurements from the literature including the most recent determination by \cite{Engle2023}, $P_{\rm rot} = 28.72 \pm 0.19$~d.

Using the star's equatorial radius, $R$, from Table~\ref{tab:parameters} and $P_{\rm rot} =29.98_{-1.06}^{+1.07}$~days, we estimated a $v_{\rm rot}$ of $0.98^{+0.09}_{-0.08}$~km~s$^{-1}$. This value is consistent at the 1-$\sigma$ level with the spectroscopically derived $v \sin{i} = 1.02 \pm 0.05$~km~s$^{-1}$ in Section~\ref{sec:param_this_work} and is compliant with our conservative upper limit of $v \sin{i}$ < 1.75~km~s$^{-1}$. Due to the small value of $v_{\rm rot}$, we can assume that GJ\,9827 is likely seen near equator on with a spin axis inclination angle close to 90~deg. The knowledge of the star's spin axis angle is relevant to address the true 3D architecture of the multiple planetary system \citep[e.g.,][]{Albrecht2021}.


\subsection{Radial velocity}
\label{sec:RV-method}


After assessing the stellar activity in the spectroscopic data, we performed a GP analysis with FWHM and RVs including a three-planet model. To do that, we used \textsf{celerite2}, an update of the \textsf{celerite} package by \cite{ForemanMackey2017}, a fast and scalable 1-D GP regression. With that, stellar rotation is modeled with a SHO kernel at $P_\textrm{rot}$ and $P_\textrm{rot}$/2 and a normal distributed prior centered around the expected rotation period of the star, that is, $30 \pm 5$~d. 

We investigated circular and eccentric models and used narrow normal distributed priors for the orbital period and time of periastron T0, as derived by \cite{Rodriguez2018} and presented in Table~\ref{tab:GP_priors}. The priors for the planet amplitudes were set uniformly between 0 and 5 times the standard deviation (stdev). The priors for the offsets for each instrument were normally distributed, centered around 0 with a sigma of 3x stdev.
We did not fit any long-term cycle, but we included a linear trend with normally distributed priors around 0, with a sigma equal to the flux range peak-to peak divided by the observation baseline. 
The jitter prior is log-uniform ranging from --10 to log(5 x stdev). 
The log likelihood of the eccentric and circular model are almost indistinguishable with ln~Z$_{ecc} = -1063.89 \pm 0.39$ compared to ln~Z$_{circ} = -1063.93 \pm 0.39$ and $\Delta$~ln~Z = 0.04. Therefore, we favored the circular model. The resulting planetary parameters for the circular model are summarized in Table~\ref{tab:results_combined}. 


\subsection{Photometry}
Before combining radial velocity and transit data, we analyzed the TESS and K2 data separately. For both, we used the PDCSAP flux and estimated the jitter of the data from the standard deviation as described in Sect.~\ref{sec:photometry}. This gave a jitter term of 0.15~ppt for K2 and 0.32~ppt for TESS.  
We defined normal and log-normal distributed priors for the offsets and linear trends of both datasets, respectively. The rotation periods and timescales of evolution got log-normal priors as well, whereas the GP amplitude priors were defined by uniform log-priors. 

We tested several kernels, such as the exponential kernel, SHO kernel, quasi-periodic kernel, and the Mat{\'e}rn kernels 1/2 to 5/2 \citep{Genton2002}. We found that the Mat{\'e}rn 3/2 kernel offers the best model for the PDCSAP flux of the TESS light curve without affecting the planetary transits due to overfitting. The kernel has the form 


\begin{equation} 
\label{eq:matern32}
    k(\Delta t) = \sigma^2 \mathrm{e}^{-\sqrt{3}\frac{\Delta t}{\rho}}
\left(1 + \sqrt{3}\frac{\Delta t}{\rho}\right) ,
\end{equation}

where $\sigma$ is the amplitude of the signal and $\rho$ is the scale. The SHO kernels are best suited for modeling the stellar activity signal in the K2 light curve. 

As can be seen from Fig.~\ref{fig:TESS}, the signal of planet c is very small, even in the binned TESS data; however, it is clearly detectable in K2, which helps to guide the GP. We incorporated the PyTransit algorithm \citep{Parviainen2015} with the quadratic limb-darkening transit model from \cite{MandelAgol2002}. The linear and quadratic limb-darkening coefficients were characterized by uniform priors, and the stellar radius by a normal distributed prior centered around 0.592~R$_{\odot}$ with a width of 0.049, calculated from the weighted average of literature values (see Sect.~\ref{sec:param_this_work}). Additionally to the priors of the orbital periods and T0, which were defined in the same way as for the RVs in Sect.~\ref{sec:RV-method}, we set uniform priors for the planetary radius and impact parameter, and, for the eccentric model, normal distributed priors for the values of $\sqrt{e}\cdot \cos{\omega}$ and $\sqrt{e}\cdot \sin{\omega}$, where $e$ is the eccentricity of the orbit and $\omega$ is the argument of periapsis. 

Also here, the log likelihoods for the eccentric and circular case are almost the same, resulting in ln~Z$_{ecc} = 41457.24 \pm 0.40$ and ln~Z$_{circ} = 41457.01 \pm 0.45$ with $\Delta$~ln~Z = 0.23. Table ~\ref{tab:results_combined} presents the planetary parameters for the circular model. 
The rotational period is modeled very well from the K2 light curve and gives a rotation period of $29.23^{+3.72}_{-3.21}$~days, consistent with the value we determined from FWHM. 

\subsection{Combined fit}
\label{sec:combined}

Finally, we combined the spectroscopic and photometric datasets and ran a combined fit using S+LEAF and Dynesty. Again, we investigated the circular and eccentric case. As described above, although we combined all RV data into a single dataset, the offsets, trends, and jitter terms were allowed to be different for each instrument, to account for uncorrelated trends and noise. The same applies to FWHM and photometry. All priors and best fit parameters for eccentric and circular cases are collected in Table~\ref{tab:GP_priors}. The priors for the offsets and trends were normal distributed, whereas for the jitters we used uniform priors. We used three GPs for RV+FWHM, K2, and TESS. All GP amplitudes used uniform priors, where K2 and TESS have one amplitude each, and RV and FWHM have four amplitudes each at $P_\textrm{rot}$, $P_\textrm{rot}$/2, and their corresponding derivatives, according to 

\begin{equation}
\label{eq:GP_amplitudes}    
    \textrm{GP} = A_{11} \cdot G_1 + A_{12} \cdot \frac{\partial}{\partial t} G_1+ A_{21} \cdot G_2 + A_{22} \cdot \frac{\partial}{\partial t} G_2 .
\end{equation}

The priors for $P_\textrm{rot}$ and the timescale of evolution were normal distributed, centered around 30 and 30x2, respectively \citep[see][]{Giles2017}. 
We sampled the planet mass and radius explicitly, and transformed them into RV amplitude and transit depth within the likelihood function.
Both, the Keplerian model for RV and the transit model for photometry shared the same planetary priors. 
We accounted for the longer exposure time of K2 by using a higher supersampling rate in the K2 time series \citep[see][]{Parviainen2015}. 
The resulting log likelihoods for the circular and eccentric fit are very similar, with ln~Z$_{ecc} =40420.0682 \pm 0.3856$ and ln~Z$_{circ} = 40420.0689 \pm 0.4039$. 
Although neither the circular nor the eccentric case is favored by the model, and the RV rms is slightly larger for ESPRESSO, HARPS, and HARPS-N for circular orbits (rms$_{\rm ESP} = 0.30$~ms$^{-1}$, rms$_{\rm H-S} = 0.83$~ms$^{-1}$, rms$_{\rm H-N} = 1.34$~ms$^{-1}$) compared to eccentric orbits (rms$_{\rm ESP} = 0.27$~ms$^{-1}$, rms$_{\rm H-S} = 0.82$~ms$^{-1}$, rms$_{\rm H-N} = 1.31$~ms$^{-1}$), we consider the circular model in the following. It is expected that, due to the small distances between the star and the planets, the orbits have circularized over the lifetime of the system \citep{VanEylen2015}. Furthermore, the uncertainties of the majority of planetary parameters improve when considering a circular orbit (see Table~\ref{tab:GP_priors}).
For $P_\textrm{rot}$, we derived consistent periods from the radial velocity and K2 data, that is, $28.16^{+3.38}_{-2.66}$~d and $29.52_{-3.25}^{+3.42}$~d, respectively. As a final value, we adopted $P_\textrm{rot} = 28.16^{+3.38}_{-2.66}$~d (see Table~\ref{tab:parameters}).

\section{Results and discussion}
\label{sec:results}

\begin{figure*}[!ht]
\centering
\includegraphics[width=0.9\textwidth]{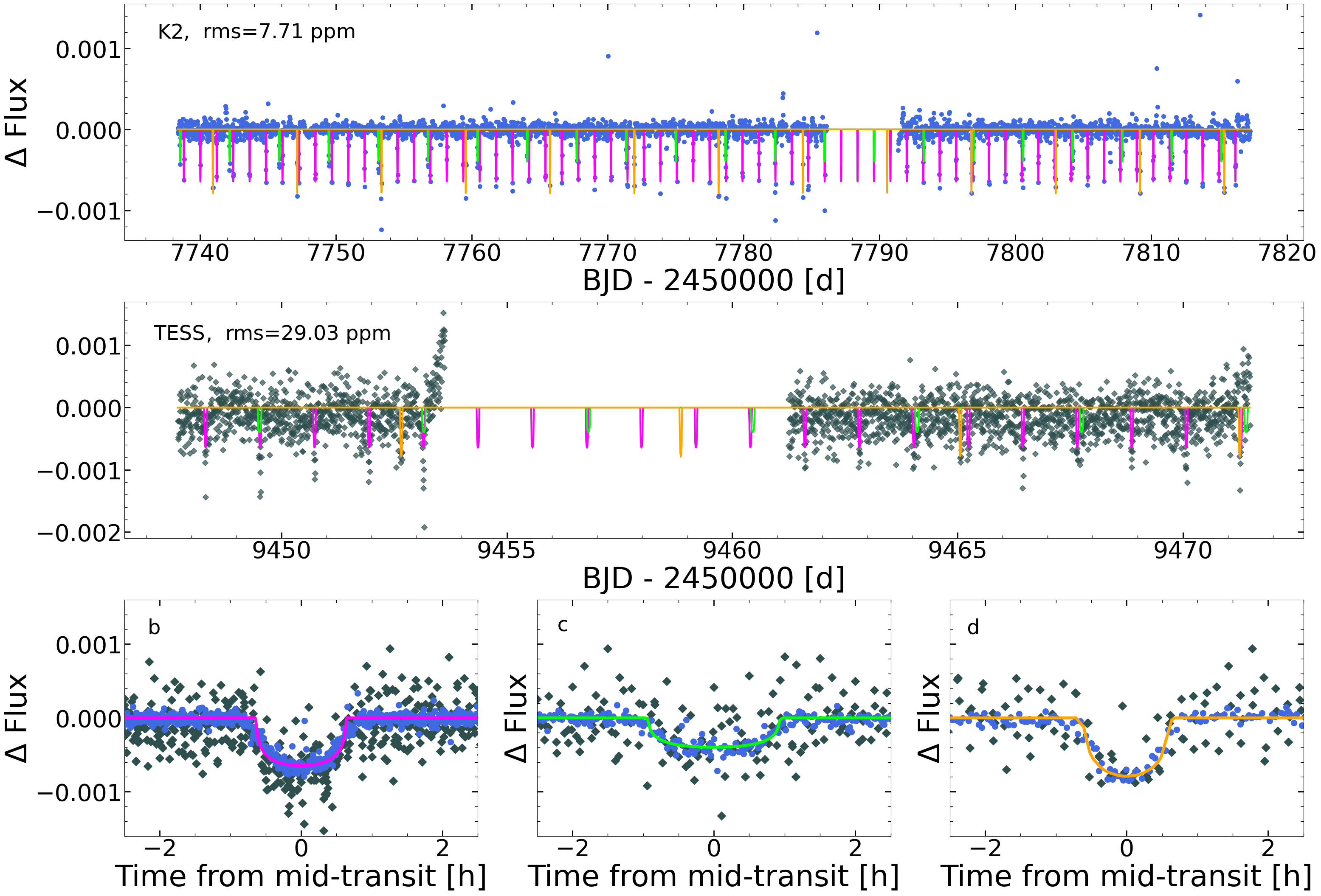}
\caption{Best-fit model from combined fit, showing K2 (top) and TESS data (middle). The three planets are illustrated with different colors. The bottom panel presents phase folds of K2 and TESS data with the transit model.}
\label{fig:K2+TESS_fit}
\end{figure*}

\begin{figure*}[!ht]
\centering
\includegraphics[width=0.90\textwidth]{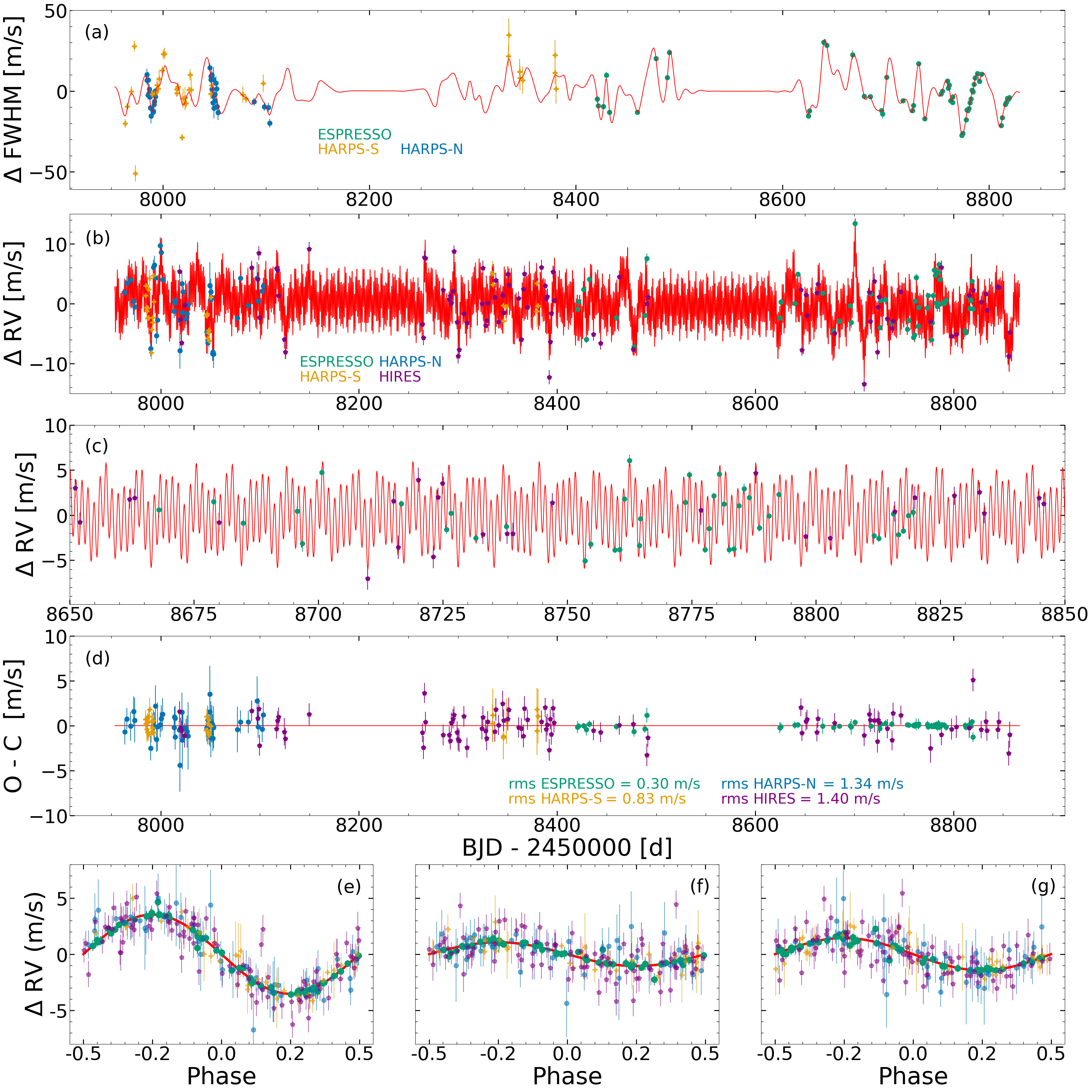}
\caption{Best-fit circular model from combined fit. (a) FWHM with activity model. (b) RV with combined activity and eccentric model. (c) Zoom-in of panel (b). (d) RV residuals after subtraction of activity and planet model. (e)--(g) Phase folds of RVs for all three planets. }
\label{fig:spec_fit}
\end{figure*}

\begin{table*}[htb]
\caption{Derived planetary parameters for GJ~9827 for the circular case.}
\label{tab:results_GJ9827}
\centering 
\renewcommand{\arraystretch}{1.4}
\begin{tabular}{l ccc}
    \hline 
    \hline 
    \noalign{\smallskip}
    Parameter & GJ~9827 b & GJ~9827 c & GJ~9827 d \\
\noalign{\smallskip}
\hline
\noalign{\smallskip}
$R_p$ [R$_{\oplus}$] &	$1.44_{-0.07}^{+0.09}$ & 		$1.13_{-0.05}^{+0.07}$ &		$1.89_{-0.14}^{+0.16}$ \\
$M_p$ [M$_{\oplus}$] &	$4.28_{-0.33}^{+0.35}$ &		$1.86_{-0.39}^{+0.37}$ &		$3.02_{-0.57}^{+0.58}$ \\
$T0_p$ [d-2450000] &		$7738.8259_{-0.0005}^{+0.0005}$ &		$7742.2000_{-0.0014}^{+0.0014}$ &		$7740.9588_{-0.0009}^{+0.0008}$ \\
$P_p$ [d]	&		$1.208974_{-0.000001}^{+0.000001}$ &		$3.648103_{-0.000010}^{+0.000013}$ &		$6.201812_{-0.000009}^{+0.000009}$ \\
$b_p$	&		$0.30_{-0.16}^{+0.14}$ &		$0.23_{-0.15}^{+0.17}$ &		$0.85_{-0.03}^{+0.02}$ \\
$i_p$ [deg] &		$87.60_{-1.27}^{+1.31}$ &		$89.09_{-0.68}^{+0.60}$ &		$87.66_{-0.16}^{+0.13}$ \\
$a_p$ [AU] &		$0.0189_{-0.0004}^{+0.0004}$ &		$0.0395_{-0.0009}^{+0.0009}$ &		$0.0563_{-0.0012}^{+0.0012}$ \\
$K_p$ [ms$^{-1}$] &	$3.53_{-0.22}^{+0.22}$ &		$1.06_{-0.21}^{+0.21}$ &		$1.44_{-0.27}^{+0.27}$ \\
$T_{eq}$ [K] & $1035.01 \pm 29.07$ & $715.94 \pm 20.33$  &	$599.68 \pm 16.86$ \\
$\rho$ [$\rho_\oplus$] & $1.43 \pm 0.27$ & $1.30 \pm 0.34$ & $0.45 \pm 0.13$ \\

\noalign{\smallskip}
\hline

\end{tabular}
\end{table*}

In this section we present our results from the combined GP fit. Figure~\ref{fig:K2+TESS_fit} shows the photometric data with the best model from the combined fit and Fig.~\ref{fig:spec_fit} shows the same model with the spectroscopic data. 
The derived parameters for each planet are presented in Table~\ref{tab:results_GJ9827} and discussed in more detail in the following. 

\subsection{Comparison with literature}

\begin{figure*}[ht]
\centering
\includegraphics[width=0.95\textwidth]{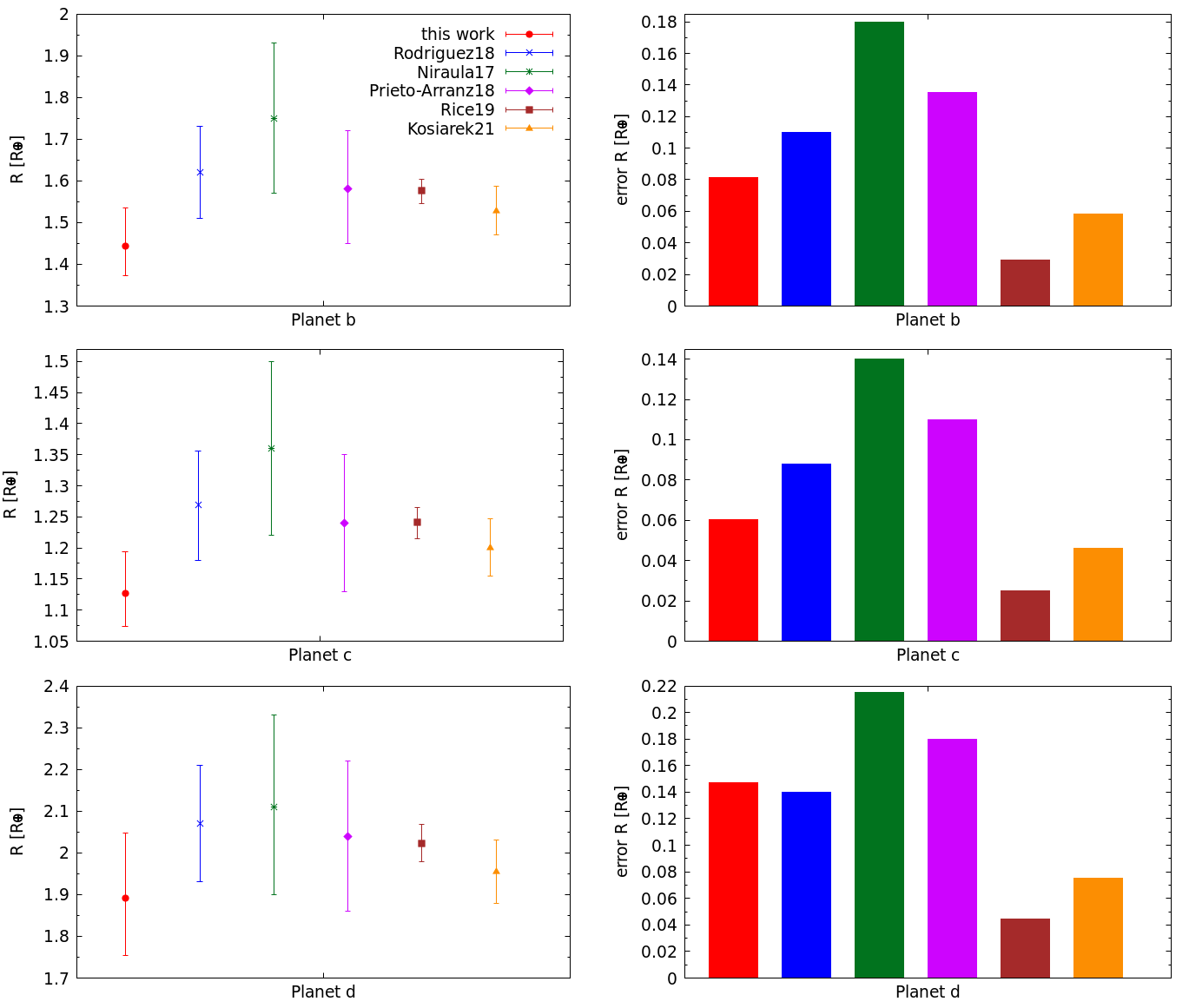}
\caption{Literature comparison for planetary radius (left column) and error (right column) for each planet.}
\label{fig:radius_lit}
\end{figure*}

\begin{figure*}[ht]
\centering
\includegraphics[width=0.95\textwidth]{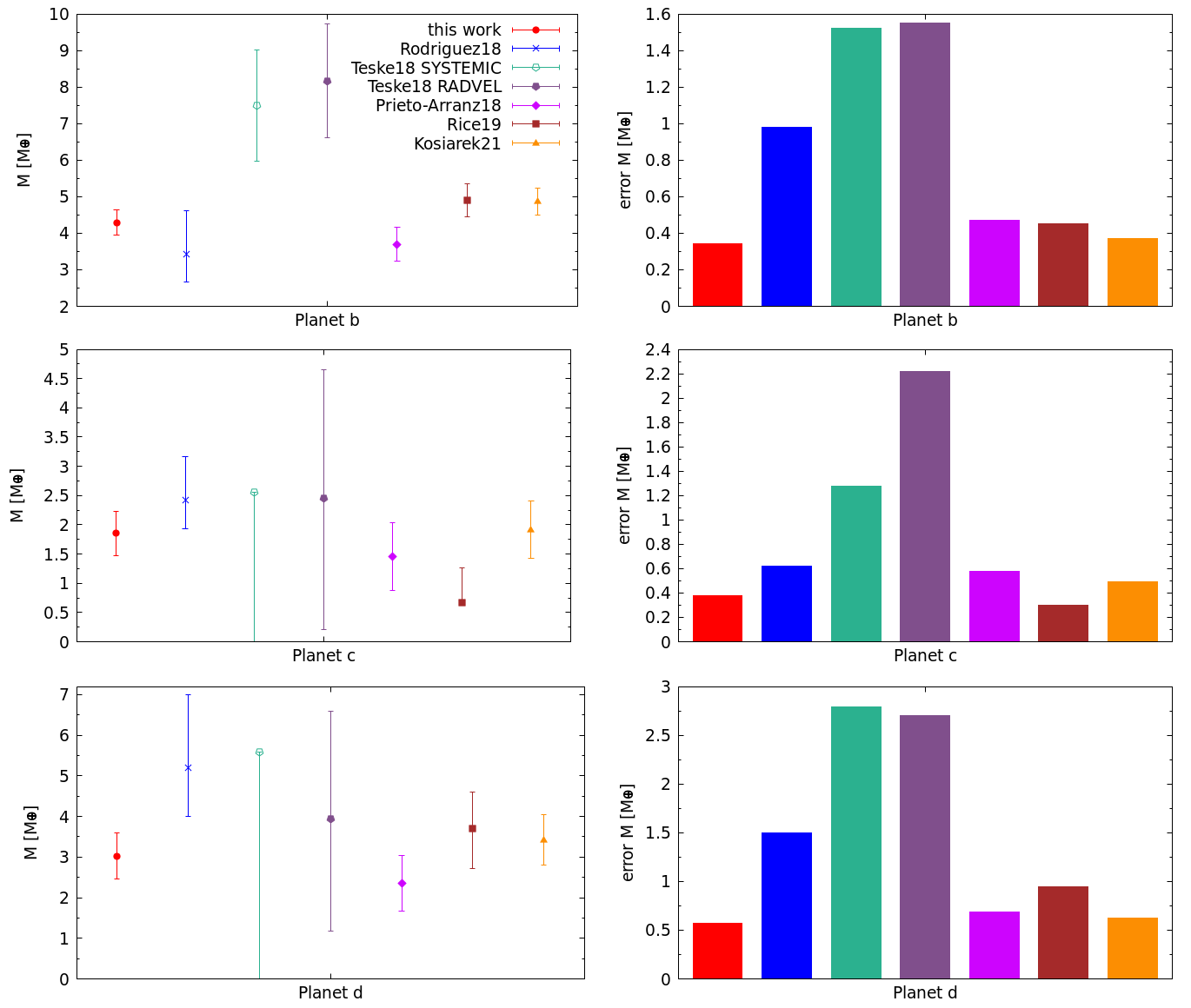}
\caption{Literature comparison for planetary mass (left column) and error (right column) for each planet.}
\label{fig:mass_lit}
\end{figure*}

\begin{figure*}[ht]
\centering
\includegraphics[width=0.95\textwidth]{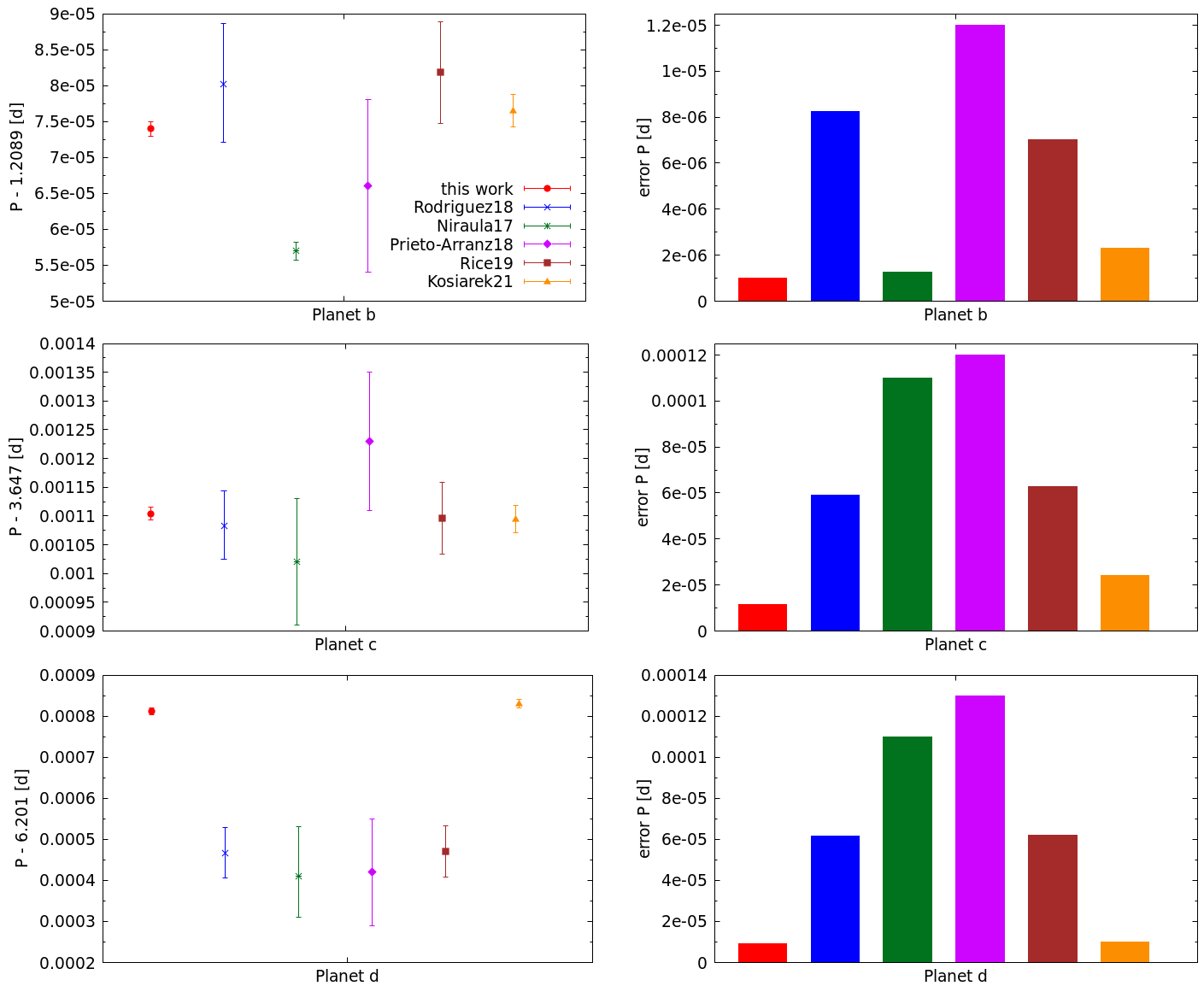}
\caption{Literature comparison for planetary period (left column) and error (right column) for each planet.}
\label{fig:period_lit}
\end{figure*}

\begin{figure*}[ht]
\centering
\includegraphics[width=0.95\textwidth]{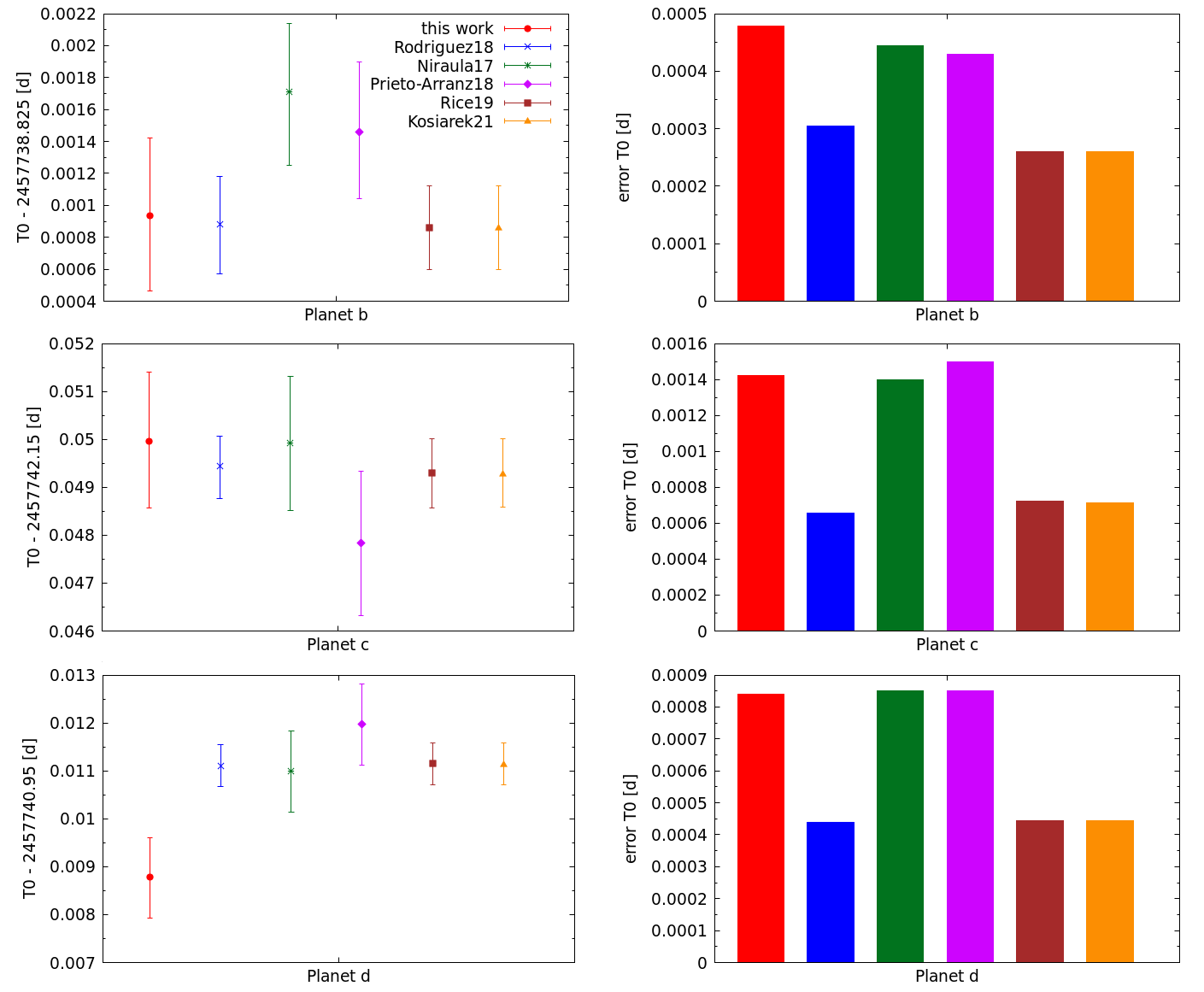}
\caption{Literature comparison for time of periastron (left column) and error (right column) for each planet. Note that for planet c we added one orbital period to T0 from \cite{Niraula2017} and \cite{PrietoArranz2018} to make their values comparable.}
\label{fig:T0_lit}
\end{figure*}

We compared our results from the combined eccentric fit to previous results in the literature, focusing on radius, mass, period, and T0, because these parameters are most critical for accurate transit spectroscopy and atmospheric characterization. 
Figure~\ref{fig:radius_lit} presents the comparison for planetary radius and shows that all measurements are consistent within their uncertainties. We derived $1.44_{-0.07}^{+0.09}$~R$_{\oplus}$ for planet b, $1.13_{-0.05}^{+0.07}$~R$_{\oplus}$ for planet c, and $1.89_{-0.14}^{+0.16}$~R$_{\oplus}$ for planet d. 
\cite{Niraula2017} determined tentatively larger radii, which is most likely attributed to the larger stellar radius they assumed (0.651 R$_{\odot}$). Although our uncertainties in radius for all three planets are less than 10\%, \cite{Rice2019} and \cite{Kosiarek2021} derived significantly smaller uncertainties, as is shown in the right column of Fig.~\ref{fig:radius_lit}. \cite{Rice2019} inferred one magnitude smaller uncertainties for the stellar radius ($0.602^{+0.005}_{-0.004}$), because they used the $isochrones$ Python package \citep{Morton2015} with $Gaia$ parallaxes as priors, which in turn influences the uncertainty in planetary radius. We let the model find the stellar mass and radius with priors based on the weighted average of all available literature values, which explains the higher, but more realistic uncertainties on the stellar parameters. \cite{Tayar2022} demonstrate that systematic uncertainties can dominate the error of stellar parameters, and errors of $\approx$ 4\% in stellar radius and $\approx$ 5\% in stellar mass are a realistic assumption. 
\cite{Kosiarek2021} analyzed the RV data from HIRES, FIES, PFS, HARPS, and HARPS-N using {\tt radvel} \citep{Fulton2018}. They determined hyperparameters for stellar activity from the K2 light curve with a quasi-periodic GP kernel, and then used these posteriors as priors to model stellar activity in RV data and derived planetary parameters. The errors were determined from an MCMC analysis. 
They derived stellar effective temperature and metallicity from HIRES spectra using the SpecMatch-Emp tool \citep{Yee2017}. Together with multi-band stellar photometry (Gaia G and Two Micron All-Sky Survey JHK) and the Gaia DR2 parallax, they input these parameters into the {\tt isoclassify} tool using the grid-mode option \citep{Huber2017} to derive stellar mass and radius. Then, they calculated the planetary radii from the radius ratios given in \cite{Rodriguez2018}. Therefore, their smaller uncertainties in planetary radii can be attributed to their smaller uncertainties in stellar radius.


To further explore our higher uncertainties in radius, we compared our results from different GP fits involving different datasets, as presented in Fig.~\ref{fig:GP_radius}. We show results derived from all datasets (i.e., ESPRESSO, HARPS, HARPS-N, HIRES, K2, TESS), from the RV datasets and K2, from K2 and TESS, and from K2 and TESS separately. In general, the values agree very well within their uncertainties, except for planet d, which has some larger spread. The uncertainties (right panel) are mostly comparable as well, however, we find slightly larger uncertainties when using K2 or TESS data separately. Including RVs helps to better constrain parameters such as, for example, orbital period, which in turn better constrains other parameters such as the radius. For planets c and d the uncertainties are smallest when using all datasets, for planet b the smallest uncertainties are found when using K2+RV followed by all datasets. However, these are still slightly higher than those derived by \cite{Kosiarek2021}, who provide an uncertainty of 0.058~R$_{\oplus}$ for planet b, whereas we derived 0.062~R$_{\oplus}$ using K2+RV. 

For planetary masses we determined M$_{b} = 4.28_{-0.33}^{+0.35}$~M$_{\oplus}$, M$_c = 1.86_{-0.39}^{+0.37}$~M$_{\oplus}$, and M$_d = 3.02_{-0.57}^{+0.58}$~M$_{\oplus}$.
Figure~\ref{fig:mass_lit} shows the comparison with the literature. The mass of planet b is generally well defined, however, \cite{Teske2018} measured almost twice the mass than other literature studies. Their uncertainties are very large, for planet c and d the values are only upper limits. This is probably due the low number of RV observations available at that time (36 measurements from PFS). For all three planets, we derived the smallest uncertainties compared to literature, which range at 8\%, 20\%, and 19\% of the derived planetary mass, respectively. Consequently, we also determined the smallest uncertainties in RV amplitude, which are $K_b$ = $3.53 \pm 0.22$~ms$^{-1}$, $K_c$ = $1.06 \pm 0.21$~ms$^{-1}$, and $K_d$ = $1.44 \pm 0.27$~ms$^{-1}$. This is slightly smaller than the uncertainties from \cite{Kosiarek2021}, who determined 0.30~ms$^{-1}$ for planets b and d, and 0.29~ms$^{-1}$ for planet c.

Looking at the comparison of orbital periods in Fig.~\ref{fig:period_lit}, we derived a period of $1.208974_{-0.000001}^{+0.000001}$ days for planet b, $3.648103_{-0.000010}^{+0.000013}$ days for planet c, and $6.201812_{-0.000009}^{+0.000009}$ days for planet d. It might seem that the periods of planet b do not agree very well in literature, with \cite{Niraula2017} measuring a much smaller period than \cite{Rice2019}. However, the difference between these two values is 2.5e-5~d, which are only 2.15~seconds. For all three planets, we derived the smallest uncertainties, which are less than 0.1~second for planet b and less than 1~second for planets c and d. Therefore, the period of planet b is determined with high accuracy and precision. Our period for planet d is consistent with \cite{Kosiarek2021}, but significantly different from the rest of the literature by about 35~seconds. However, our work and \cite{Kosiarek2021} are the most recent works and include most recent data, including high-precision RV data from ESPRESSO, therefore we believe that our value is more accurate than previous works. 

Figure~\ref{fig:T0_lit} shows the literature comparison for T0. We measured T0$_b = 2457738.825934_{-0.000472}^{+0.000486}$~d, T0$_c = 2457742.199967_{-0.001402}^{+0.001441}$~d, and T0$_d = 2457740.958775_{-0.000854}^{+0.000828}$~d. In general, the values agree with each other within their uncertainties. For planet c, T0 from \cite{Niraula2017} and \cite{PrietoArranz2018} are significantly different from our result and other literature values, amounting to $2457738.5519 \pm 0.0014$~d and $2457738.5496 \pm 0.0015$~d, respectively. Both studies find the first transit of planet c at the very beginning of the K2 light curve and right before the first transit of planet b. This transit is visible in our light curve (see top panel of Fig.~\ref{fig:K2+TESS_fit}), but because we used narrow priors based on results from \cite{Rodriguez2018} it was not found. The first transit of planet c is also visible in Fig.~1 of \cite{Rodriguez2018}, it is not clear why it was not found. To be able to compare T0 for all three planets, we adjusted T0 from \cite{Niraula2017} and \cite{PrietoArranz2018} by adding one orbital period of planet c. 
For all three planets our uncertainties are larger than other literature uncertainties, except \cite{PrietoArranz2018} for planet c, which is even slightly larger. Compared to the smallest uncertainties (\citealt{Kosiarek2021} and \citealt{Rice2019} for planet b, \citealt{Rodriguez2018} for planets c and d), our uncertainties are larger by 19~s, 66~s, and 35~s for each planet, respectively. As for the radius, we compared results from different GP runs (see Fig.~\ref{fig:GP_T0}) and found that the largest uncertainties arise when using only RV, whereas uncertainties decrease when including K2 and TESS. Therefore, we derive the smallest uncertainties when using the whole dataset. 

Overall, our parameters agree very well with those of \cite{Kosiarek2021}, also the uncertainties are comparable in most cases. We derive the smallest uncertainties for planetary mass, RV amplitude and orbital period.

\subsection{Orbital configuration}

As mentioned in Sec.~\ref{sec:combined}, we performed GP fits with circular and eccentric orbits. In the eccentric case, the eccentricities of all three planets are essentially consistent with zero within their uncertainties, with $e_b = 0.03_{-0.02}^{+0.03}$, $e_c = 0.05_{-0.03}^{+0.06}$, and $e_d = 0.13_{-0.09}^{+0.12}$. \cite{Kosiarek2021} evaluated if the system is stable over $10^9$ orbits for different values of eccentricity and the argument of periapse, $\omega$, using $spock$ \citep{Tamayo2020}. Averaging over the $\omega$ grid shows that the orbital configuration is stable if $e_b \leq 0.4$, $e_c \leq 0.2$, $e_d \leq 0.1$, which is consistent with the eccentricity values we derived for the system in the eccentric case. \cite{PrietoArranz2018} also performed two types of stability analyses. First, they integrated their MCMC samples using the SWIFT N-body package \citep{Duncan1998} and the MVS integrator \citep{WisdomHolman1991} with a maximum integration time of 1~Myr. Additionally, they did a stability test using mercury6 \citep{Chambers1999} over 100,000 years. In both cases, they assumed tidal circularization and find that the semi-major axes are constant within 0.1\% and the eccentricity variation is less than $10^{-3}$. 

All planets have a similar inclination ($i_b = 87.60_{-1.27}^{+1.31}$~deg, $i_c = 89.09_{-0.68}^{+0.60}$~deg, $i_b = 87.66_{-0.16}^{+0.13}$~deg), however, they cannot be considered coplanar, because of the unknown angle $\Omega$, the longitude of the ascending node. Concerning the impact factor $b$, it is notable that planet d ($0.85_{-0.03}^{+0.02}$) has a significantly higher value than planet b ($0.30_{-0.16}^{+0.14}$) and planet c ($0.23_{-0.15}^{+0.17}$). Herewith, we confirm the values published by \citet[][$0.91_{-0.013}^{+0.011}$]{Niraula2017}, \citet[][$0.896_{-0.016}^{+0.012}$]{Rodriguez2018}, \citet[][$0.864_{-0.013}^{+0.022}$]{PrietoArranz2018}, and \citet[][$0.8927_{-0.009}^{+0.0071}$]{Rice2019}. \cite{Niraula2017} first noted the higher impact parameter of planet d and argued that this might suggest additional non-transiting planets in the system. We investigated this possibility in the following section. 

No further discussion on the impact parameter is given by the other authors. \cite{PrietoArranz2018} discuss how the planets could have reached a near 5:3:1 resonance, which could be either due to inward migration and lining up in a resonant chain \citep{Izidoro2017}, or due to in situ formation as suggested by \cite{ChiangLaughlin2013}, which reproduces many of the observed properties of close-in super-Earths. In the latter case, the planets should have retained their primordial hydrogen envelopes and their atmospheres should not show any water features due to the proximity to the star. 


\subsection{Potential additional planet}

\begin{figure*}[ht]
\centering
\includegraphics[width=0.95\textwidth]{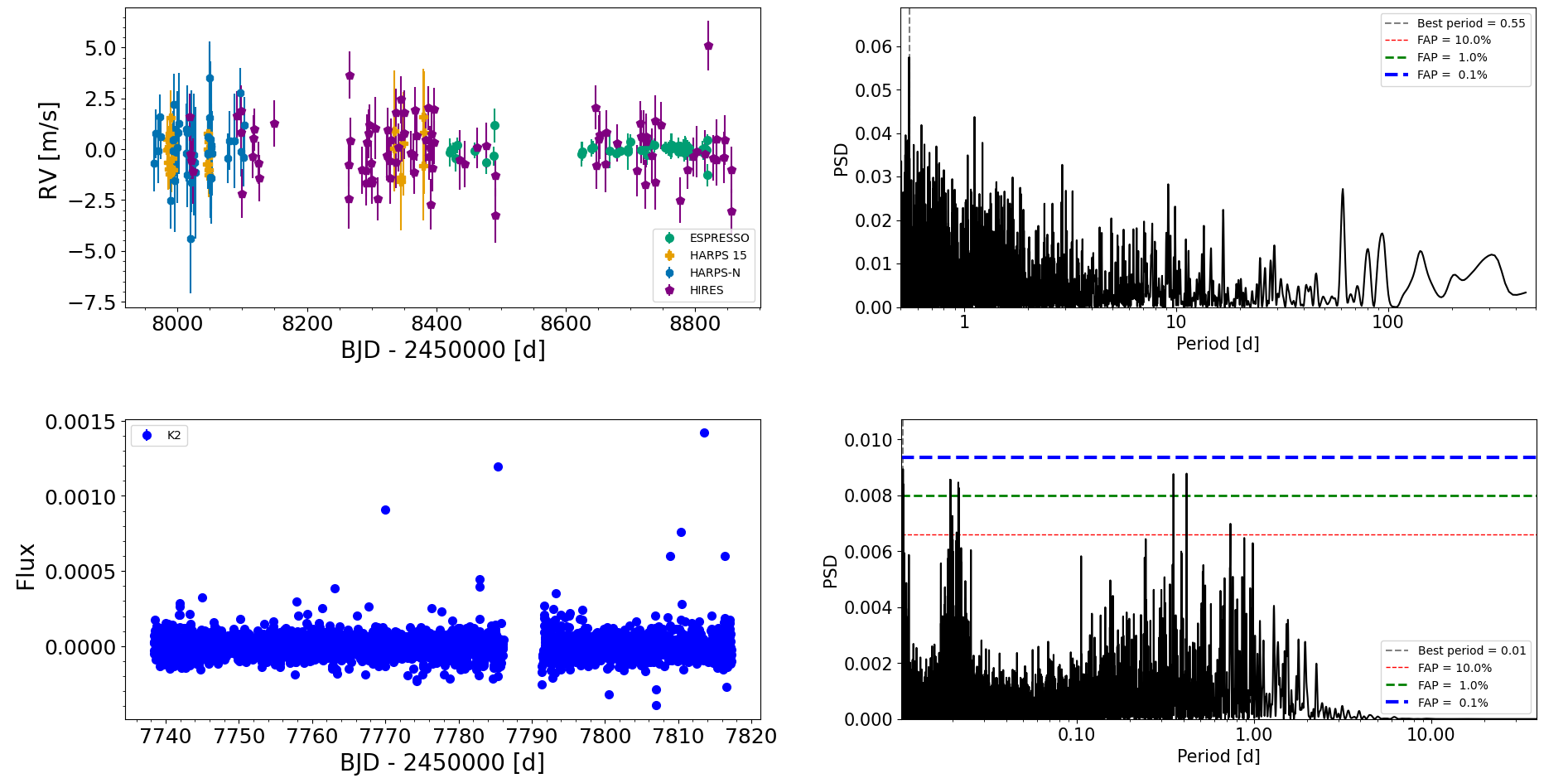}
\caption{$Left$: Residuals of the RV and K2 time series after subtracting stellar activity and the three planetary signals. $Right$: Correpsonding GLS periodograms do not show any remaining significant signals.}
\label{fig:residuals}
\end{figure*}

We investigated the possibility of an undiscovered additional planetary signal. As suggested by \cite{Niraula2017}, this planet is likely to be non-transiting. Figure~\ref{fig:residuals} shows the residuals of the RV and K2 time series after subtracting stellar activity and the three planetary signals. The GLS periodograms of both datasets show no significant signals. However, we performed a GP analysis on the RV data as described in Section~\ref{sec:RV-method}, using our results of the three planet model as priors for the three known planets and wide uniform priors for the potential fourth planet.
The posterior distributions of the parameters of the fourth planet do not reveal a significant signal, indicating that no other planetary signal can be found in the data. The corresponding corner plot is shown in Fig.~\ref{fig:planet4}.

\subsection{Future multi transits}

Multiple transits are an interesting phenomenon. Besides the fact that these events are quite rare, they are predicted and observed in several exoplanet systems \citep{Luger2017,Muller2022}. The great advantage of such events is given by the fact that, if they exist, the observer can study, for instance, transit timing variations (TTVs), and therefore learn something about the orbital dynamics of such systems and find potential undetected planets in the system. \cite{Niraula2017} do not find TTVs greater than three minutes for any of the three planets in the system. Recently, \cite{Roy2023} analyzed data from Wide Field Camera 3 on the Hubble Space Telescope (HST/WFC3) and find TTVs in the order of 5-10 minutes for planet d, which is not surprising for a near-resonant system.

\begin{figure}[ht]
\centering
\includegraphics[width=0.45\textwidth]{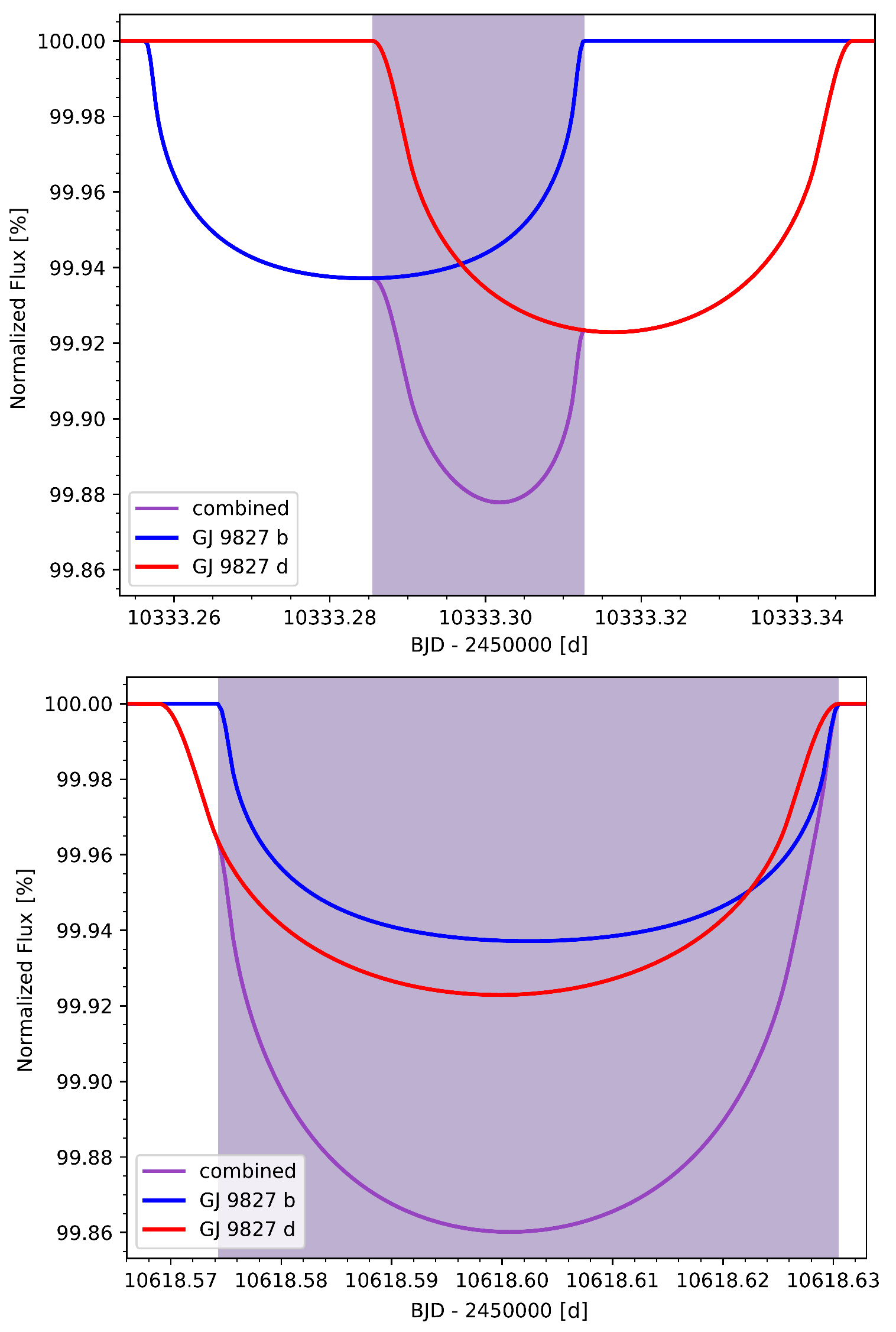}
\caption{Simulated light curves of multi transits of GJ~9827~b and GJ~9827~d starting at BJD 2460333.25~d (top) and 2460618.57~d (bottom). The latter event is the deepest in this series.}
\label{fig:multitransit}
\end{figure}

Additionally, these multi transits offer the possibility of the occurrence of extremely rare events called planet-planet occultations (PPOs), where one planet passes in front of another one during its transit passage leaving a well defined imprint to the light curve \citep[e.g.,][]{Hirano2012}. If present, such an event allows to pin down the orbital parameters of the planets to a very high precision. Nevertheless, both cases, ordinary multi transits or those with PPOs, are of special interest for planet observers.
For the year 2024, we predict double transit events for planets in the GJ~9827 system. We carried out a simulation of transit light curves using the {\tt occultquad} routine from \cite{MandelAgol2002} together with the planetary and orbital parameters of all planets determined in Sect.~\ref{sec:combined}, not taking into account uncertainties in $P$ and $T0$. 
We found seven double transits of the planets b and c, and six double transits for planets b and d. However, we did not find any double transits of planets c and d until end of March 2051.
In Table~\ref{tab:future-transits}, we list double transit events for planets b+c and b+d, including transit start and end time of the individual planets, duration of the whole event, and duration of the double transit.
Due to the smaller size of planet c, the double transits of planet b and c are shallower compared to those of planet b and d. In Fig.~\ref{fig:multitransit}, we show as an example of simulated light curves of the first and fifth multi transit of planets b and d occurring at BJD 2460333.25~d and 2460618.57~d, respectively.  

For the sake of completeness, we also tried to find a triple transit where all three planets transit in front of the stellar disk at the same time and which is of course a very rare event. We found that the first triple transit occurs in January 2057.
Since the multi transits of planets c+d and b+c+d are far in the future and given the uncertainties of the orbital periods of all planets together with possible periastron drifts, these events might occur earlier or later, or maybe not at all.

\begin{table*}[htb]
\caption{Future multi transits of GJ~9827 for the year 2024.}
\label{tab:future-transits}
\centering 
\begin{tabular}{l cccc}
    \hline 
    \hline 
    \noalign{\smallskip}
    Planets & Planet 1 & Planet 2 & \multicolumn{2}{c}{Duration of}\\
      & start - end & start - end & whole event & multi transit \\
      & [BJD - 2450000~d] & [BJD - 2450000~d] & [h] & [h] \\
\noalign{\smallskip}
\hline
\noalign{\smallskip}

 b + c & 10481.9603 - 10482.0165	& 10481.8843 - 10481.9661 & 3.172 & 0.139 \\
 & 10485.5871 - 10485.6435	& 10485.5325 - 10485.6143 & 2.664 & 0.653 \\
 & 10489.2141 - 10489.2703	& 10489.1805 - 10489.2623 & 2.154 & 1.157 \\
 & 10492.8411 - 10492.8971	& 10492.8287 - 10492.9105 & 1.963 & 1.344 \\
 & 10496.4677 - 10496.5241	& 10496.4769 - 10496.5587 & 2.184 & 1.133 \\
 & 10500.0947 - 10500.1509	& 10500.1249 - 10500.2068 & 2.691 & 0.624 \\
 & 10503.7217 - 10503.7779	& 10503.7731 - 10503.8549 & 3.197 & 0.115 \\
  \\
 b + d & 10333.2562 - 10333.3126 & 10333.2856 - 10333.3470 & 2.179 & 0.648 \\
 & 10426.3475 - 10426.4037 & 10426.3128 - 10426.3743 & 2.179 & 0.643 \\
 & 10475.9153 - 10475.9715 & 10475.9273 - 10475.9887 & 1.762 & 1.061 \\
 & 10569.0063 - 10569.0626 & 10568.9544 - 10569.0159 & 2.597 & 0.230 \\
 & 10618.5744 - 10618.6304 & 10618.5690 - 10618.6304 & 1.474 & 1.344 \\
 & 10668.1422 - 10668.1984 & 10668.1834 - 10668.2450 & 2.467 & 0.360 \\

\noalign{\smallskip}
\hline

\end{tabular}

\end{table*}

\subsection{Interior bulk composition}

\begin{figure}[!ht]
\centering
\includegraphics[width=0.49\textwidth]{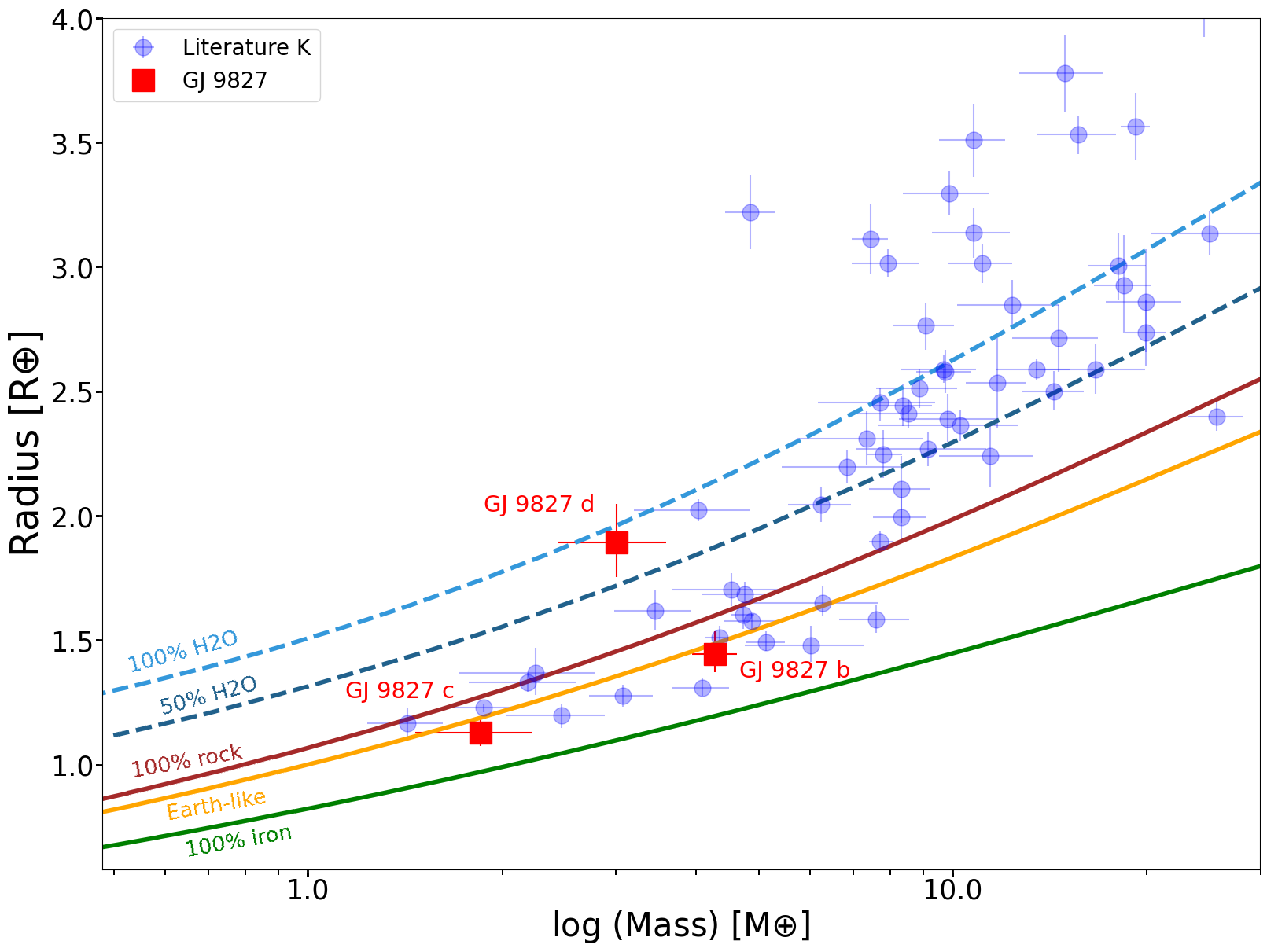}
\caption{Mass-radius diagram showing the positions of the three planets around GJ~9827 and planets detected around K dwarfs from literature. The different lines present different planetary compositions from 100\% iron to 100\% H$_2$O.}
\label{fig:mass-radius}
\end{figure}

\begin{figure}[!ht]
\centering
\includegraphics[width=0.49\textwidth]{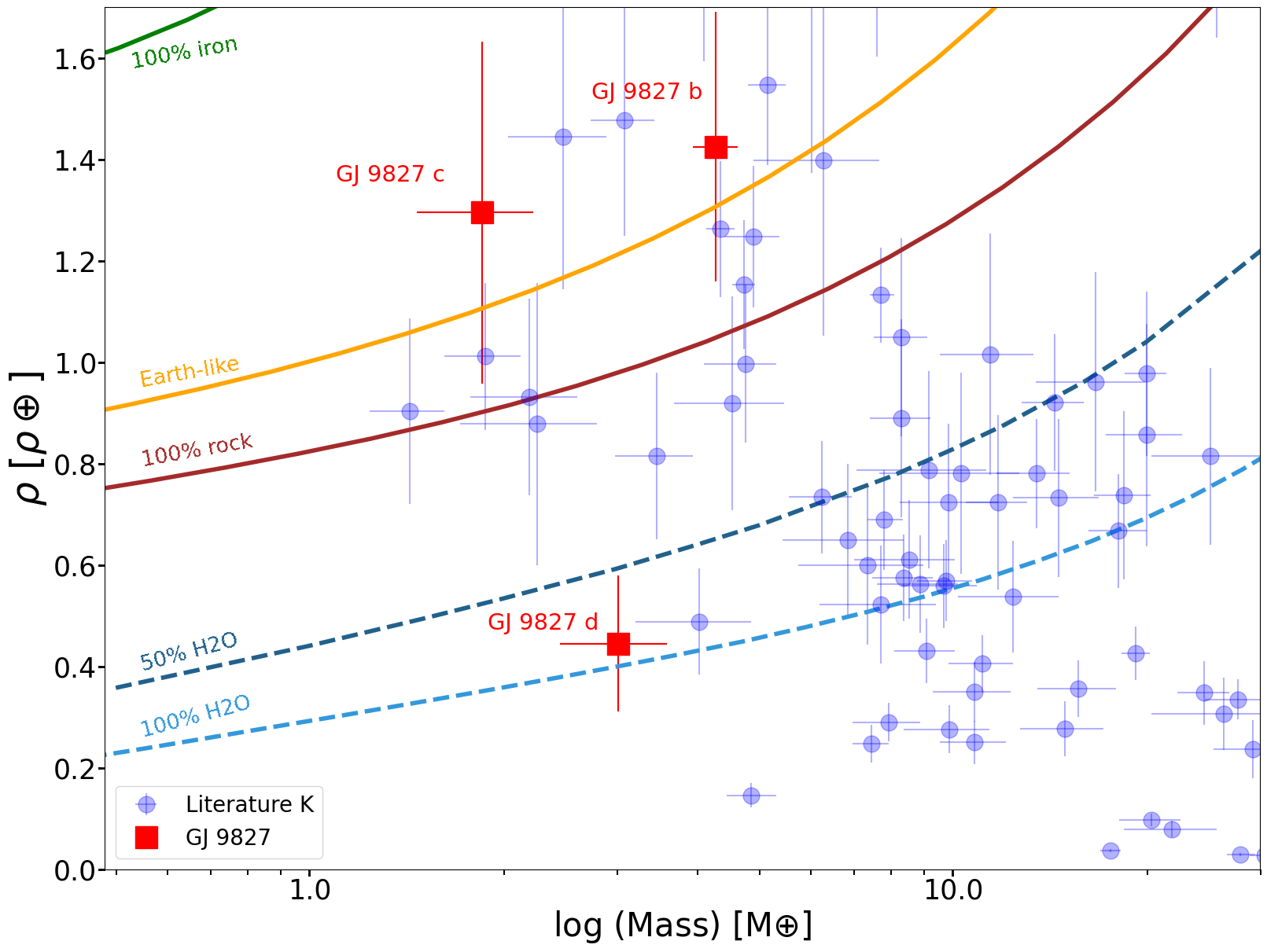}
\caption{Mass-density diagram showing the positions of the three planets around GJ~9827 and planets detected around K dwarfs from literature. The different lines present different planetary compositions from 100\% iron to 100\% H$_2$O.}
\label{fig:mass-density}
\end{figure}

\begin{figure*}[!ht]
\centering
\includegraphics[width=0.95\textwidth]{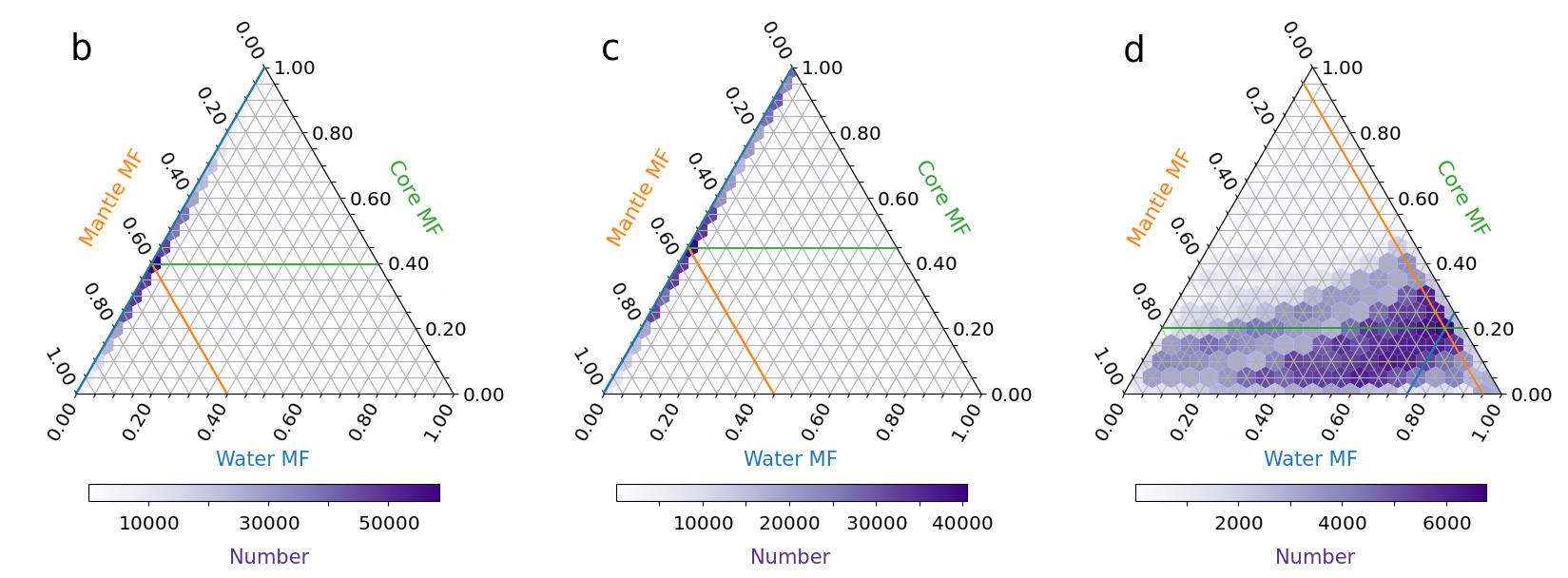}
\caption{Ternary diagrams of the mass fractions of GJ~9827~b, c, and d derived with ExoMDN.}

\label{fig:interior_mf}
\end{figure*}

\begin{figure*}[!ht]
\centering
\includegraphics[width=0.95\textwidth]{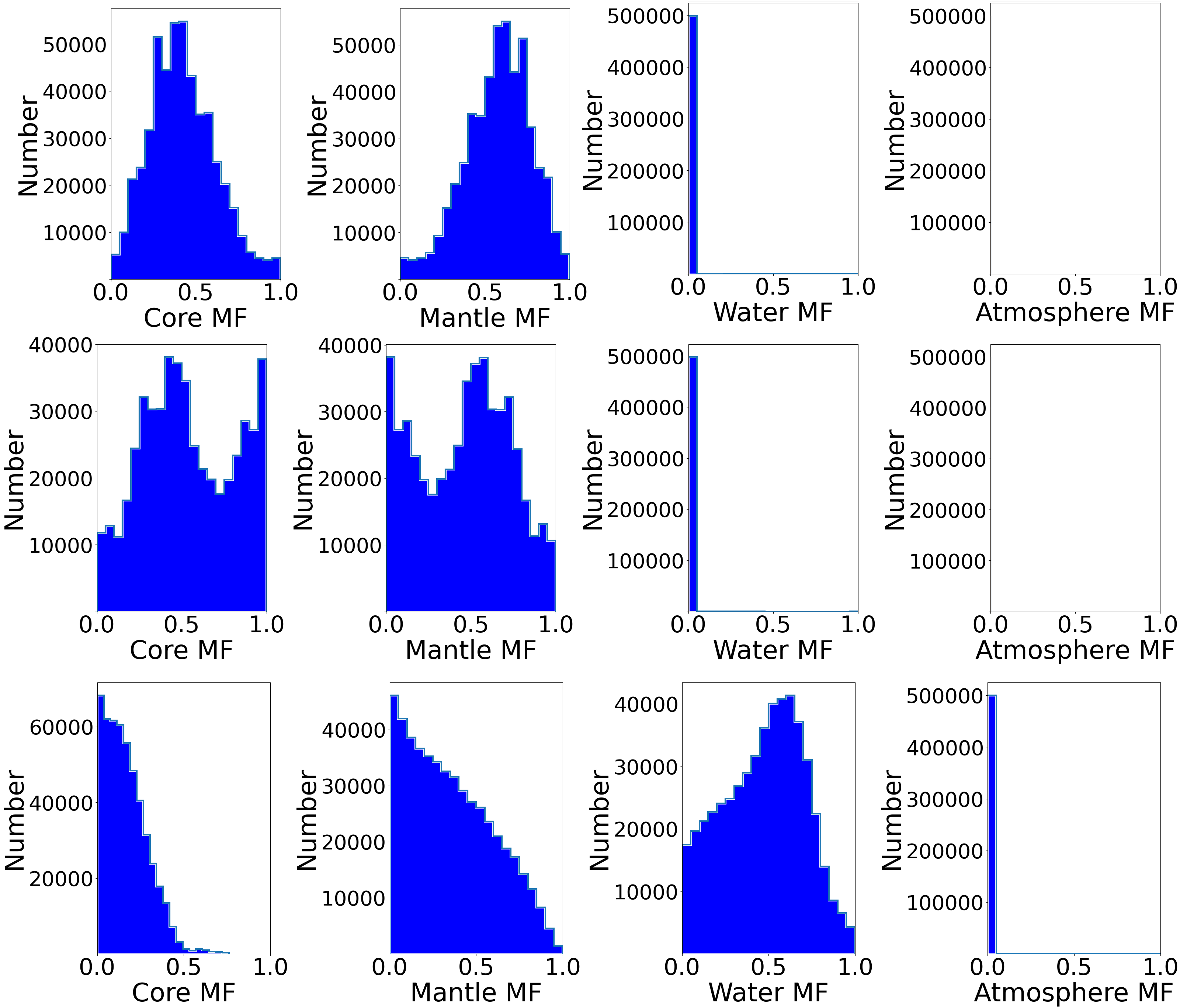}
\caption{Histograms of mass fractions for core, mantle, water, and atmosphere of GJ~9827~b (top), c (middle), and d (bottom) derived with ExoMDN.}

\label{fig:histogram_mf}
\end{figure*}

\begin{figure*}[!ht]
\centering
\includegraphics[width=0.94\textwidth]{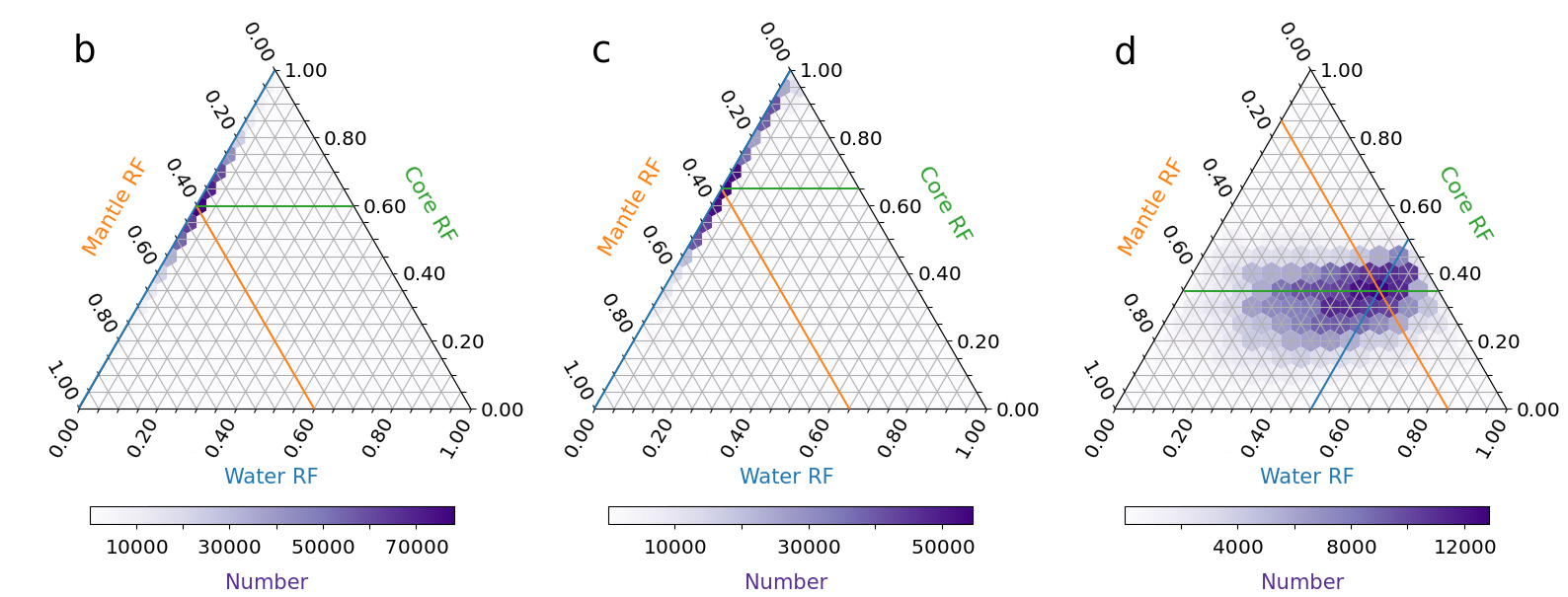}
\caption{Ternary diagrams of the radius fractions of GJ~9827~b, c, and d derived with ExoMDN.}

\label{fig:interior_rf}
\end{figure*}

\begin{figure*}[!ht]
\centering
\includegraphics[width=0.93\textwidth]{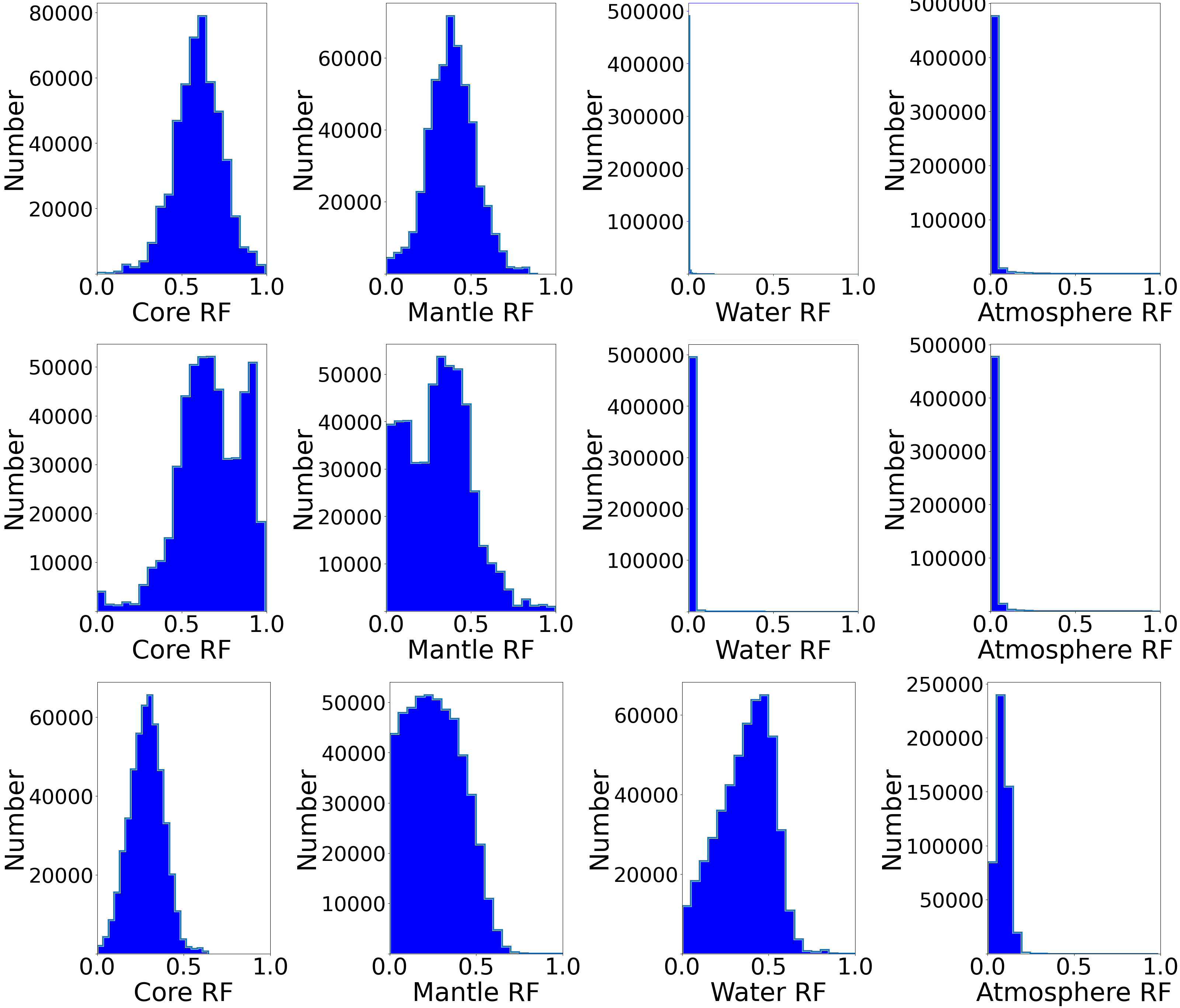}
\caption{Histograms of radius fractions for core, mantle, water, and atmosphere of GJ~9827~b (top), c (middle), and d (bottom) derived with ExoMDN.}

\label{fig:histogram_rf}
\end{figure*}

\begin{table*}[htb]
\caption{Planetary densities for GJ~9827 from the literature.}
\label{tab:densities}
\centering 
\renewcommand{\arraystretch}{1.4}
\begin{tabular}{l ccc}
    \hline 
    \hline 
    \noalign{\smallskip}
    Source & $\rho_b$ [g cm$^{-3}$] & $\rho_c$ [g cm$^{-3}$] & $\rho_d$ [g cm$^{-3}$] \\
\noalign{\smallskip}
\hline
\noalign{\smallskip}
 \cite{Rodriguez2018} & $4.50^{+1.50}_{-0.98}$ & $6.4^{+2.0}_{-1.1}$ & $3.23^{+1.10}_{-0.72}$ \\
 \cite{PrietoArranz2018} & $5.11^{+1.74}_{-1.27}$ & $4.13^{+2.31}_{-1.77}$ & $1.51^{+0.71}_{-0.53}$ \\
 \cite{Rice2019}$^{(a)}$ & $7.157 \pm 0.769$ & $1.933 \pm 1.735$ & $2.474 \pm 0.651$ \\
 \cite{Kosiarek2021} & $7.47^{+1.1}_{-0.95}$ & $6.1^{+1.8}_{-1.6}$ & $2.51^{+0.57}_{-0.51}$ \\
 This work &  $7.858 \pm 1.467$ &  $7.142 \pm 1.857$ &  $2.454 \pm 0.739$ \\
 
\noalign{\smallskip}
\hline

\end{tabular}
\tablefoot{
\tablefoottext{a}{calculated from M and R, error from error propagation assuming symmetrical errors.}
}
\end{table*}

The composition of the planets is discussed extensively in \cite{Teske2018}, \cite{Rice2019}, \cite{PrietoArranz2018}, and \cite{Kosiarek2021}. \cite{Teske2018} note that planet b is one of the densest and most massive terrestrial planets found so far, which is due to the high mass of 8~M$_{\oplus}$ they derive. Considering our mass measurement of 4.28~M$_{\oplus}$, planet b is not so anomalously dense anymore. Because of its small mass and large errors, planet c is the least constrained in \cite{Teske2018}, however, for planet d they find a water mass fraction of 50--100\%, maybe indicating an H$_2$/He-dominated envelope. 

\cite{Rice2019} also cannot constrain the mass of planet~c very well, therefore its composition remains unclear. They speculate that it could be rocky with a substantial water mass fraction. From their mass and radius measurements, they conclude that planet b is Earth-like, and planet d must have a substantial atmosphere and water mass fraction. 
They explain that planet b and c receive 316 and 73 times more flux than Earth, respectively, therefore these planets could have formed with a similar composition as planet d and then lost their atmospheres due to photoevaporation. 

\cite{PrietoArranz2018} present better constraints for planet c and conclude that both planet b and c have rocky nuclei with traces of lighter elements, whereas planet d has a large gaseous H/He-rich envelope with a nucleus similar to the other two planets. They also suggest that the inner planets might have lost their envelopes because of photoevaporation. 

\cite{Kosiarek2021} used the three-component H$_2$O/MgSiO$_3$/Fe model \citep{ZengSasselov2013,Zeng2016} to draw ternary diagrams. They show that planet b and c have a H$_2$O fraction of $\leq 40$\% and a wide range of possible MgSiO$_3$ and Fe compositions. Planet d is consistent with a higher H$_2$O fraction of 50--100\% and smaller fractions of MgSiO$_3$ ($\le 50$\%) and Fe ($\leq 30$\%). 
In Table~\ref{tab:densities}, we summarize all planetary densities from literature, including results from this work, which show that planets b and c have a larger density than Earth ($\rho_{\oplus} = 5.51$~g~cm$^{-3}$) and planet d has a density larger than Neptune ($\rho_{\rm Neptune}$ = 1.64~g~cm$^{-3}$). 

We investigated the inner structure of all three planets and first produced mass-radius and mass-density diagrams, presented in Figs.~\ref{fig:mass-radius} and \ref{fig:mass-density} with the position of the planets around GJ~9827 clearly marked. We overplotted composition curves taken from \cite{Zeng2019}. Earth-like composition corresponds to 32.5\% Fe and 67.5\% MgSiO$_3$, 100\% rock is equal to 100\% MgSiO$_3$, and 50\% H$_2$O assumes a composition of 50\% water and 50\% Earth-like. We included other confirmed planets around K dwarfs from the Transiting Extrasolar Planets Catalogue \citep[TEPCat,][]{Southworth2011}, selecting only those planets with host star effective temperatures between 4000 and 5200~K, planetary radius precisions better than 8\%, and mass precisions better than 25\%. From these plots, we can see that planets b and c have an almost Earth-like composition, while the outer planet d must have a substantial water fraction.

Comparing these plots to Fig.~S19 (A-B) of \cite{Luque2022}, who used TEPCat with the same constraints, we can also see distinct groups of planets, the rocky planets and the water- and gas-rich planets. The density gap proposed by \cite{Luque2022} does not appear so clearly in our plots, which could be due to an update of the planet database. However, GJ~9827~b and c can be assigned to the rocky family of planets with Earth-like composition, while GJ~9827~d could be consistent with both a large water mass fraction (water world) and a H/He-rich sub-Neptune scenarios. 

Then, we calculated the equilibrium temperature $T_{eq}$ for each planet following Equ.~\ref{eq:Teq}, where $a$ is the semi-major axis of the planet and $A_b$ the albedo. For each planet, we adopted an albedo of 0.3 from \cite{Niraula2017}. For error propagation (Equ.~\ref{eq:error_Teq}), we assumed symmetrical errors for each parameter by calculating the average error: 

\begin{equation} 
\label{eq:Teq}
    T_{eq} = T_{star} \cdot \sqrt{\frac{R_{star}}{2 a}} \cdot \left( 1-A_b \right) ^{1/4}
\end{equation}

\begin{equation}
\label{eq:error_Teq}    
    \Delta T_{eq} = {T_{eq}} \cdot \sqrt{\left( \frac{\Delta T_{star}}{T_{star}} \right) ^2 + \left( \frac{\Delta R_{star}}{2 R_{star}} \right) ^2 + \left (- \frac{\Delta a}{2 a} \right) ^2} .
\end{equation}

We modeled the interior structure of the three planets using ExoMDN \citep{Baumeister2023}. ExoMDN is a machine-learning-based exoplanet interior inference model trained on more than 5.6 million synthetic planet interior structures consisting of an iron core, a silicate mantle, a water and high-pressure ice layer, and a H/He atmosphere. It uses Mixture Density Networks \citep{Bishop1994}, which approximate the parameters with a linear combination of Gaussian kernels to predict their probability distributions. ExoMDN provides a full inference of the interior structure of exoplanets below 25~M$_\oplus$, given mass, radius, and equilibrium temperature of the planet, in under a second. \cite{Baumeister2023} show that mass, radius, and equilibrium temperature alone are not always sufficient to fully constrain the interior of a planet. Therefore, they provide a second trained model including the second degree fluid Love number k2 \citep{Love1909}. The Love number depends on the density distribution of the planet's interior and describes the shape of a rotating planet in hydrostatic equilibrium \citep{Padovan2018,Kellermann2018,Baumeister2020}. For example, k2 = 0 corresponds to a body with all mass concentrated in the center and k2 = 1.5 characterizes a fully homogeneous body. 

Although it is possible to measure k2 for exoplanets, it is rather difficult as it requires the measurement of second-order effects in the shape of the transit light curve \citep{Akinsanmi2019,Hellard2019,Barros2022} or the apsidal precession of the orbit \citep{Csizmadia2019}. Therefore, we assumed the Love number of Earth \citep[k2 = 0.933,][]{Lambeck1980} for planets b and c and the Love number of Neptune \citep[k2 = 0.392,][]{Baumeister2023} for planet d. This imposes an internal structure model similar to Earth and Neptune, respectively, which might not necessarily be true, but is a legitimate assumption, based on the planets' positions in the mass-radius diagram of Fig.~\ref{fig:mass-radius}.

Together with the planetary masses, radii, and equilibrium temperatures shown in Table~\ref{tab:results_GJ9827}, we ran ExoMDN. To account for uncertainties in the planetary parameters we drew normal distributions with $\sigma$ equal to the uncertainties of each parameter. For the Love numbers, we assumed an uncertainty of 10\%. \cite{Baumeister2023} demonstrate that with a k2 uncertainty of 10\%, the core and mantle thickness of Earth can be constrained to about 13\% of their actual value. We randomly picked planetary parameters from these distributions and ran 500 predictions with 1000 samples each, amounting to a total of 500,000 predictions. We present the results for the mass and radius fraction as ternary diagrams in Fig.~\ref{fig:interior_mf} and Fig.~\ref{fig:interior_rf} and as histograms in Fig.~\ref{fig:histogram_mf} and Fig.~\ref{fig:histogram_rf}.

It can be seen that the radius fraction for core and mantle are very similar for planets b and c, 60\% of the radius is attributed to the core and 40\% to the mantle for planet b, for planet c it is 65\% for the core and 35\% for the mantle. Planet b has a core mass fraction (CMF) of 40\% and a mantle mass fraction (MMF) of 60\%, while planet c has a CMF of 45\% and an MMF of 55\%. This distribution is similar to Earth, which has a CMF of 32.5\% and an MMF of 67\%. 
For planet c, the histograms of the CMF and MMF are two-fold, indicating that a significant amount of models points toward a planet with a higher CMF and smaller MMF.
Both planets have no water and no atmosphere. This is consistent with the non-detection of Ly$\alpha$, H$\alpha$, and He I transitions by \cite{Carleo2021} using HST and CARMENES transit observations. With an estimated mass-loss rate of 1.9~M$_{\oplus}$~Gyr$^{-1}$ for planet b and 0.5~M$_{\oplus}$~Gyr$^{-1}$ for planet c \citep{Carleo2021}, both planets are very likely to have lost their primordial atmospheres. 

As expected, planet d is different and consists of a significant amount of water. It has a CMF of 20\%, a MMF of 5\%, and a water mass fraction (WMF) of 75\%. The histograms of the MMF and WMF show a tail toward higher MMF and lower WMF, which is also visible in the ternary diagram. The position of the planet in the mass-radius and mass-density diagrams, on the other hand, supports a large water composition. 
The radius histograms for planet d are better constrained (Fig.~\ref{fig:histogram_rf}, bottom panel), showing a core radius fraction of 30\%, a mantle radius fraction of 20\%, a water radius fraction of 45\%, and an atmosphere radius fraction of 5\%. 
From this, we conclude that planet d has a rocky core surrounded by a significant hydrogen envelope, placing it in the mini-Neptune regime. This is in line with transmission spectroscopy from \cite{Roy2023} using HST/WFC3. They report a hint of an absorption feature at $1.4 \mu$m in its transit spectrum due to water vapor in the atmosphere of planet d. This points toward a water-dominated envelope, although a cloudy H2/He atmosphere with a small amount of water vapor can also explain the observed features \citep{Roy2023}. However, as for planet b, no H$\alpha$ and He features are found in the atmosphere of GJ~9827~d \citep{Carleo2021,Krishnamurthy2023}. Due to the planet's proximity to the host star, its low mass, and an estimated mass-loss rate of >~0.2~M$_{\oplus}$~Gyr$^{-1}$ \citep[\citet{Krishnamurthy2023}; 0.3~M$_{\oplus}$~Gyr$^{-1}$:][]{Carleo2021} it is unlikely that the planet could retain an H-dominated atmosphere.


\section{Summary and conclusions}
\label{Summary}

GJ~9827 hosts three transiting planets, which have first been detected using K2 data. The system was observed and characterized with different spectrographs in the recent years. Several studies derive planetary masses and radii, however, the results span a wide range of values with large uncertainties. In this study, we used data from ESPRESSO, HARPS, HARPS-N, HIRES, K2, and TESS to improve the accuracy and precision of planetary parameters. To do this, we employed a combined GP analysis using high-precision spectroscopy and photometry. 
 
With this, we could model the stellar activity signal from the radial velocity and K2 data, and derive a consistent rotation period of $28.16_{-2.66}^{+3.38}$~d and $29.52_{-3.25}^{+3.42}$~d, respectively. We used two SHO kernels at $P_\textrm{rot}$ and $P_\textrm{rot}/2$ for the RV and K2 datasets and a Mat{\'e}rn 3/2 kernel for TESS from the S+LEAF code, together with nested sampling using Dynesty. We investigated circular and eccentric orbits and find that both models are indistinguishable with almost the same likelihood. Therefore, we favor the simpler circular model. Due to the proximity of the planets to their host star, circularization of the orbits can be expected over the lifetime of the system. Planet d has a higher impact parameter of $b_d = 0.85_{-0.03}^{+0.02}$ compared to planet b with $b_b = 0.30_{-0.16}^{+0.14}$ and planet c $b_c = 0.23_{-0.15}^{+0.17}$, which is also found by previous studies. On the other hand, all planets have very similar orbital inclination between 87 and 89~deg, which could give interesting insights in the evolution of the system. 

Regarding planetary mass and radius, we derive $R_b = 1.44_{-0.07}^{+0.09}$~R$_{\oplus}$ and $M_b = 4.28_{-0.33}^{+0.35}$~M${_\oplus}$, $R_c = 1.13_{-0.05}^{+0.07}$~R$_{\oplus}$ and $M_c = 1.86_{-0.39}^{+0.37}$~M${_\oplus}$, and $R_d = 1.89_{-0.14}^{+0.16}$~R$_{\oplus}$ and $M_d = 3.02_{-0.57}^{+0.58}$~M${_\oplus}$. 
We compared our results to those of \cite{Niraula2017}, \cite{Rodriguez2018}, \cite{Teske2018}, \cite{Rice2019}, \cite{PrietoArranz2018}, and \cite{Kosiarek2021}. Our values are mostly consistent with the results of \cite{Kosiarek2021} and we improve uncertainties in RV amplitude, mass, and period. 
We drew mass-radius and mass-density diagrams and modeled the inner structure of all three planets with ExoMDN. We confirm previous composition studies and find that GJ~9827~b and c have an Earth-like composition, whereas the outer planet d consists of a rocky core with an hydrogen envelope. It can also be seen that the planets lie on both sides of the radius gap. 

The planetary system around GJ~9827 is a good candidate for transit spectroscopy. Our improvement on the precision of period and mass will help in an accurate atmospheric characterization. Additionally, we predict times of future multi transit events, which allow further analysis of orbital dynamics and, in case of planet-planet-occultations, the determination of orbital parameters to very high precision. 

\begin{acknowledgements}
V.M.P., A.S.M., J.I.G.H., and R.R. acknowledge financial support from the Spanish Ministry 
of Science and Innovation (MICINN) project PID2020-117493GB-I00 and from the Government of the Canary Islands project ProID2020010129. This work has been carried out within the framework of the NCCR PlanetS supported by the Swiss National Science Foundation under grants 51NF40\_182901 and 51NF40\_205606. R.A. is a Trottier Postdoctoral Fellow and acknowledges support from the Trottier Family Foundation. This work was funded by the Institut Trottier de Recherche sur les Exoplanètes (iREx) and supported in part through a grant from the Fonds de Recherche du Québec - Nature et Technologies (FRQNT).
This project has received funding from the European Research Council (ERC) under the European Union's Horizon 2020 research and innovation programme (grant agreement SCORE No 851555). F.P. and C.L. would like to acknowledge the Swiss National Science Foundation (SNSF) for supporting research with ESPRESSO through the SNSF grants nr. 140649, 152721, 166227, 184618 and 215190. The ESPRESSO Instrument Project was partially funded through SNSF's FLARE Programme for large infrastructures. 
This work was supported by FCT - Funda\c{c}{\~a}o para a Ci\^encia e a Tecnologia through national funds in the framework of the project 2022.04048.PTDC (Phi in the Sky) and by FEDER through COMPETE2020 - Programa Operacional Competitividade e Internacionalização by these grants: UIDB/04434/2020; UIDP/04434/2020. 
A.M.S acknowledges support from the FCT through the Fellowship 2020.05387.BD. C.J.A.P.M. also acknowledges FCT and POCH/FSE (EC) support through Investigador FCT Contract 2021.01214.CEECIND/CP1658/CT0001.
A.C.-G. is funded by the Spanish Ministry of Science through MCIN/AEI/10.13039/501100011033 grant PID2019-107061GB-C61. 
We acknowledge financial support from the Agencia Estatal de Investigaci\'on (AEI/10.13039/501100011033) of the Ministerio de Ciencia e Innovaci\'on and the ERDF "A way of making Europe'' through projects  PID2019-109522GB-C51 and PID2019-109522GB-C54.
The research leading to these results has received funding from the European Research Council through the grant agreement 101052347 (FIERCE). Views and opinions expressed are however those of the author(s) only and do not necessarily reflect those of the European Union or the European Research Council. Neither the European Union nor the granting authority can be held responsible for them. 
\newline
The manuscript was written using \texttt{Overleaf}. 
Main analysis performed in \texttt{Python3} \citep{Python3} running on \texttt{Ubuntu} \citep{Ubuntu} systems.
Extensive use of the DACE platform \footnote{\url{https://dace.unige.ch/}}. 
Extensive usage of \texttt{Numpy} \citep{Numpy}.
Extensive usage of \texttt{Scipy} \citep{Scipy}.
All figures built with \texttt{Matplotlib} \citep{Matplotlib} and \texttt{gnuplot} \citep{gnuplot}.
The bulk of the analysis was performed on a server hosting 2x Intel$^{\rm (R)}$ Xeon$^{\rm (R)}$ Gold 5218 (2x16 cores, 2 threads per core, 2.3--3.9 GHz).

\end{acknowledgements}
\bibliographystyle{aa}  
\bibliography{GJ9827.bib}

\begin{thebibliography}{134}
\expandafter\ifx\csname natexlab\endcsname\relax\def\natexlab#1{#1}\fi

\bibitem[{{Aigrain} {et~al.}(2012){Aigrain}, {Pont}, \& {Zucker}}]{Aigrain2012}
{Aigrain}, S., {Pont}, F., \& {Zucker}, S. 2012, \mnras, 419, 3147

\bibitem[{{Akinsanmi} {et~al.}(2019){Akinsanmi}, {Barros}, {Santos}, {Correia},
  {Maxted}, {Bou{\'e}}, \& {Laskar}}]{Akinsanmi2019}
{Akinsanmi}, B., {Barros}, S.~C.~C., {Santos}, N.~C., {et~al.} 2019, \aap, 621,
  A117

\bibitem[{{Albrecht} {et~al.}(2021){Albrecht}, {Marcussen}, {Winn}, {Dawson},
  \& {Knudstrup}}]{Albrecht2021}
{Albrecht}, S.~H., {Marcussen}, M.~L., {Winn}, J.~N., {Dawson}, R.~I., \&
  {Knudstrup}, E. 2021, \apjl, 916, L1

\bibitem[{Ambikasaran {et~al.}(2016)Ambikasaran, Foreman-Mackey, Greengard,
  Hogg, \& O’Neil}]{Ambikasaran2016}
Ambikasaran, S., Foreman-Mackey, D., Greengard, L., Hogg, D.~W., \& O’Neil,
  M. 2016, IEEE Transactions on Pattern Analysis and Machine Intelligence, 38,
  252

\bibitem[{{Astudillo-Defru} {et~al.}(2017){Astudillo-Defru}, {Forveille},
  {Bonfils}, {S{\'e}gransan}, {Bouchy}, {Delfosse}, {Lovis}, {Mayor}, {Murgas},
  {Pepe}, {Santos}, {Udry}, \& {W{\"u}nsche}}]{Astudillo2017}
{Astudillo-Defru}, N., {Forveille}, T., {Bonfils}, X., {et~al.} 2017, \aap,
  602, A88

\bibitem[{{Barrag{\'a}n} {et~al.}(2022){Barrag{\'a}n}, {Aigrain}, {Rajpaul}, \&
  {Zicher}}]{Barragan2022}
{Barrag{\'a}n}, O., {Aigrain}, S., {Rajpaul}, V.~M., \& {Zicher}, N. 2022,
  \mnras, 509, 866

\bibitem[{{Barrag{\'a}n} {et~al.}(2017){Barrag{\'a}n}, {Gandolfi}, \&
  {Antoniciello}}]{Barragan2017}
{Barrag{\'a}n}, O., {Gandolfi}, D., \& {Antoniciello}, G. 2017, {pyaneti:
  Multi-planet radial velocity and transit fitting}, Astrophysics Source Code
  Library, record ascl:1707.003

\bibitem[{{Barros} {et~al.}(2022){Barros}, {Akinsanmi}, {Bou{\'e}}, {Smith},
  {Laskar}, {Ulmer-Moll}, {Lillo-Box}, {Queloz}, {Cameron}, {Sousa},
  {Ehrenreich}, {Hooton}, {Bruno}, {Demory}, {Correia}, {Demangeon}, {Wilson},
  {Bonfanti}, {Hoyer}, {Alibert}, {Alonso}, {Escud{\'e}}, {Barbato},
  {B{\'a}rczy}, {Barrado}, {Baumjohann}, {Beck}, {Beck}, {Benz}, {Bergomi},
  {Billot}, {Bonfils}, {Bouchy}, {Brandeker}, {Broeg}, {Cabrera}, {Cessa},
  {Charnoz}, {Damme}, {Davies}, {Deleuil}, {Deline}, {Delrez}, {Erikson},
  {Fortier}, {Fossati}, {Fridlund}, {Gandolfi}, {Mu{\~n}oz}, {Gillon},
  {G{\"u}del}, {Isaak}, {Heng}, {Kiss}, {des Etangs}, {Lendl}, {Lovis},
  {Magrin}, {Nascimbeni}, {Maxted}, {Olofsson}, {Ottensamer}, {Pagano},
  {Pall{\'e}}, {Parviainen}, {Peter}, {Piotto}, {Pollacco}, {Ragazzoni},
  {Rando}, {Rauer}, {Ribas}, {Santos}, {Scandariato}, {S{\'e}gransan}, {Simon},
  {Steller}, {Szab{\'o}}, {Thomas}, {Udry}, {Ulmer}, {Van Grootel}, \&
  {Walton}}]{Barros2022}
{Barros}, S.~C.~C., {Akinsanmi}, B., {Bou{\'e}}, G., {et~al.} 2022, \aap, 657,
  A52

\bibitem[{Baumeister {et~al.}(2020)Baumeister, Padovan, Tosi, Montavon,
  Nettelmann, MacKenzie, \& Godolt}]{Baumeister2020}
Baumeister, P., Padovan, S., Tosi, N., {et~al.} 2020, The Astrophysical
  Journal, 889, 42

\bibitem[{{Baumeister} \& {Tosi}(2023)}]{Baumeister2023}
{Baumeister}, P. \& {Tosi}, N. 2023, \aap, 676, A106

\bibitem[{Bishop(1994)}]{Bishop1994}
Bishop, C. 1994, Mixture density networks, Workingpaper, Aston University

\bibitem[{{Borucki} {et~al.}(2010){Borucki}, {Koch}, {Basri}, {Batalha},
  {Brown}, {Caldwell}, {Caldwell}, {Christensen-Dalsgaard}, {Cochran},
  {DeVore}, {Dunham}, {Dupree}, {Gautier}, {Geary}, {Gilliland}, {Gould},
  {Howell}, {Jenkins}, {Kondo}, {Latham}, {Marcy}, {Meibom}, {Kjeldsen},
  {Lissauer}, {Monet}, {Morrison}, {Sasselov}, {Tarter}, {Boss}, {Brownlee},
  {Owen}, {Buzasi}, {Charbonneau}, {Doyle}, {Fortney}, {Ford}, {Holman},
  {Seager}, {Steffen}, {Welsh}, {Rowe}, {Anderson}, {Buchhave}, {Ciardi},
  {Walkowicz}, {Sherry}, {Horch}, {Isaacson}, {Everett}, {Fischer}, {Torres},
  {Johnson}, {Endl}, {MacQueen}, {Bryson}, {Dotson}, {Haas}, {Kolodziejczak},
  {Van Cleve}, {Chandrasekaran}, {Twicken}, {Quintana}, {Clarke}, {Allen},
  {Li}, {Wu}, {Tenenbaum}, {Verner}, {Bruhweiler}, {Barnes}, \&
  {Prsa}}]{Borucki2010}
{Borucki}, W.~J., {Koch}, D., {Basri}, G., {et~al.} 2010, Science, 327, 977

\bibitem[{{Bressan} {et~al.}(2012){Bressan}, {Marigo}, {Girardi}, {Salasnich},
  {Dal Cero}, {Rubele}, \& {Nanni}}]{Bressan2012}
{Bressan}, A., {Marigo}, P., {Girardi}, L., {et~al.} 2012, \mnras, 427, 127

\bibitem[{{Brewer} {et~al.}(2016){Brewer}, {Fischer}, {Valenti}, \&
  {Piskunov}}]{Brewer2016}
{Brewer}, J.~M., {Fischer}, D.~A., {Valenti}, J.~A., \& {Piskunov}, N. 2016,
  \apjs, 225, 32

\bibitem[{Carleo {et~al.}(2021)Carleo, Youngblood, Redfield, Barris, Ayres,
  Vannier, Fossati, Palle, Livingston, Lanza, Niraula, Alvarado-Gómez, Chen,
  Gandolfi, Guenther, Linsky, Nagel, Narita, Nortmann, Shkolnik, \&
  Stangret}]{Carleo2021}
Carleo, I., Youngblood, A., Redfield, S., {et~al.} 2021, The Astronomical
  Journal, 161, 136

\bibitem[{{Castro-Gonz{\'a}lez} {et~al.}(2023){Castro-Gonz{\'a}lez},
  {Demangeon}, {Lillo-Box}, {Lovis}, {Lavie}, {Adibekyan}, {Acu{\~n}a},
  {Deleuil}, {Aguichine}, {Zapatero Osorio}, {Tabernero}, {Davoult}, {Alibert},
  {Santos}, {Sousa}, {Antoniadis-Karnavas}, {Borsa}, {Winn}, {Allende Prieto},
  {Figueira}, {Jenkins}, {Sozzetti}, {Damasso}, {Silva}, {Astudillo-Defru},
  {Barros}, {Bonfils}, {Cristiani}, {Di Marcantonio}, {Gonz{\'a}lez
  Hern{\'a}ndez}, {Curto}, {Martins}, {Nunes}, {Palle}, {Pepe}, {Seager}, \&
  {Su{\'a}rez Mascare{\~n}o}}]{Castro-Gonzalez2023}
{Castro-Gonz{\'a}lez}, A., {Demangeon}, O.~D.~S., {Lillo-Box}, J., {et~al.}
  2023, \aap, 675, A52

\bibitem[{{Chambers}(1999)}]{Chambers1999}
{Chambers}, J.~E. 1999, \mnras, 304, 793

\bibitem[{{Chen} \& {Kipping}(2017)}]{ChenKipping2017}
{Chen}, J. \& {Kipping}, D. 2017, \apj, 834, 17

\bibitem[{{Chiang} \& {Laughlin}(2013)}]{ChiangLaughlin2013}
{Chiang}, E. \& {Laughlin}, G. 2013, \mnras, 431, 3444

\bibitem[{{Coffinet} {et~al.}(2019){Coffinet}, {Lovis}, {Dumusque}, \&
  {Pepe}}]{Coffinet2019}
{Coffinet}, A., {Lovis}, C., {Dumusque}, X., \& {Pepe}, F. 2019, \aap, 629, A27

\bibitem[{{Cosentino} {et~al.}(2012){Cosentino}, {Lovis}, {Pepe}, {Collier
  Cameron}, {Latham}, {Molinari}, {Udry}, {Bezawada}, {Black}, {Born},
  {Buchschacher}, {Charbonneau}, {Figueira}, {Fleury}, {Galli}, {Gallie},
  {Gao}, {Ghedina}, {Gonzalez}, {Gonzalez}, {Guerra}, {Henry}, {Horne},
  {Hughes}, {Kelly}, {Lodi}, {Lunney}, {Maire}, {Mayor}, {Micela}, {Ordway},
  {Peacock}, {Phillips}, {Piotto}, {Pollacco}, {Queloz}, {Rice}, {Riverol},
  {Riverol}, {San Juan}, {Sasselov}, {Segransan}, {Sozzetti}, {Sosnowska},
  {Stobie}, {Szentgyorgyi}, {Vick}, \& {Weber}}]{Cosentino2012}
{Cosentino}, R., {Lovis}, C., {Pepe}, F., {et~al.} 2012, in Society of
  Photo-Optical Instrumentation Engineers (SPIE) Conference Series, Vol. 8446,
  Ground-based and Airborne Instrumentation for Astronomy IV, ed. I.~S.
  {McLean}, S.~K. {Ramsay}, \& H.~{Takami}, 84461V

\bibitem[{{Crane} {et~al.}(2006){Crane}, {Shectman}, \& {Butler}}]{Crane2006}
{Crane}, J.~D., {Shectman}, S.~A., \& {Butler}, R.~P. 2006, in Society of
  Photo-Optical Instrumentation Engineers (SPIE) Conference Series, Vol. 6269,
  Society of Photo-Optical Instrumentation Engineers (SPIE) Conference Series,
  ed. I.~S. {McLean} \& M.~{Iye}, 626931

\bibitem[{{Crane} {et~al.}(2010){Crane}, {Shectman}, {Butler}, {Thompson},
  {Birk}, {Jones}, \& {Burley}}]{Crane2010}
{Crane}, J.~D., {Shectman}, S.~A., {Butler}, R.~P., {et~al.} 2010, in Society
  of Photo-Optical Instrumentation Engineers (SPIE) Conference Series, Vol.
  7735, Ground-based and Airborne Instrumentation for Astronomy III, ed. I.~S.
  {McLean}, S.~K. {Ramsay}, \& H.~{Takami}, 773553

\bibitem[{{Crane} {et~al.}(2008){Crane}, {Shectman}, {Butler}, {Thompson}, \&
  {Burley}}]{Crane2008}
{Crane}, J.~D., {Shectman}, S.~A., {Butler}, R.~P., {Thompson}, I.~B., \&
  {Burley}, G.~S. 2008, in Society of Photo-Optical Instrumentation Engineers
  (SPIE) Conference Series, Vol. 7014, Ground-based and Airborne
  Instrumentation for Astronomy II, ed. I.~S. {McLean} \& M.~M. {Casali},
  701479

\bibitem[{{Csizmadia} {et~al.}(2019){Csizmadia}, {Hellard}, \&
  {Smith}}]{Csizmadia2019}
{Csizmadia}, S., {Hellard}, H., \& {Smith}, A.~M.~S. 2019, \aap, 623, A45

\bibitem[{{Cutri} {et~al.}(2003){Cutri}, {Skrutskie}, {van Dyk}, {Beichman},
  {Carpenter}, {Chester}, {Cambresy}, {Evans}, {Fowler}, {Gizis}, {Howard},
  {Huchra}, {Jarrett}, {Kopan}, {Kirkpatrick}, {Light}, {Marsh}, {McCallon},
  {Schneider}, {Stiening}, {Sykes}, {Weinberg}, {Wheaton}, {Wheelock}, \&
  {Zacarias}}]{Cutri2003}
{Cutri}, R.~M., {Skrutskie}, M.~F., {van Dyk}, S., {et~al.} 2003, VizieR Online
  Data Catalog, II/246

\bibitem[{{da Silva} {et~al.}(2006){da Silva}, {Girardi}, {Pasquini},
  {Setiawan}, {von der L{\"u}he}, {de Medeiros}, {Hatzes}, {D{\"o}llinger}, \&
  {Weiss}}]{daSilva2006}
{da Silva}, L., {Girardi}, L., {Pasquini}, L., {et~al.} 2006, \aap, 458, 609

\bibitem[{{Damasso} {et~al.}(2023){Damasso}, {Rodrigues},
  {Castro-Gonz{\'a}lez}, {Lavie}, {Davoult}, {Zapatero Osorio}, {Dou}, {Sousa},
  {Owen}, {Sossi}, {Adibekyan}, {Osborn}, {Leinhardt}, {Alibert}, {Lovis},
  {Delgado Mena}, {Sozzetti}, {Barros}, {Bossini}, {Ziegler}, {Ciardi},
  {Matthews}, {Carter}, {Lillo-Box}, {Su{\'a}rez Mascare{\~n}o}, {Cristiani},
  {Pepe}, {Rebolo}, {Santos}, {Allende Prieto}, {Benatti}, {Bouchy},
  {Brice{\~n}o}, {Di Marcantonio}, {D'Odorico}, {Dumusque}, {Egger},
  {Ehrenreich}, {Faria}, {Figueira}, {G{\'e}nova Santos}, {Gonzales},
  {Gonz{\'a}lez Hern{\'a}ndez}, {Law}, {Lo Curto}, {Mann}, {Martins}, {Mehner},
  {Micela}, {Molaro}, {Nunes}, {Palle}, {Poretti}, {Schlieder}, \&
  {Udry}}]{Damasso2023}
{Damasso}, M., {Rodrigues}, J., {Castro-Gonz{\'a}lez}, A., {et~al.} 2023, \aap,
  679, A33

\bibitem[{{Damasso} {et~al.}(2020){Damasso}, {Sozzetti}, {Lovis}, {Barros},
  {Sousa}, {Demangeon}, {Faria}, {Lillo-Box}, {Cristiani}, {Pepe}, {Rebolo},
  {Santos}, {Zapatero Osorio}, {Gonz{\'a}lez Hern{\'a}ndez}, {Amate},
  {Pasquini}, {Zerbi}, {Adibekyan}, {Abreu}, {Affolter}, {Alibert}, {Aliverti},
  {Allart}, {Allende Prieto}, {{\'A}lvarez}, {Alves}, {Avila}, {Baldini},
  {Bandy}, {Benz}, {Bianco}, {Borsa}, {Bossini}, {Bourrier}, {Bouchy}, {Broeg},
  {Cabral}, {Calderone}, {Cirami}, {Coelho}, {Conconi}, {Coretti}, {Cumani},
  {Cupani}, {D'Odorico}, {Deiries}, {Dekker}, {Delabre}, {Di Marcantonio},
  {Dumusque}, {Ehrenreich}, {Figueira}, {Fragoso}, {Genolet}, {Genoni},
  {G{\'e}nova Santos}, {Hughes}, {Iwert}, {Kerber}, {Knudstrup}, {Landoni},
  {Lavie}, {Lizon}, {Lo Curto}, {Maire}, {Martins}, {M{\'e}gevand}, {Mehner},
  {Micela}, {Modigliani}, {Molaro}, {Monteiro}, {Monteiro}, {Moschetti},
  {Mueller}, {Murphy}, {Nunes}, {Oggioni}, {Oliveira}, {Oshagh}, {Pall{\'e}},
  {Pariani}, {Poretti}, {Rasilla}, {Rebord{\~a}o}, {Redaelli}, {Riva}, {Santana
  Tschudi}, {Santin}, {Santos}, {S{\'e}gransan}, {Schmidt}, {Segovia},
  {Sosnowska}, {Span{\`o}}, {Su{\'a}rez Mascare{\~n}o}, {Tabernero}, {Tenegi},
  {Udry}, \& {Zanutta}}]{Damasso2020}
{Damasso}, M., {Sozzetti}, A., {Lovis}, C., {et~al.} 2020, \aap, 642, A31

\bibitem[{{Delisle} {et~al.}(2020){Delisle}, {Hara}, \&
  {S{\'e}gransan}}]{Delisle2020}
{Delisle}, J.~B., {Hara}, N., \& {S{\'e}gransan}, D. 2020, \aap, 638, A95

\bibitem[{{Delisle} {et~al.}(2022){Delisle}, {Unger}, {Hara}, \&
  {S{\'e}gransan}}]{Delisle2022}
{Delisle}, J.~B., {Unger}, N., {Hara}, N.~C., \& {S{\'e}gransan}, D. 2022,
  \aap, 659, A182

\bibitem[{{Demangeon} {et~al.}(2021){Demangeon}, {Zapatero Osorio}, {Alibert},
  {Barros}, {Adibekyan}, {Tabernero}, {Antoniadis-Karnavas}, {Camacho},
  {Su{\'a}rez Mascare{\~n}o}, {Oshagh}, {Micela}, {Sousa}, {Lovis}, {Pepe},
  {Rebolo}, {Cristiani}, {Santos}, {Allart}, {Allende Prieto}, {Bossini},
  {Bouchy}, {Cabral}, {Damasso}, {Di Marcantonio}, {D'Odorico}, {Ehrenreich},
  {Faria}, {Figueira}, {G{\'e}nova Santos}, {Haldemann}, {Hara}, {Gonz{\'a}lez
  Hern{\'a}ndez}, {Lavie}, {Lillo-Box}, {Lo Curto}, {Martins}, {M{\'e}gevand},
  {Mehner}, {Molaro}, {Nunes}, {Pall{\'e}}, {Pasquini}, {Poretti}, {Sozzetti},
  \& {Udry}}]{Demangeon2021}
{Demangeon}, O.~D.~S., {Zapatero Osorio}, M.~R., {Alibert}, Y., {et~al.} 2021,
  \aap, 653, A41

\bibitem[{{Dressing} {et~al.}(2019){Dressing}, {Hardegree-Ullman}, {Schlieder},
  {Newton}, {Vanderburg}, {Feinstein}, {Duvvuri}, {Arnold}, {Bristow},
  {Thackeray}, {Schwab Abrahams}, {Ciardi}, {Crossfield}, {Yu}, {Martinez},
  {Christiansen}, {Crepp}, \& {Isaacson}}]{Dressing2019}
{Dressing}, C.~D., {Hardegree-Ullman}, K., {Schlieder}, J.~E., {et~al.} 2019,
  \aj, 158, 87

\bibitem[{{Dumusque} {et~al.}(2012){Dumusque}, {Pepe}, {Lovis},
  {S{\'e}gransan}, {Sahlmann}, {Benz}, {Bouchy}, {Mayor}, {Queloz}, {Santos},
  \& {Udry}}]{Dumusque2012}
{Dumusque}, X., {Pepe}, F., {Lovis}, C., {et~al.} 2012, \nat, 491, 207

\bibitem[{{Duncan} {et~al.}(1998){Duncan}, {Levison}, \& {Lee}}]{Duncan1998}
{Duncan}, M.~J., {Levison}, H.~F., \& {Lee}, M.~H. 1998, \aj, 116, 2067

\bibitem[{{Eastman}(2017)}]{Eastman2017}
{Eastman}, J. 2017, {EXOFASTv2: Generalized publication-quality exoplanet
  modeling code}, Astrophysics Source Code Library, record ascl:1710.003

\bibitem[{{Eastman} {et~al.}(2019){Eastman}, {Rodriguez}, {Agol}, {Stassun},
  {Beatty}, {Vanderburg}, {Gaudi}, {Collins}, \& {Luger}}]{Eastman2019}
{Eastman}, J.~D., {Rodriguez}, J.~E., {Agol}, E., {et~al.} 2019, arXiv
  e-prints, arXiv:1907.09480

\bibitem[{{Engle} \& {Guinan}(2023)}]{Engle2023}
{Engle}, S.~G. \& {Guinan}, E.~F. 2023, \apjl, 954, L50

\bibitem[{{Faria} {et~al.}(2022){Faria}, {Su{\'a}rez Mascare{\~n}o},
  {Figueira}, {Silva}, {Damasso}, {Demangeon}, {Pepe}, {Santos}, {Rebolo},
  {Cristiani}, {Adibekyan}, {Alibert}, {Allart}, {Barros}, {Cabral},
  {D'Odorico}, {Di Marcantonio}, {Dumusque}, {Ehrenreich}, {Gonz{\'a}lez
  Hern{\'a}ndez}, {Hara}, {Lillo-Box}, {Lo Curto}, {Lovis}, {Martins},
  {M{\'e}gevand}, {Mehner}, {Micela}, {Molaro}, {Nunes}, {Pall{\'e}},
  {Poretti}, {Sousa}, {Sozzetti}, {Tabernero}, {Udry}, \& {Zapatero
  Osorio}}]{Faria2022}
{Faria}, J.~P., {Su{\'a}rez Mascare{\~n}o}, A., {Figueira}, P., {et~al.} 2022,
  \aap, 658, A115

\bibitem[{{Foreman-Mackey} {et~al.}(2017){Foreman-Mackey}, {Agol},
  {Ambikasaran}, \& {Angus}}]{ForemanMackey2017}
{Foreman-Mackey}, D., {Agol}, E., {Ambikasaran}, S., \& {Angus}, R. 2017,
  {celerite: Scalable 1D Gaussian Processes in C++, Python, and Julia},
  Astrophysics Source Code Library, record ascl:1709.008

\bibitem[{{Foreman-Mackey} {et~al.}(2013){Foreman-Mackey}, {Hogg}, {Lang}, \&
  {Goodman}}]{ForemanMackey2013}
{Foreman-Mackey}, D., {Hogg}, D.~W., {Lang}, D., \& {Goodman}, J. 2013, \pasp,
  125, 306

\bibitem[{{Frandsen} \& {Lindberg}(1999)}]{Frandsen1999}
{Frandsen}, S. \& {Lindberg}, B. 1999, in Astrophysics with the NOT, ed.
  H.~{Karttunen} \& V.~{Piirola}, 71

\bibitem[{Fulton {et~al.}(2018)Fulton, Petigura, Blunt, \&
  Sinukoff}]{Fulton2018}
Fulton, B.~J., Petigura, E.~A., Blunt, S., \& Sinukoff, E. 2018, Publications
  of the Astronomical Society of the Pacific, 130, 044504

\bibitem[{{Fulton} {et~al.}(2017){Fulton}, {Petigura}, {Howard}, {Isaacson},
  {Marcy}, {Cargile}, {Hebb}, {Weiss}, {Johnson}, {Morton}, {Sinukoff},
  {Crossfield}, \& {Hirsch}}]{Fulton2017}
{Fulton}, B.~J., {Petigura}, E.~A., {Howard}, A.~W., {et~al.} 2017, \aj, 154,
  109

\bibitem[{{Gaia Collaboration}(2020)}]{Gaia2020}
{Gaia Collaboration}. 2020, VizieR Online Data Catalog, I/350

\bibitem[{{Gaia Collaboration}(2022)}]{Gaia2022}
{Gaia Collaboration}. 2022, VizieR Online Data Catalog, I/355

\bibitem[{{Gardner} {et~al.}(2006){Gardner}, {Mather}, {Clampin}, {Doyon},
  {Greenhouse}, {Hammel}, {Hutchings}, {Jakobsen}, {Lilly}, {Long}, {Lunine},
  {McCaughrean}, {Mountain}, {Nella}, {Rieke}, {Rieke}, {Rix}, {Smith},
  {Sonneborn}, {Stiavelli}, {Stockman}, {Windhorst}, \& {Wright}}]{Gardner2006}
{Gardner}, J.~P., {Mather}, J.~C., {Clampin}, M., {et~al.} 2006, \ssr, 123, 485

\bibitem[{Genton(2002)}]{Genton2002}
Genton, M.~G. 2002, J. Mach. Learn. Res., 2, 299–312

\bibitem[{{Giles} {et~al.}(2017){Giles}, {Collier Cameron}, \&
  {Haywood}}]{Giles2017}
{Giles}, H. A.~C., {Collier Cameron}, A., \& {Haywood}, R.~D. 2017, \mnras,
  472, 1618

\bibitem[{{Handley} {et~al.}(2015{\natexlab{a}}){Handley}, {Hobson}, \&
  {Lasenby}}]{Handley2015a}
{Handley}, W.~J., {Hobson}, M.~P., \& {Lasenby}, A.~N. 2015{\natexlab{a}},
  \mnras, 450, L61

\bibitem[{{Handley} {et~al.}(2015{\natexlab{b}}){Handley}, {Hobson}, \&
  {Lasenby}}]{Handley2015b}
{Handley}, W.~J., {Hobson}, M.~P., \& {Lasenby}, A.~N. 2015{\natexlab{b}},
  \mnras, 453, 4384

\bibitem[{{Haywood} {et~al.}(2014){Haywood}, {Collier Cameron}, {Queloz},
  {Barros}, {Deleuil}, {Fares}, {Gillon}, {Lanza}, {Lovis}, {Moutou}, {Pepe},
  {Pollacco}, {Santerne}, {S{\'e}gransan}, \& {Unruh}}]{Haywood2014}
{Haywood}, R.~D., {Collier Cameron}, A., {Queloz}, D., {et~al.} 2014, \mnras,
  443, 2517

\bibitem[{{Hellard} {et~al.}(2019){Hellard}, {Csizmadia}, {Padovan}, {Rauer},
  {Cabrera}, {Sohl}, {Spohn}, \& {Breuer}}]{Hellard2019}
{Hellard}, H., {Csizmadia}, S., {Padovan}, S., {et~al.} 2019, \apj, 878, 119

\bibitem[{{Henden} {et~al.}(2016){Henden}, {Templeton}, {Terrell}, {Smith},
  {Levine}, \& {Welch}}]{Henden2016}
{Henden}, A.~A., {Templeton}, M., {Terrell}, D., {et~al.} 2016, VizieR Online
  Data Catalog, II/336

\bibitem[{{Hirano} {et~al.}(2012){Hirano}, {Narita}, {Sato}, {Takahashi},
  {Masuda}, {Takeda}, {Aoki}, {Tamura}, \& {Suto}}]{Hirano2012}
{Hirano}, T., {Narita}, N., {Sato}, B., {et~al.} 2012, \apjl, 759, L36

\bibitem[{{Houdebine} {et~al.}(2017){Houdebine}, {Mullan}, {Paletou}, \&
  {Gebran}}]{Houdebine2017}
{Houdebine}, E.~R., {Mullan}, D.~J., {Paletou}, F., \& {Gebran}, M. 2017,
  VizieR Online Data Catalog, J/ApJ/822/97

\bibitem[{{Howell} {et~al.}(2014){Howell}, {Sobeck}, {Haas}, {Still},
  {Barclay}, {Mullally}, {Troeltzsch}, {Aigrain}, {Bryson}, {Caldwell},
  {Chaplin}, {Cochran}, {Huber}, {Marcy}, {Miglio}, {Najita}, {Smith},
  {Twicken}, \& {Fortney}}]{Howell2014}
{Howell}, S.~B., {Sobeck}, C., {Haas}, M., {et~al.} 2014, \pasp, 126, 398

\bibitem[{{Huber} {et~al.}(2017){Huber}, {Zinn}, {Bojsen-Hansen},
  {Pinsonneault}, {Sahlholdt}, {Serenelli}, {Silva Aguirre}, {Stassun},
  {Stello}, {Tayar}, {Bastien}, {Bedding}, {Buchhave}, {Chaplin}, {Davies},
  {Garc{\'\i}a}, {Latham}, {Mathur}, {Mosser}, \& {Sharma}}]{Huber2017}
{Huber}, D., {Zinn}, J., {Bojsen-Hansen}, M., {et~al.} 2017, \apj, 844, 102

\bibitem[{Hunter(2007)}]{Matplotlib}
Hunter, J.~D. 2007, Computing in Science \& Engineering, 9, 90

\bibitem[{{Izidoro} {et~al.}(2017){Izidoro}, {Ogihara}, {Raymond},
  {Morbidelli}, {Pierens}, {Bitsch}, {Cossou}, \& {Hersant}}]{Izidoro2017}
{Izidoro}, A., {Ogihara}, M., {Raymond}, S.~N., {et~al.} 2017, \mnras, 470,
  1750

\bibitem[{{Jenkins}(2002)}]{Jenkins2002}
{Jenkins}, J.~M. 2002, \apj, 575, 493

\bibitem[{{Jenkins} {et~al.}(2010){Jenkins}, {Chandrasekaran}, {McCauliff},
  {Caldwell}, {Tenenbaum}, {Li}, {Klaus}, {Cote}, \& {Middour}}]{Jenkins2010}
{Jenkins}, J.~M., {Chandrasekaran}, H., {McCauliff}, S.~D., {et~al.} 2010, in
  Society of Photo-Optical Instrumentation Engineers (SPIE) Conference Series,
  Vol. 7740, Software and Cyberinfrastructure for Astronomy, ed. N.~M.
  {Radziwill} \& A.~{Bridger}, 77400D

\bibitem[{{Jenkins} {et~al.}(2016){Jenkins}, {Twicken}, {McCauliff},
  {Campbell}, {Sanderfer}, {Lung}, {Mansouri-Samani}, {Girouard}, {Tenenbaum},
  {Klaus}, {Smith}, {Caldwell}, {Chacon}, {Henze}, {Heiges}, {Latham},
  {Morgan}, {Swade}, {Rinehart}, \& {Vanderspek}}]{Jenkins2016}
{Jenkins}, J.~M., {Twicken}, J.~D., {McCauliff}, S., {et~al.} 2016, in Society
  of Photo-Optical Instrumentation Engineers (SPIE) Conference Series, Vol.
  9913, Software and Cyberinfrastructure for Astronomy IV, ed. G.~{Chiozzi} \&
  J.~C. {Guzman}, 99133E

\bibitem[{{Kellermann} {et~al.}(2018){Kellermann}, {Becker}, \&
  {Redmer}}]{Kellermann2018}
{Kellermann}, C., {Becker}, A., \& {Redmer}, R. 2018, \aap, 615, A39

\bibitem[{{Kervella} {et~al.}(2022){Kervella}, {Arenou}, \&
  {Th{\'e}venin}}]{Kervella2022}
{Kervella}, P., {Arenou}, F., \& {Th{\'e}venin}, F. 2022, \aap, 657, A7

\bibitem[{{Kosiarek} {et~al.}(2021){Kosiarek}, {Berardo}, {Crossfield},
  {Laguna}, {Piaulet}, {Akana Murphy}, {Howell}, {Henry}, {Isaacson}, {Fulton},
  {Weiss}, {Petigura}, {Behmard}, {Hirsch}, {Teske}, {Burt}, {Mills},
  {Chontos}, {Mo{\v{c}}nik}, {Howard}, {Werner}, {Livingston}, {Krick},
  {Beichman}, {Gorjian}, {Kreidberg}, {Morley}, {Christiansen}, {Morales},
  {Scott}, {Crane}, {Wang}, {Shectman}, {Rosenthal}, {Grunblatt}, {Rubenzahl},
  {Dalba}, {Giacalone}, {Villanueva}, {Liu}, {Dai}, {Hill}, {Rice}, {Kane}, \&
  {Mayo}}]{Kosiarek2021}
{Kosiarek}, M.~R., {Berardo}, D.~A., {Crossfield}, I. J.~M., {et~al.} 2021,
  \aj, 161, 47

\bibitem[{{Kov{\'a}cs} {et~al.}(2002){Kov{\'a}cs}, {Zucker}, \&
  {Mazeh}}]{Kovacs2002}
{Kov{\'a}cs}, G., {Zucker}, S., \& {Mazeh}, T. 2002, \aap, 391, 369

\bibitem[{{Kreidberg}(2015)}]{Kreidberg2015}
{Kreidberg}, L. 2015, \pasp, 127, 1161

\bibitem[{Kreidberg(2018)}]{Kreidberg2018}
Kreidberg, L. 2018, in Handbook of Exoplanets (Springer International
  Publishing), 2083--2105

\bibitem[{{Krishnamurthy} {et~al.}(2023){Krishnamurthy}, {Hirano}, {Gaidos},
  {Sato}, {Kopparapu}, {Barclay}, {Garcia-Sage}, {Harakawa}, {Hodapp},
  {Jacobson}, {Konishi}, {Kotani}, {Kudo}, {Kurokawa}, {Kuzuhara}, {Lopez},
  {Nishikawa}, {Omiya}, {Schlieder}, {Serizawa}, {Tamura}, {Ueda}, \&
  {Vievard}}]{Krishnamurthy2023}
{Krishnamurthy}, V., {Hirano}, T., {Gaidos}, E., {et~al.} 2023, \mnras, 521,
  1210

\bibitem[{{Lambeck}(1980)}]{Lambeck1980}
{Lambeck}, K. 1980, {The earth's variable rotation : geophysical causes and
  consequences}

\bibitem[{{Lavie} {et~al.}(2023){Lavie}, {Bouchy}, {Lovis}, {Zapatero Osorio},
  {Deline}, {Barros}, {Figueira}, {Sozzetti}, {Gonz{\'a}lez Hern{\'a}ndez},
  {Lillo-Box}, {Rodrigues}, {Mehner}, {Damasso}, {Adibekyan}, {Alibert},
  {Allende Prieto}, {Cristiani}, {D'Odorico}, {Di Marcantonio}, {Ehrenreich},
  {G{\'e}nova Santos}, {Lo Curto}, {Martins}, {Micela}, {Molaro}, {Nunes},
  {Palle}, {Pepe}, {Poretti}, {Rebolo}, {Santos}, {Sousa}, {Su{\'a}rez
  Mascare{\~n}o}, {Tabrenero}, \& {Udry}}]{Lavie2023}
{Lavie}, B., {Bouchy}, F., {Lovis}, C., {et~al.} 2023, \aap, 673, A69

\bibitem[{{Lo Curto} {et~al.}(2015){Lo Curto}, {Pepe}, {Avila}, {Boffin},
  {Bovay}, {Chazelas}, {Coffinet}, {Fleury}, {Hughes}, {Lovis}, {Maire},
  {Manescau}, {Pasquini}, {Rihs}, {Sinclaire}, \& {Udry}}]{LoCurto2015}
{Lo Curto}, G., {Pepe}, F., {Avila}, G., {et~al.} 2015, The Messenger, 162, 9

\bibitem[{{Love}(1909)}]{Love1909}
{Love}, A.~E.~H. 1909, \mnras, 69, 476

\bibitem[{{Luger} {et~al.}(2016){Luger}, {Agol}, {Kruse}, {Barnes}, {Becker},
  {Foreman-Mackey}, \& {Deming}}]{Luger2016}
{Luger}, R., {Agol}, E., {Kruse}, E., {et~al.} 2016, \aj, 152, 100

\bibitem[{{Luger} {et~al.}(2018){Luger}, {Kruse}, {Foreman-Mackey}, {Agol}, \&
  {Saunders}}]{Luger2018}
{Luger}, R., {Kruse}, E., {Foreman-Mackey}, D., {Agol}, E., \& {Saunders}, N.
  2018, \aj, 156, 99

\bibitem[{{Luger} {et~al.}(2017){Luger}, {Lustig-Yaeger}, \&
  {Agol}}]{Luger2017}
{Luger}, R., {Lustig-Yaeger}, J., \& {Agol}, E. 2017, \apj, 851, 94

\bibitem[{{Luque} \& {Pall{\'e}}(2022)}]{Luque2022}
{Luque}, R. \& {Pall{\'e}}, E. 2022, Science, 377, 1211

\bibitem[{{Malavolta} {et~al.}(2016){Malavolta}, {Nascimbeni}, {Piotto},
  {Quinn}, {Borsato}, {Granata}, {Bonomo}, {Marzari}, {Bedin}, {Rainer},
  {Desidera}, {Lanza}, {Poretti}, {Sozzetti}, {White}, {Latham}, {Cunial},
  {Libralato}, {Nardiello}, {Boccato}, {Claudi}, {Cosentino}, {Covino},
  {Gratton}, {Maggio}, {Micela}, {Molinari}, {Pagano}, {Smareglia}, {Affer},
  {Andreuzzi}, {Aparicio}, {Benatti}, {Bignamini}, {Borsa}, {Damasso}, {Di
  Fabrizio}, {Harutyunyan}, {Esposito}, {Fiorenzano}, {Gandolfi}, {Giacobbe},
  {Gonz{\'a}lez Hern{\'a}ndez}, {Maldonado}, {Masiero}, {Molinaro}, {Pedani},
  \& {Scandariato}}]{Malavolta2016}
{Malavolta}, L., {Nascimbeni}, V., {Piotto}, G., {et~al.} 2016, \aap, 588, A118

\bibitem[{{Mandel} \& {Agol}(2002)}]{MandelAgol2002}
{Mandel}, K. \& {Agol}, E. 2002, \apjl, 580, L171

\bibitem[{Marconi {et~al.}(2022)Marconi, Abreu, Adibekyan, Alberti, Albrecht,
  Alcaniz, Aliverti, Prieto, G{\'o}mez, Amado, Amate, Andersen, Artigau, Baker,
  Baldini, Balestra, Barnes, Baron, Barros, Bauer, Beaulieu, Bellido-Tirado,
  Benneke, Bensby, Bergin, Biazzo, Bik, Birkby, Blind, Boisse, Bolmont,
  Bonaglia, Bonfils, Borsa, Brandeker, Brandner, Broeg, Brogi, Brousseau,
  Brucalassi, Brynnel, Buchhave, Buscher, Cabral, Calderone, Calvo-Ortega,
  Martins, Cantalloube, Carbonaro, Chauvin, Chazelas, Cheffot, Cheng,
  Chiavassa, Christensen, Cirami, Cook, Cooke, Coretti, Covino, Cowan, Cresci,
  Cristiani, Parro, Cupani, D'Odorico, de~Castro~Le{\~a}o, Cia, Medeiros,
  Debras, Debus, Demangeon, Dessauges-Zavadsky, Marcantonio, Dionies, Doyon,
  Dunn, Ehrenreich, Faria, Feruglio, Fisher, Fontana, Fumagalli, Fusco, Fynbo,
  Gabella, Gaessler, Gallo, Gao, Genolet, Genoni, Giacobbe, Giro,
  Gon{\c{c}}alves, Gonzalez, Hern{\'a}ndez, T{\'e}mich, Haehnelt, Haniff,
  Hatzes, Helled, Hoeijmakers, Huke, J{\"a}rvinen, J{\"a}rvinen, Kaminski,
  Korn, Kouach, Kowzan, Kreidberg, Landoni, Lanotte, Lavail, Li, Liske, Lovis,
  Lucatello, Lunney, MacIntosh, Madhusudhan, Magrini, Maiolino, Malo, Man,
  Marquart, Marques, Martins, Martins, Maslowski, Mason, Mason, McCracken,
  Mergo, Micela, Mitchell, Molli{\`e}re, Monteiro, Montgomery, Mordasini,
  Morin, Mucciarelli, Murphy, N'Diaye, Neichel, Niedzielski, Niemczura,
  Nortmann, Noterdaeme, Nunes, Oggioni, Oliva, {\"O}nel, Origlia, {\"O}stlin,
  Palle, Papaderos, Pariani, Castro, Pepe, Levasseur, Petit, Pino, Piqueras,
  Pollo, Poppenhaeger, Quirrenbach, Rauscher, Rebolo, Redaelli, Reffert, Reid,
  Reiners, Richter, Riva, Rivoire, Rodr{\'i}guez-L{\'o}pez, Roederer, Romano,
  Rousseau, Rowe, Salvadori, Santos, Diaz, Sanz-Forcada, Sarajlic, Sauvage,
  Sch{\"a}fer, Schiavon, Schmidt, Selmi, Sivanandam, Sordet, Sordo, Sortino,
  Sosnowska, Sousa, Stempels, Strassmeier, Mascare{\~n}o, Sulich, Sun, Tanvir,
  Tenegi-Sangin{\'e}s, Thibault, Thompson, Tozzi, Turbet, Vall{\'e}e, Varas,
  Venn, V{\'e}ran, Verma, Viel, Wade, Waring, Weber, Weder, Wehbe, Weingrill,
  Woche, Xompero, Zackrisson, Zanutta, Osorio, Zechmeister, \&
  Zimara}]{Marconi2022}
Marconi, A., Abreu, M., Adibekyan, V., {et~al.} 2022, in Ground-based and
  Airborne Instrumentation for Astronomy IX, ed. C.~J. Evans, J.~J. Bryant, \&
  K.~Motohara, Vol. 12184, International Society for Optics and Photonics
  (SPIE), 1218424

\bibitem[{{Marfil} {et~al.}(2021){Marfil}, {Tabernero}, {Montes}, {Caballero},
  {L{\'a}zaro}, {Gonz{\'a}lez Hern{\'a}ndez}, {Nagel}, {Passegger},
  {Schweitzer}, {Ribas}, {Reiners}, {Quirrenbach}, {Amado}, {Cifuentes},
  {Cort{\'e}s-Contreras}, {Dreizler}, {Duque-Arribas},
  {Galad{\'\i}-Enr{\'\i}quez}, {Henning}, {Jeffers}, {Kaminski}, {K{\"u}rster},
  {Lafarga}, {L{\'o}pez-Gallifa}, {Morales}, {Shan}, \&
  {Zechmeister}}]{Marfil2021}
{Marfil}, E., {Tabernero}, H.~M., {Montes}, D., {et~al.} 2021, \aap, 656, A162

\bibitem[{{Mayor} {et~al.}(2011){Mayor}, {Marmier}, {Lovis}, {Udry},
  {S{\'e}gransan}, {Pepe}, {Benz}, {Bertaux}, {Bouchy}, {Dumusque}, {Lo Curto},
  {Mordasini}, {Queloz}, \& {Santos}}]{Mayor2011}
{Mayor}, M., {Marmier}, M., {Lovis}, C., {et~al.} 2011, arXiv e-prints,
  arXiv:1109.2497

\bibitem[{{Mayor} {et~al.}(2003){Mayor}, {Pepe}, {Queloz}, {Bouchy},
  {Rupprecht}, {Lo Curto}, {Avila}, {Benz}, {Bertaux}, {Bonfils}, {Dall},
  {Dekker}, {Delabre}, {Eckert}, {Fleury}, {Gilliotte}, {Gojak}, {Guzman},
  {Kohler}, {Lizon}, {Longinotti}, {Lovis}, {Megevand}, {Pasquini}, {Reyes},
  {Sivan}, {Sosnowska}, {Soto}, {Udry}, {van Kesteren}, {Weber}, \&
  {Weilenmann}}]{Mayor2003}
{Mayor}, M., {Pepe}, F., {Queloz}, D., {et~al.} 2003, The Messenger, 114, 20

\bibitem[{{Meschiari} {et~al.}(2009){Meschiari}, {Wolf}, {Rivera}, {Laughlin},
  {Vogt}, \& {Butler}}]{Meschiari2009}
{Meschiari}, S., {Wolf}, A.~S., {Rivera}, E., {et~al.} 2009, \pasp, 121, 1016

\bibitem[{{Mortier} {et~al.}(2020){Mortier}, {Zapatero Osorio}, {Malavolta},
  {Alibert}, {Rice}, {Lillo-Box}, {Vanderburg}, {Oshagh}, {Buchhave},
  {Adibekyan}, {Delgado Mena}, {Lopez-Morales}, {Charbonneau}, {Sousa},
  {Lovis}, {Affer}, {Allende Prieto}, {Barros}, {Benatti}, {Bonomo}, {Boschin},
  {Bouchy}, {Cabral}, {Collier Cameron}, {Cosentino}, {Cristiani}, {Demangeon},
  {Di Marcantonio}, {D'Odorico}, {Dumusque}, {Ehrenreich}, {Figueira},
  {Fiorenzano}, {Ghedina}, {Gonz{\'a}lez Hern{\'a}ndez}, {Haldemann},
  {Harutyunyan}, {Haywood}, {Latham}, {Lavie}, {Lo Curto}, {Maldonado},
  {Manescau}, {Martins}, {Mayor}, {M{\'e}gevand}, {Mehner}, {Micela}, {Molaro},
  {Molinari}, {Nunes}, {Pepe}, {Palle}, {Phillips}, {Piotto}, {Pinamonti},
  {Poretti}, {Riva}, {Rebolo}, {Santos}, {Sasselov}, {Sozzetti}, {Su{\'a}rez
  Mascare{\~n}o}, {Udry}, {West}, {Watson}, \& {Wilson}}]{Mortier2020}
{Mortier}, A., {Zapatero Osorio}, M.~R., {Malavolta}, L., {et~al.} 2020,
  \mnras, 499, 5004

\bibitem[{{Morton}(2015)}]{Morton2015}
{Morton}, T.~D. 2015, {isochrones: Stellar model grid package}, Astrophysics
  Source Code Library, record ascl:1503.010

\bibitem[{{M{\"u}ller} {et~al.}(2022){M{\"u}ller}, {Ioannidis}, \&
  {Schmitt}}]{Muller2022}
{M{\"u}ller}, H.~M., {Ioannidis}, P., \& {Schmitt}, J.~H.~M.~M. 2022, \aap,
  657, A37

\bibitem[{{Murgas} {et~al.}(2023){Murgas}, {Castro-Gonz{\'a}lez}, {Pall{\'e}},
  {Pozuelos}, {Millholland}, {Foo}, {Korth}, {Marfil}, {Amado}, {Caballero},
  {Christiansen}, {Ciardi}, {Collins}, {Di Sora}, {Fukui}, {Gan}, {Gonzales},
  {Henning}, {Herrero}, {Isopi}, {Jenkins}, {Lillo-Box}, {Lodieu}, {Luque},
  {Mallia}, {Morales}, {Morello}, {Narita}, {Orell-Miquel}, {Parviainen},
  {P{\'e}rez-Torres}, {Quirrenbach}, {Reiners}, {Ribas}, {Safonov}, {Seager},
  {Schwarz}, {Schweitzer}, {Schlecker}, {Strakhov}, {Vanaverbeke}, {Watanabe},
  {Winn}, \& {Zechmeister}}]{Murgas2023}
{Murgas}, F., {Castro-Gonz{\'a}lez}, A., {Pall{\'e}}, E., {et~al.} 2023, \aap,
  677, A182

\bibitem[{{Niraula} {et~al.}(2017){Niraula}, {Redfield}, {Dai}, {Barrag{\'a}n},
  {Gandolfi}, {Cauley}, {Hirano}, {Korth}, {Smith}, {Prieto-Arranz}, {Grziwa},
  {Fridlund}, {Persson}, {Justesen}, {Winn}, {Albrecht}, {Cochran},
  {Csizmadia}, {Duvvuri}, {Endl}, {Hatzes}, {Livingston}, {Narita}, {Nespral},
  {Nowak}, {P{\"a}tzold}, {Palle}, \& {Van Eylen}}]{Niraula2017}
{Niraula}, P., {Redfield}, S., {Dai}, F., {et~al.} 2017, \aj, 154, 266

\bibitem[{{Padovan} {et~al.}(2018){Padovan}, {Spohn}, {Baumeister}, {Tosi},
  {Breuer}, {Csizmadia}, {Hellard}, \& {Sohl}}]{Padovan2018}
{Padovan}, S., {Spohn}, T., {Baumeister}, P., {et~al.} 2018, \aap, 620, A178

\bibitem[{Parviainen(2015)}]{Parviainen2015}
Parviainen, H. 2015, MNRAS, 450, 3233

\bibitem[{{Pepe} {et~al.}(2021){Pepe}, {Cristiani}, {Rebolo}, {Santos},
  {Dekker}, {Cabral}, {Di Marcantonio}, {Figueira}, {Lo Curto}, {Lovis},
  {Mayor}, {M{\'e}gevand}, {Molaro}, {Riva}, {Zapatero Osorio}, {Amate},
  {Manescau}, {Pasquini}, {Zerbi}, {Adibekyan}, {Abreu}, {Affolter}, {Alibert},
  {Aliverti}, {Allart}, {Allende Prieto}, {{\'A}lvarez}, {Alves}, {Avila},
  {Baldini}, {Bandy}, {Barros}, {Benz}, {Bianco}, {Borsa}, {Bourrier},
  {Bouchy}, {Broeg}, {Calderone}, {Cirami}, {Coelho}, {Conconi}, {Coretti},
  {Cumani}, {Cupani}, {D'Odorico}, {Damasso}, {Deiries}, {Delabre},
  {Demangeon}, {Dumusque}, {Ehrenreich}, {Faria}, {Fragoso}, {Genolet},
  {Genoni}, {G{\'e}nova Santos}, {Gonz{\'a}lez Hern{\'a}ndez}, {Hughes},
  {Iwert}, {Kerber}, {Knudstrup}, {Landoni}, {Lavie}, {Lillo-Box}, {Lizon},
  {Maire}, {Martins}, {Mehner}, {Micela}, {Modigliani}, {Monteiro}, {Monteiro},
  {Moschetti}, {Murphy}, {Nunes}, {Oggioni}, {Oliveira}, {Oshagh}, {Pall{\'e}},
  {Pariani}, {Poretti}, {Rasilla}, {Rebord{\~a}o}, {Redaelli}, {Santana
  Tschudi}, {Santin}, {Santos}, {S{\'e}gransan}, {Schmidt}, {Segovia},
  {Sosnowska}, {Sozzetti}, {Sousa}, {Span{\`o}}, {Su{\'a}rez Mascare{\~n}o},
  {Tabernero}, {Tenegi}, {Udry}, \& {Zanutta}}]{Pepe2021}
{Pepe}, F., {Cristiani}, S., {Rebolo}, R., {et~al.} 2021, \aap, 645, A96

\bibitem[{{Pepe} {et~al.}(2013){Pepe}, {Cristiani}, {Rebolo}, {Santos},
  {Dekker}, {M{\'e}gevand}, {Zerbi}, {Cabral}, {Molaro}, {Di Marcantonio},
  {Abreu}, {Affolter}, {Aliverti}, {Allende Prieto}, {Amate}, {Avila},
  {Baldini}, {Bristow}, {Broeg}, {Cirami}, {Coelho}, {Conconi}, {Coretti},
  {Cupani}, {D'Odorico}, {De Caprio}, {Delabre}, {Dorn}, {Figueira}, {Fragoso},
  {Galeotta}, {Genolet}, {Gomes}, {Gonz{\'a}lez Hern{\'a}ndez}, {Hughes},
  {Iwert}, {Kerber}, {Landoni}, {Lizon}, {Lovis}, {Maire}, {Mannetta},
  {Martins}, {Monteiro}, {Oliveira}, {Poretti}, {Rasilla}, {Riva}, {Santana
  Tschudi}, {Santos}, {Sosnowska}, {Sousa}, {Span{\`o}}, {Tenegi}, {Toso},
  {Vanzella}, {Viel}, \& {Zapatero Osorio}}]{Pepe2013}
{Pepe}, F., {Cristiani}, S., {Rebolo}, R., {et~al.} 2013, The Messenger, 153, 6

\bibitem[{{Prieto-Arranz} {et~al.}(2018){Prieto-Arranz}, {Palle}, {Gandolfi},
  {Barrag{\'a}n}, {Guenther}, {Dai}, {Fridlund}, {Hirano}, {Livingston},
  {Luque}, {Niraula}, {Persson}, {Redfield}, {Albrecht}, {Alonso},
  {Antoniciello}, {Cabrera}, {Cochran}, {Csizmadia}, {Deeg}, {Eigm{\"u}ller},
  {Endl}, {Erikson}, {Everett}, {Fukui}, {Grziwa}, {Hatzes}, {Hidalgo},
  {Hjorth}, {Korth}, {Lorenzo-Oliveira}, {Murgas}, {Narita}, {Nespral},
  {Nowak}, {P{\"a}tzold}, {Monta{\~n}ez Rodr{\'\i}guez}, {Rauer}, {Ribas},
  {Smith}, {Trifonov}, {Van Eylen}, \& {Winn}}]{PrietoArranz2018}
{Prieto-Arranz}, J., {Palle}, E., {Gandolfi}, D., {et~al.} 2018, \aap, 618,
  A116

\bibitem[{{Rajpaul} {et~al.}(2015){Rajpaul}, {Aigrain}, {Osborne}, {Reece}, \&
  {Roberts}}]{Rajpaul2015}
{Rajpaul}, V., {Aigrain}, S., {Osborne}, M.~A., {Reece}, S., \& {Roberts}, S.
  2015, \mnras, 452, 2269

\bibitem[{{Rasmussen} \& {Williams}(2006)}]{RasmussenWilliams2006}
{Rasmussen}, C.~E. \& {Williams}, C. K.~I. 2006, {Gaussian Processes for
  Machine Learning}

\bibitem[{{Rice} {et~al.}(2019){Rice}, {Malavolta}, {Mayo}, {Mortier},
  {Buchhave}, {Affer}, {Vanderburg}, {Lopez-Morales}, {Poretti}, {Zeng},
  {Collier Cameron}, {Damasso}, {Coffinet}, {Latham}, {Bonomo}, {Bouchy},
  {Charbonneau}, {Dumusque}, {Figueira}, {Martinez Fiorenzano}, {Haywood},
  {Johnson}, {Lopez}, {Lovis}, {Mayor}, {Micela}, {Molinari}, {Nascimbeni},
  {Nava}, {Pepe}, {Phillips}, {Piotto}, {Sasselov}, {S{\'e}gransan},
  {Sozzetti}, {Udry}, \& {Watson}}]{Rice2019}
{Rice}, K., {Malavolta}, L., {Mayo}, A., {et~al.} 2019, \mnras, 484, 3731

\bibitem[{{Ricker} {et~al.}(2014){Ricker}, {Winn}, {Vanderspek}, {Latham},
  {Bakos}, {Bean}, {Berta-Thompson}, {Brown}, {Buchhave}, {Butler}, {Butler},
  {Chaplin}, {Charbonneau}, {Christensen-Dalsgaard}, {Clampin}, {Deming},
  {Doty}, {De Lee}, {Dressing}, {Dunham}, {Endl}, {Fressin}, {Ge}, {Henning},
  {Holman}, {Howard}, {Ida}, {Jenkins}, {Jernigan}, {Johnson}, {Kaltenegger},
  {Kawai}, {Kjeldsen}, {Laughlin}, {Levine}, {Lin}, {Lissauer}, {MacQueen},
  {Marcy}, {McCullough}, {Morton}, {Narita}, {Paegert}, {Palle}, {Pepe},
  {Pepper}, {Quirrenbach}, {Rinehart}, {Sasselov}, {Sato}, {Seager},
  {Sozzetti}, {Stassun}, {Sullivan}, {Szentgyorgyi}, {Torres}, {Udry}, \&
  {Villasenor}}]{Ricker2014}
{Ricker}, G.~R., {Winn}, J.~N., {Vanderspek}, R., {et~al.} 2014, in Society of
  Photo-Optical Instrumentation Engineers (SPIE) Conference Series, Vol. 9143,
  Space Telescopes and Instrumentation 2014: Optical, Infrared, and Millimeter
  Wave, ed. J.~{Oschmann}, Jacobus~M., M.~{Clampin}, G.~G. {Fazio}, \& H.~A.
  {MacEwen}, 914320

\bibitem[{{Roberts} {et~al.}(2012){Roberts}, {Osborne}, {Ebden}, {Reece},
  {Gibson}, \& {Aigrain}}]{Roberts2012}
{Roberts}, S., {Osborne}, M., {Ebden}, M., {et~al.} 2012, Philosophical
  Transactions of the Royal Society of London Series A, 371, 20110550

\bibitem[{{Rodriguez} {et~al.}(2018){Rodriguez}, {Vanderburg}, {Eastman},
  {Mann}, {Crossfield}, {Ciardi}, {Latham}, \& {Quinn}}]{Rodriguez2018}
{Rodriguez}, J.~E., {Vanderburg}, A., {Eastman}, J.~D., {et~al.} 2018, \aj,
  155, 72

\bibitem[{{Roy} {et~al.}(2023){Roy}, {Benneke}, {Piaulet}, {Gully-Santiago},
  {Crossfield}, {Morley}, {Kreidberg}, {Mikal-Evans}, {Brande}, {Delisle},
  {Greene}, {Hardegree-Ullman}, {Barman}, {Christiansen}, {Dragomir},
  {Fortney}, {Howard}, {Kosiarek}, \& {Lothringer}}]{Roy2023}
{Roy}, P.-A., {Benneke}, B., {Piaulet}, C., {et~al.} 2023, \apjl, 954, L52

\bibitem[{{Silva} {et~al.}(2022){Silva}, {Faria}, {Santos, N. C.}, {Sousa, S.
  G.}, {Viana, P. T. P.}, {Martins, J. H. C.}, {Figueira, P.}, {Lovis, C.},
  {Pepe, F.}, {Cristiani, S.}, {Rebolo, R.}, {Allart, R.}, {Cabral, A.},
  {Mehner, A.}, {Sozzetti, A.}, {Mascare\~no, A. Su\'arez}, {Martins, C. J. A.
  P.}, {Ehrenreich, D.}, {M\'egevand, D.}, {Palle, E.}, {Curto, G. Lo},
  {Tabernero, H. M.}, {Lillo-Box, J.}, {Hern\'andez, J. I. Gonz\'alez},
  {Osorio, M. R. Zapatero}, {Hara, N. C.}, {Nunes, N. J.}, {Di Marcantonio,
  P.}, {Udry, S.}, {Adibekyan, V.}, \& {Dumusque, X.}}]{Silva2022}
{Silva}, A.~M., {Faria}, J.~P., {Santos, N. C.}, {et~al.} 2022, A\&A, 663, A143

\bibitem[{{Skilling}(2004)}]{Skilling2004}
{Skilling}, J. 2004, in American Institute of Physics Conference Series, Vol.
  735, Bayesian Inference and Maximum Entropy Methods in Science and
  Engineering: 24th International Workshop on Bayesian Inference and Maximum
  Entropy Methods in Science and Engineering, ed. R.~{Fischer}, R.~{Preuss}, \&
  U.~V. {Toussaint}, 395--405

\bibitem[{Skilling(2006)}]{Skilling2006}
Skilling, J. 2006, Bayesian Analysis, 1, 833

\bibitem[{Sobell(2015)}]{Ubuntu}
Sobell, M.~G. 2015, A practical guide to Ubuntu Linux (Pearson Education)

\bibitem[{{Southworth}(2011)}]{Southworth2011}
{Southworth}, J. 2011, \mnras, 417, 2166

\bibitem[{{Sozzetti} {et~al.}(2021){Sozzetti}, {Damasso}, {Bonomo}, {Alibert},
  {Sousa}, {Adibekyan}, {Zapatero Osorio}, {Gonz{\'a}lez Hern{\'a}ndez},
  {Barros}, {Lillo-Box}, {Stassun}, {Winn}, {Cristiani}, {Pepe}, {Rebolo},
  {Santos}, {Allart}, {Barclay}, {Bouchy}, {Cabral}, {Ciardi}, {Di
  Marcantonio}, {D'Odorico}, {Ehrenreich}, {Fasnaugh}, {Figueira}, {Haldemann},
  {Jenkins}, {Latham}, {Lavie}, {Lo Curto}, {Lovis}, {Martins}, {M{\'e}gevand},
  {Mehner}, {Micela}, {Molaro}, {Nunes}, {Oshagh}, {Otegi}, {Pall{\'e}},
  {Poretti}, {Ricker}, {Rodriguez}, {Seager}, {Su{\'a}rez Mascare{\~n}o},
  {Twicken}, \& {Udry}}]{Sozzetti2021}
{Sozzetti}, A., {Damasso}, M., {Bonomo}, A.~S., {et~al.} 2021, \aap, 648, A75

\bibitem[{{Speagle}(2020)}]{Speagle2020}
{Speagle}, J.~S. 2020, \mnras, 493, 3132

\bibitem[{{Su{\'a}rez Mascare{\~n}o} {et~al.}(2020){Su{\'a}rez Mascare{\~n}o},
  {Faria}, {Figueira}, {Lovis}, {Damasso}, {Gonz{\'a}lez Hern{\'a}ndez},
  {Rebolo}, {Cristiani}, {Pepe}, {Santos}, {Zapatero Osorio}, {Adibekyan},
  {Hojjatpanah}, {Sozzetti}, {Murgas}, {Abreu}, {Affolter}, {Alibert},
  {Aliverti}, {Allart}, {Allende Prieto}, {Alves}, {Amate}, {Avila}, {Baldini},
  {Bandi}, {Barros}, {Bianco}, {Benz}, {Bouchy}, {Broeng}, {Cabral},
  {Calderone}, {Cirami}, {Coelho}, {Conconi}, {Coretti}, {Cumani}, {Cupani},
  {D'Odorico}, {Deiries}, {Delabre}, {Di Marcantonio}, {Dumusque},
  {Ehrenreich}, {Fragoso}, {Genolet}, {Genoni}, {G{\'e}nova Santos}, {Hughes},
  {Iwert}, {Kerber}, {Knusdstrup}, {Landoni}, {Lavie}, {Lillo-Box}, {Lizon},
  {Lo Curto}, {Maire}, {Manescau}, {Martins}, {M{\'e}gevand}, {Mehner},
  {Micela}, {Modigliani}, {Molaro}, {Monteiro}, {Monteiro}, {Moschetti},
  {Mueller}, {Nunes}, {Oggioni}, {Oliveira}, {Pall{\'e}}, {Pariani},
  {Pasquini}, {Poretti}, {Rasilla}, {Redaelli}, {Riva}, {Santana Tschudi},
  {Santin}, {Santos}, {Segovia}, {Sosnowska}, {Sousa}, {Span{\`o}}, {Tenegi},
  {Udry}, {Zanutta}, \& {Zerbi}}]{Suarez2020}
{Su{\'a}rez Mascare{\~n}o}, A., {Faria}, J.~P., {Figueira}, P., {et~al.} 2020,
  \aap, 639, A77

\bibitem[{{Su{\'a}rez Mascare{\~n}o} {et~al.}(2023){Su{\'a}rez Mascare{\~n}o},
  {Gonz{\'a}lez-{\'A}lvarez}, {Zapatero Osorio}, {Lillo-Box}, {Faria},
  {Passegger}, {Gonz{\'a}lez Hern{\'a}ndez}, {Figueira}, {Sozzetti}, {Rebolo},
  {Pepe}, {Santos}, {Cristiani}, {Lovis}, {Silva}, {Ribas}, {Amado},
  {Caballero}, {Quirrenbach}, {Reiners}, {Zechmeister}, {Adibekyan}, {Alibert},
  {B{\'e}jar}, {Benatti}, {D'Odorico}, {Damasso}, {Delisle}, {Di Marcantonio},
  {Dreizler}, {Ehrenreich}, {Hatzes}, {Hara}, {Henning}, {Kaminski},
  {L{\'o}pez-Gonz{\'a}lez}, {Martins}, {Micela}, {Montes}, {Pall{\'e}},
  {Pedraz}, {Rodr{\'\i}guez}, {Rodr{\'\i}guez-L{\'o}pez}, {Tal-Or}, {Sousa}, \&
  {Udry}}]{Suarez2023}
{Su{\'a}rez Mascare{\~n}o}, A., {Gonz{\'a}lez-{\'A}lvarez}, E., {Zapatero
  Osorio}, M.~R., {et~al.} 2023, \aap, 670, A5

\bibitem[{{Su{\'a}rez Mascare{\~n}o} {et~al.}(2016){Su{\'a}rez Mascare{\~n}o},
  {Rebolo}, \& {Gonz{\'a}lez Hern{\'a}ndez}}]{Suarez2016}
{Su{\'a}rez Mascare{\~n}o}, A., {Rebolo}, R., \& {Gonz{\'a}lez Hern{\'a}ndez},
  J.~I. 2016, \aap, 595, A12

\bibitem[{{Tabernero} {et~al.}(2022){Tabernero}, {Marfil}, {Montes}, \&
  {Gonz{\'a}lez Hern{\'a}ndez}}]{Tabernero2022}
{Tabernero}, H.~M., {Marfil}, E., {Montes}, D., \& {Gonz{\'a}lez
  Hern{\'a}ndez}, J.~I. 2022, \aap, 657, A66

\bibitem[{{Tamayo} {et~al.}(2020){Tamayo}, {Cranmer}, {Hadden}, {Rein},
  {Battaglia}, {Obertas}, {Armitage}, {Ho}, {Spergel}, {Gilbertson}, {Hussain},
  {Silburt}, {Jontof-Hutter}, \& {Menou}}]{Tamayo2020}
{Tamayo}, D., {Cranmer}, M., {Hadden}, S., {et~al.} 2020, Proceedings of the
  National Academy of Science, 117, 18194

\bibitem[{{Tayar} {et~al.}(2022){Tayar}, {Claytor}, {Huber}, \& {van
  Saders}}]{Tayar2022}
{Tayar}, J., {Claytor}, Z.~R., {Huber}, D., \& {van Saders}, J. 2022, \apj,
  927, 31

\bibitem[{{Telting} {et~al.}(2014){Telting}, {Avila}, {Buchhave}, {Frandsen},
  {Gandolfi}, {Lindberg}, {Stempels}, {Prins}, \& {NOT staff}}]{Telting2014}
{Telting}, J.~H., {Avila}, G., {Buchhave}, L., {et~al.} 2014, Astronomische
  Nachrichten, 335, 41

\bibitem[{{Teske} {et~al.}(2018){Teske}, {Wang}, {Wolfgang}, {Dai}, {Shectman},
  {Butler}, {Crane}, \& {Thompson}}]{Teske2018}
{Teske}, J.~K., {Wang}, S., {Wolfgang}, A., {et~al.} 2018, \aj, 155, 148

\bibitem[{{Toledo-Padr{\'o}n} {et~al.}(2020){Toledo-Padr{\'o}n}, {Lovis},
  {Su{\'a}rez Mascare{\~n}o}, {Barros}, {Gonz{\'a}lez Hern{\'a}ndez},
  {Sozzetti}, {Bouchy}, {Zapatero Osorio}, {Rebolo}, {Cristiani}, {Pepe},
  {Santos}, {Sousa}, {Tabernero}, {Lillo-Box}, {Bossini}, {Adibekyan},
  {Allart}, {Damasso}, {D'Odorico}, {Figueira}, {Lavie}, {Lo Curto}, {Mehner},
  {Micela}, {Modigliani}, {Nunes}, {Pall{\'e}}, {Abreu}, {Affolter}, {Alibert},
  {Aliverti}, {Allende Prieto}, {Alves}, {Amate}, {Avila}, {Baldini}, {Bandy},
  {Benatti}, {Benz}, {Bianco}, {Broeg}, {Cabral}, {Calderone}, {Cirami},
  {Coelho}, {Conconi}, {Coretti}, {Cumani}, {Cupani}, {Deiries}, {Dekker},
  {Delabre}, {Demangeon}, {Di Marcantonio}, {Ehrenreich}, {Fragoso}, {Genolet},
  {Genoni}, {G{\'e}nova Santos}, {Hughes}, {Iwert}, {Knudstrup}, {Landoni},
  {Lizon}, {Maire}, {Manescau}, {Martins}, {M{\'e}gevand}, {Molaro},
  {Monteiro}, {Monteiro}, {Moschetti}, {Mueller}, {Oggioni}, {Oliveira},
  {Oshagh}, {Pariani}, {Pasquini}, {Poretti}, {Rasilla}, {Redaelli}, {Riva},
  {Santana Tschudi}, {Santin}, {Santos}, {Segovia}, {Sosnowska}, {Span{\`o}},
  {Tenegi}, {Udry}, {Zanutta}, \& {Zerbi}}]{Toledo2020}
{Toledo-Padr{\'o}n}, B., {Lovis}, C., {Su{\'a}rez Mascare{\~n}o}, A., {et~al.}
  2020, \aap, 641, A92

\bibitem[{{Unger} {et~al.}(2021){Unger}, {S{\'e}gransan}, {Queloz}, {Udry},
  {Lovis}, {Mordasini}, {Ahrer}, {Benz}, {Bouchy}, {Delisle}, {D{\'\i}az},
  {Dumusque}, {Lo Curto}, {Marmier}, {Mayor}, {Pepe}, {Santos}, {Stalport},
  {Alonso}, {Collier Cameron}, {Deleuil}, {Figueira}, {Gillon}, {Moutou},
  {Pollacco}, \& {Pompei}}]{Unger2021}
{Unger}, N., {S{\'e}gransan}, D., {Queloz}, D., {et~al.} 2021, \aap, 654, A104

\bibitem[{{van der Walt} {et~al.}(2011){van der Walt}, {Colbert}, \&
  {Varoquaux}}]{Numpy}
{van der Walt}, S., {Colbert}, S.~C., \& {Varoquaux}, G. 2011, Computing in
  Science and Engineering, 13, 22

\bibitem[{{Van Eylen} \& {Albrecht}(2015)}]{VanEylen2015}
{Van Eylen}, V. \& {Albrecht}, S. 2015, \apj, 808, 126

\bibitem[{Van~Rossum \& Drake(2009)}]{Python3}
Van~Rossum, G. \& Drake, F.~L. 2009, Python 3 Reference Manual (Scotts Valley,
  CA: CreateSpace)

\bibitem[{{Vaughan} {et~al.}(1978){Vaughan}, {Preston}, \&
  {Wilson}}]{Vaughan1978}
{Vaughan}, A.~H., {Preston}, G.~W., \& {Wilson}, O.~C. 1978, \pasp, 90, 267

\bibitem[{Virtanen {et~al.}(2020)Virtanen, Gommers, Oliphant, Haberland, Reddy,
  Cournapeau, Burovski, Peterson, Weckesser, Bright, {van der Walt}, Brett,
  Wilson, Millman, Mayorov, Nelson, Jones, Kern, Larson, Carey, Polat, Feng,
  Moore, {VanderPlas}, Laxalde, Perktold, Cimrman, Henriksen, Quintero, Harris,
  Archibald, Ribeiro, Pedregosa, {van Mulbregt}, \& {SciPy 1.0
  Contributors}}]{Scipy}
Virtanen, P., Gommers, R., Oliphant, T.~E., {et~al.} 2020, Nature Methods, 17,
  261

\bibitem[{{Vogt} {et~al.}(1994){Vogt}, {Allen}, {Bigelow}, {Bresee}, {Brown},
  {Cantrall}, {Conrad}, {Couture}, {Delaney}, {Epps}, {Hilyard}, {Hilyard},
  {Horn}, {Jern}, {Kanto}, {Keane}, {Kibrick}, {Lewis}, {Osborne},
  {Pardeilhan}, {Pfister}, {Ricketts}, {Robinson}, {Stover}, {Tucker}, {Ward},
  \& {Wei}}]{Vogt1994}
{Vogt}, S.~S., {Allen}, S.~L., {Bigelow}, B.~C., {et~al.} 1994, in Society of
  Photo-Optical Instrumentation Engineers (SPIE) Conference Series, Vol. 2198,
  Instrumentation in Astronomy VIII, ed. D.~L. {Crawford} \& E.~R. {Craine},
  362

\bibitem[{{Wildi} {et~al.}(2010){Wildi}, {Pepe}, {Chazelas}, {Lo Curto}, \&
  {Lovis}}]{Wildi2010}
{Wildi}, F., {Pepe}, F., {Chazelas}, B., {Lo Curto}, G., \& {Lovis}, C. 2010,
  in Society of Photo-Optical Instrumentation Engineers (SPIE) Conference
  Series, Vol. 7735, Ground-based and Airborne Instrumentation for Astronomy
  III, ed. I.~S. {McLean}, S.~K. {Ramsay}, \& H.~{Takami}, 77354X

\bibitem[{Williams {et~al.}(2020)Williams, Kelley, \& {many others}}]{gnuplot}
Williams, T., Kelley, C., \& {many others}. 2020, Gnuplot 5.4: an interactive
  plotting program, \url{http://gnuplot.sourceforge.net/}

\bibitem[{{Wisdom} \& {Holman}(1991)}]{WisdomHolman1991}
{Wisdom}, J. \& {Holman}, M. 1991, \aj, 102, 1528

\bibitem[{{Yee} {et~al.}(2017){Yee}, {Petigura}, \& {von Braun}}]{Yee2017}
{Yee}, S.~W., {Petigura}, E.~A., \& {von Braun}, K. 2017, \apj, 836, 77

\bibitem[{{Zacharias} {et~al.}(2012){Zacharias}, {Finch}, {Girard}, {Henden},
  {Bartlett}, {Monet}, \& {Zacharias}}]{Zacharias2012}
{Zacharias}, N., {Finch}, C.~T., {Girard}, T.~M., {et~al.} 2012, VizieR Online
  Data Catalog, I/322A

\bibitem[{{Zechmeister} \& {K{\"u}rster}(2009)}]{Zechmeister2009}
{Zechmeister}, M. \& {K{\"u}rster}, M. 2009, \aap, 496, 577

\bibitem[{{Zeng} {et~al.}(2019){Zeng}, {Jacobsen}, {Sasselov}, {Petaev},
  {Vanderburg}, {Lopez-Morales}, {Perez-Mercader}, {Mattsson}, {Li}, {Heising},
  {Bonomo}, {Damasso}, {Berger}, {Cao}, {Levi}, \& {Wordsworth}}]{Zeng2019}
{Zeng}, L., {Jacobsen}, S.~B., {Sasselov}, D.~D., {et~al.} 2019, Proceedings of
  the National Academy of Science, 116, 9723

\bibitem[{{Zeng} \& {Sasselov}(2013)}]{ZengSasselov2013}
{Zeng}, L. \& {Sasselov}, D. 2013, \pasp, 125, 227

\bibitem[{{Zeng} {et~al.}(2016){Zeng}, {Sasselov}, \& {Jacobsen}}]{Zeng2016}
{Zeng}, L., {Sasselov}, D.~D., \& {Jacobsen}, S.~B. 2016, \apj, 819, 127

\end{thebibliography}

\begin{appendix}
\section{Additional tables and plots}

\begin{landscape}
\begin{table}[]
\caption{Collection of planetary parameters from the literature.}
\label{tab:all_literature}
\centering 
\renewcommand{\arraystretch}{1.4}
\begin{tabular}{l ccccccc}
    \hline 
    \hline 
    \noalign{\smallskip}
     & \cite{Rodriguez2018}	&	\cite{Niraula2017}	&	\multicolumn{2}{c}{\cite{Teske2018}} &	\cite{PrietoArranz2018}	&	\cite{Rice2019}	&	\cite{Kosiarek2021}		\\
     &                        &                       & SYSTEMIC    & {\tt radvel} & & & \\
     \hline 
    \noalign{\smallskip}
$R_{star}$ [R$_{\odot}$] & $0.613_{-0.034}^{+0.033}$	&	$0.651 \pm 0.065$	&	$0.63 \pm 0.19$	&	...	&	$0.622 \pm 0.051$	&	$0.602_{-0.004}^{+0.005}$		&	$0.579 \pm 0.018$ \\
$M_{star}$ [M$_{\odot}$] & $0.614_{-0.029}^{+0.030}$	&	$0.659 \pm 0.06$	&	$0.62 \pm 0.08$	&	...	&	$0.637 \pm 0.051$	&	$0.606_{-0.014}^{+0.020}$		&	$0.593 \pm 0.018$ \\
		
$R_{b}$ [R$_{\oplus}$] & $1.62 \pm 0.11$		&	$1.75 \pm 0.18$		&	...	&	...	&	$1.58_{-0.13}^{+0.14}$		&	$1.577_{-0.031}^{+0.027}$		&	$1.529 \pm 0.058$ \\
$M_{b}$ [M$_{\oplus}$]& $3.42_{-0.76}^{+1.20}$		&	...		&	$7.50 \pm 1.52$	&	$8.16_{-1.54}^{+1.56}$	&	$3.69_{-0.46}^{+0.48}$		&	$4.90 \pm 0.45$			&	$4.87 \pm 0.37$ \\
$T0_{b}$ [d] & $7738.82588_{-3.1e-4}^{+3.0e-4}$ &	$7738.82671_{-4.6e-4}^{+4.3e-4}$ &	...	&	...	&	$7738.82646_{-4.2e-4}^{+4.4e-4}$ &	$7738.82586 \pm $2.6e-4	&	$7738.82586 \pm $2.6e-4 \\
$P_{b}$ [d] & $1.2089802_{-8.1e-6}^{+8.4e-6}$	& $1.208957_{-1.3e-6}^{+1.2e-6}$ & ...	&	...	&	$1.208966 \pm $1.2e-5 &	$1.2089819_{-7.14e-6}^{+6.93e-6}$ &	$1.2089765 \pm $2.3e-6 \\
$b_{b}$ & $0.493_{-0.120}^{+0.080}$	&	$0.595_{-0.070}^{+0.056}$	&	...	&	...	&	$0.21_{-0.14}^{+0.23}$		&	$0.4602_{-0.0443}^{+0.0352}$		&	... \\
$i_{b}$ [deg] & $85.73_{-0.96}^{+1.20}$		&	$84.86 \pm 0.54$	&	...	&	...	&	$88.33_{-2.10}^{+1.15}$		&	$86.07_{-0.34}^{+0.41}$		&	... \\
$a_{b}$ [AU] & $0.01888_{-3.1e-4}^{+3.0e-4}$	&	$0.020 \pm 0.002$	&	...	&	...	&	$0.0206_{-0.0023}^{+0.0020}$	&	$0.01880_{-1.4e-4}^{+2.0e-4}$		&	$0.01866 \pm $1.9e-4 \\
$K_{b}$ [ms$^{-1}$] & $2.84_{-0.64}^{+0.97}$		&	...		&	$5.90 \pm 1.20$	&	$6.41_{-1.20}^{+1.22}$	&	$3.00 \pm 0.35$		&	$4.10 \pm 0.37$			&	$4.1 \pm 0.3$ \\
\\
		
$R_{c}$ [R$_{\oplus}$] & $1.269_{-0.089}^{+0.087}$	&	$1.36 \pm 0.14$		&	...	&	...	&	$1.24 \pm 0.11$		&	$1.241_{-0.026}^{+0.024}$		&	$1.201 \pm 0.046$ \\
$M_{c}$ [M$_{\oplus}$] & $2.42_{-0.49}^{+0.75}$		&	...		&	$<2.56$	&	$2.45_{-2.24}^{+2.20}$	&	$1.45_{-0.57}^{+0.58}$		&	$0.67^{+0.60}$				&	$1.92 \pm 0.49$ \\
$T0_{c}$ [d] & $7742.19944_{-6.8e-4}^{+6.3e-4}$ &	$7738.5519 \pm 0.0014$ &	...	&	...	&	$7738.5496 \pm 0.0015$ &	$7742.1993_{-7.3e-4}^{+7.2e-4}$	&	$7742.19929_{-7.1e-4}^{+7.2e-4}$ \\
$P_{c}$ [d] & $3.648083_{-5.8e-5}^{+6.0e-5}$ &	$3.64802 \pm $1.1e-4 &	...	&	...	&	$3.64823 \pm $1.2e-4 &	$3.6480957_{-6.21e-5}^{+6.33e-5}$	&	$3.648095 \pm $2.4e-5 \\
$b_{c}$ & $0.469_{-0.13}^{+0.085}$	&	$0.558_{-0.096}^{+0.068}$	&	...	&	...	&	$0.25_{-0.16}^{+0.21}$		&	$0.4428_{-0.0483}^{+0.0415}$		&	... \\
$i_{c}$ [deg] & $88.05_{-0.48}^{+0.64}$	&	$87.66_{-0.31}^{+0.30}$		&	...	&	...	&	$89.07_{-0.92}^{+0.59}$	&	$88.19_{-0.18}^{+0.21}$		&	... \\
$a_{c}$ [AU] & $0.03942_{-6.4e-4}^{+6.2e-4}$ &	$0.041 \pm 0.003$	&	...	&	...	&	$0.0429_{-0.0048}^{+0.0042}$	&	$0.03925_{-2.9e-4}^{+4.2e-4}$	&	$0.03896_{-4.0e-4}^{+3.9e-4}$ \\
$K_{c}$ [ms$^{-1}$] & $1.39_{-0.29}^{+0.44}$		&	...		&	$<1.39$	&	$1.33_{-1.22}^{+1.20}$	&	$0.82 \pm 0.32$		&	$0.39^{+0.35}$				&	$1.13 \pm 0.29$ \\
\\

$R_{d}$ [R$_{\oplus}$] & $2.07 \pm 0.14$		&	$2.11_{-0.21}^{+0.22}$		&	...	&	...	&	$2.04 \pm 0.18$		&	$2.022_{-0.043}^{+0.046}$		&	$1.955 \pm 0.075$ \\
$M_{d}$ [M$_{\oplus}$] & $5.2_{-1.2}^{+1.8}$		&	...		&	$<5.58$ &	$3.93_{-2.76}^{+2.65}$	&	$2.35_{-0.68}^{+0.70}$		&	$3.71_{-0.99}^{+0.90}$			&	$3.42 \pm 0.62$ \\
$T0_{d}$ [d] & $7740.96111 \pm $4.4e-4 &	$7740.961_{-8.7e-4}^{+8.3e-4}$ &	...	&	...	&	$7740.96198_{-8.6e-4}^{+8.4e-4}$ &	$7740.96115_{-4.5e-4}^{+4.4e-4}$	&	$7740.96114_{-4.4e-4}^{+4.5e-4}$ \\
$P_{d}$ [d] & $6.201467_{-6.1e-5}^{+6.2e-5}$ &	$6.20141_{-1.0e-4}^{+1.2e-4}$	&	...	&	...	&	$6.20142 \pm $1.3e-4 &	$6.2014698_{-6.11e-5}^{+6.26e-5}$	&	$6.20183 \pm $1e-5 \\
$b_{d}$ & $0.896_{-0.016}^{+0.012}$	&	$0.91_{-0.013}^{+0.011}$	&	...	&	...	&	$0.864_{-0.013}^{+0.022}$	&	$0.8927_{-0.0090}^{+0.0071}$		&	... \\
$i_{d}$ [deg] & $87.39_{-0.18}^{+0.20}$		&	$87.32_{-0.13}^{+0.12}$	&	...	&	...	&	$87.7_{-0.25}^{+0.08}$		&	$87.443 \pm 0.045$		&	... \\
$a_{d}$ [AU] & $0.05615_{-9.1e-4}^{+8.9e-4}$ &	$0.059_{-0.005}^{+0.004}$	&	...	&	...	&	$0.0611_{-0.0068}^{+0.0060}$	&	$0.05591_{-4.1e-4}^{+5.9e-4}$	&	$0.0555_{-5.7e-4}^{+5.5e-4}$ \\
$K_{d}$ [ms$^{-1}$] & $2.5_{-0.58}^{+0.86}$		&	...		&	$<2.54$	&	$1.79_{-1.26}^{+1.21}$	&	$1.11 \pm 0.32$		&	$1.8_{-0.48}^{+0.43}$			&	$1.7_{-0.3}^{+0.3}$ \\
\noalign{\smallskip}
\hline

\end{tabular}
\end{table}

\end{landscape}

\newpage



\clearpage
\onecolumn
\renewcommand{\arraystretch}{1.35}
\begin{longtable}{lccc}
\label{tab:GP_priors}\\
\caption{Priors and measured parameters for GP models in the circular and eccentric case.}\\
   \noalign{\smallskip}
   \hline
   \hline
   \noalign{\smallskip}
  Parameter & Prior	&	Circular	&	Eccentric \\
\noalign{\smallskip}
    \hline
    \noalign{\smallskip}                
 \endfirsthead
\caption{continued}\\ 
  \hline
  \hline
   \noalign{\smallskip}
Parameter & Prior	&	Circular	&	Eccentric \\ 
  \noalign{\smallskip}
  \hline
  \noalign{\smallskip}
  \endhead
  \noalign{\smallskip}
  \hline
  \endfoot
  
Zero points \\
V0$_{\textrm{K2}}$ [ms$^{-1}$]  &  $\mathcal{N}$  (0 , std(flux$_{\textrm{K2}}$)) & $0.0008^{+0.0005}_{-0.0005}$& $0.0008^{+0.0005}_{-0.0005}$\\
V0$_{\textrm{TESS}}$ [ms$^{-1}$]  &  $\mathcal{N}$  (0 , std(flux$_{\textrm{TESS}}$)) & $0.0001^{+0.0002}_{-0.0003}$& $0.0001^{+0.0002}_{-0.0002}$\\
V0$_{\textrm{ESP RV}}$ [ms$^{-1}$]  & $\mathcal{N}$  (0 , std(RV$_{\textrm{ESP}}$)) & $-0.06^{+0.44}_{-0.45}$& $-0.09^{+0.47}_{-0.46}$\\
V0$_{\textrm{HARPS RV}}$ [ms$^{-1}$]  & $\mathcal{N}$  (0 , std(RV$_{\textrm{HARPS}}$)) & $1.22^{+0.71}_{-0.70}$& $1.21^{+0.69}_{-0.67}$\\
V0$_{\textrm{HARPS-N RV}}$ [ms$^{-1}$]  & $\mathcal{N}$  (0 , std(RV$_{\textrm{HARPS-N}}$)) & $-0.42^{+0.64}_{-0.65}$& $-0.40^{+0.62}_{-0.62}$\\
V0$_{\textrm{HIRES RV}}$ [ms$^{-1}$]  & $\mathcal{N}$  (0 , std(RV$_{\textrm{HARPS-N}}$)) & $-0.44^{+0.32}_{-0.34}$& $-0.45^{+0.33}_{-0.34}$\\
V0$_{\textrm{ESP FWHM}}$ [ms$^{-1}$]  & $\mathcal{N}$  (0 , std(FWHM$_{\textrm{ESP}}$)) & $4.96^{+3.17}_{-3.28}$& $5.02^{+3.37}_{-3.20}$\\
V0$_{\textrm{HARPS FWHM}}$ [ms$^{-1}$]  & $\mathcal{N}$  (0 , std(FWHM$_{\textrm{HARPS}}$)) & $1.60^{+4.56}_{-4.54}$& $1.60^{+4.45}_{-4.39}$\\
V0$_{\textrm{HARPS-N FWHM}}$ [ms$^{-1}$]  & $\mathcal{N}$  (0 , std(FWHM$_{\textrm{HARPS-N}}$)) & $-1.22^{+3.98}_{-3.71}$& $-1.20^{+3.85}_{-3.70}$\\
\\
Trends \\
ln (LIN$_{\textrm{K2}}$) [ms$^{-1}$d$^{-1}$] & $\mathcal{N}$  (-10 , 5) & $-13.15^{+2.27}_{-3.59}$& $-13.26^{+2.29}_{-3.60}$\\
ln (LIN$_{\textrm{TESS}}$) [ms$^{-1}$d$^{-1}$] & $\mathcal{N}$  (-10 , 5) & $-12.77^{+2.60}_{-3.59}$& $-12.46^{+2.30}_{-3.70}$\\
LIN$_{\textrm{RV}}$ [ms$^{-1}$d$^{-1}$] & $\mathcal{N}$  (0 , 0.01) & $-0.002^{+0.001}_{-0.001}$& $-0.002^{+0.001}_{-0.001}$\\
LIN$_{\textrm{FWHM}}$ [ms$^{-1}$d$^{-1}$] & $\mathcal{N}$  (0 , 0.01) & $-0.0006^{+0.0073}_{-0.0077}$& $-0.0002^{+0.0077}_{-0.0079}$\\
\\
Jitter\\
ln (Jit$_{\textrm{ESP RV}}$) & $\mathcal{U}$  (-5 , 10) & $-0.87^{+0.99}_{-2.46}$& $-1.01^{+1.10}_{-2.51}$\\
ln (Jit$_{\textrm{HARPS RV}}$)   & $\mathcal{U}$  (-5 , 10) & $-1.76^{+1.37}_{-1.93}$& $-1.99^{+1.51}_{-1.87}$\\
ln (Jit$_{\textrm{HARPS-N RV}}$)   & $\mathcal{U}$  (-5 , 10) & $-2.53^{+1.58}_{-1.61}$& $-2.70^{+1.62}_{-1.52}$\\
ln (Jit$_{\textrm{HIRES RV}}$)   & $\mathcal{U}$  (-5 , 10) & $0.38^{+0.28}_{-0.74}$& $0.42^{+0.26}_{-0.55}$\\
ln (Jit$_{\textrm{ESP FWHM}}$)  & $\mathcal{U}$  (-5 , 10) & $-0.94^{+1.35}_{-2.71}$& $-0.72^{+1.21}_{-2.51}$\\
ln (Jit$_{\textrm{HARPS FWHM}}$)   & $\mathcal{U}$  (-5 , 10) & $2.54^{+0.16}_{-0.16}$& $2.53^{+0.16}_{-0.16}$\\
ln (Jit$_{\textrm{HARPS-N FWHM}}$)   & $\mathcal{U}$  (-5 , 10) & $1.04^{+0.31}_{-0.51}$& $1.02^{+0.32}_{-0.57}$\\
\\
Parameters GP\\
ln (A K2)  & $\mathcal{U}$  (-10 , 20) & $-5.58^{+0.37}_{-0.31}$& $-5.55^{+0.35}_{-0.33}$\\
ln (A TESS)  & $\mathcal{U}$  (-10 , 20) & $-6.86^{+0.89}_{-0.58}$& $-6.89^{+0.81}_{-0.57}$\\
ln (A11 RV)  & $\mathcal{U}$  (-6 , 10) & $-1.95^{+1.76}_{-2.52}$ & $-2.25^{+1.84}_{-2.41}$\\
ln (A12 RV)  & $\mathcal{U}$  (-5 , 10) & $2.53^{+0.15}_{-0.21}$& $2.53^{+0.15}_{-0.20}$\\
ln (A21 RV)  & $\mathcal{U}$  (-6 , 10) & $-0.60^{+1.19}_{-2.91}$& $-0.18^{+0.83}_{-3.10}$\\
ln (A22 RV)  & $\mathcal{U}$  (-5 , 10) &$-1.76^{+1.94}_{-1.93}$ & $-2.40^{+1.96}_{-1.76}$\\
ln (A11 FWHM)  & $\mathcal{U}$  (0 , 10) & $2.36^{+0.22}_{-0.27}$& $2.39^{+0.20}_{-0.26}$\\
ln (A12 FWHM)  & $\mathcal{U}$  (-5 , 10) & $-1.39^{+2.63}_{-2.40}$& $-1.63^{+2.38}_{-2.21}$\\
ln (A21 FWHM)  & $\mathcal{U}$  (0 , 10) & $2.12^{+0.24}_{-0.44}$& $2.11^{+0.24}_{-0.50}$\\
ln (A22 FWHM) & $\mathcal{U}$  (-5 , 10) & $-0.12^{+2.14}_{-3.29}$& $0.73^{+1.55}_{-3.48}$\\
P$_{\textrm{rot}}$ K2 [d]  & $\mathcal{N}$  (30 , 5) & $29.52^{+3.42}_{-3.25}$ & $29.28^{+3.41}_{-3.07}$\\
P$_{\textrm{rot}}$ FWHM [d]  & $\mathcal{N}$  (30 , 5) & $28.16^{+3.38}_{-2.66}$& $28.38^{+3.16}_{-2.66}$\\
ln (T$_{\textrm{evol}}$ K2) [d]  & $\mathcal{N}$  (60 , 1) & $3.83^{+0.70}_{-0.60}$& $3.91^{+0.68}_{-0.64}$\\
ln (T$_{\textrm{evol}}$ TESS) [d]  & $\mathcal{N}$  (60 , 1) & $1.51^{+0.65}_{-0.49}$& $1.49^{+0.60}_{-0.46}$\\
ln (T$_{\textrm{evol}}$ FWHM) [d]  & $\mathcal{N}$  (60 , 1) & $2.07^{+0.27}_{-0.32}$ & $2.03^{+0.26}_{-0.32}$\\
\\
Planet b \\
P [d] &  $\mathcal{N}$  (1.2089802 , 0.001) & $1.2089742^{+0.0000008}_{-0.0000008}$& $1.2089742^{+0.0000009}_{-0.0000008}$\\
T0 [d] & $\mathcal{N}$  (7738.82588 , 0.01) & $7738.8259^{+0.0005}_{-0.0005}$& $7738.8259^{+0.0005}_{-0.0005}$\\
R [R${_\oplus}$] & $\mathcal{U}$ (0 , 3) & $1.44^{+0.09}_{-0.07}$& $1.44^{+0.09}_{-0.07}$\\
M [M${_\oplus}$] & $\mathcal{U}$ (0 , 10) & $4.28^{+0.35}_{-0.33}$& $4.25^{+0.31}_{-0.32}$\\
b & $\mathcal{U}$ (0 , 1) & $0.30^{+0.14}_{-0.16}$& $0.31^{+0.13}_{-0.16}$\\
$\sqrt{e}\cdot \cos{\omega}$ & $\mathcal{N}$ (0 , 0.03) & -- & $-0.06^{+0.11}_{-0.10}$\\
$\sqrt{e}\cdot \sin{\omega}$ & $\mathcal{N}$ (0 , 0.03) & -- & $-0.07^{+0.13}_{-0.13}$\\
\\
Planet c \\
P [d] &  $\mathcal{N}$  (3.648083 , 0.001) & $3.64810^{+0.00001}_{-0.00001}$& $3.64810^{+0.00001}_{-0.00001}$\\
T0 [d] & $\mathcal{N}$  (7742.19944 , 0.01) & $7742.200^{+0.001}_{-0.001}$& $7742.200^{+0.002}_{-0.001}$\\
R [R${_\oplus}$] & $\mathcal{U}$ (0 , 3) & $1.13^{+0.07}_{-0.05}$& $1.13^{+0.07}_{-0.06}$\\
M [M${_\oplus}$] & $\mathcal{U}$ (0 , 10) & $1.86^{+0.37}_{-0.39}$ & $1.84^{+0.39}_{-0.39}$\\
b & $\mathcal{U}$ (0 , 1) & $0.23^{+0.17}_{-0.15}$& $0.25^{+0.16}_{-0.16}$\\
$\sqrt{e}\cdot \cos{\omega}$ & $\mathcal{N}$ (0 , 0.03) & -- & $0.06^{+0.18}_{-0.18}$\\
$\sqrt{e}\cdot \sin{\omega}$ & $\mathcal{N}$ (0 , 0.03) & -- & $-0.03^{+0.17}_{-0.18}$\\
\\
Planet d \\
P [d] &  $\mathcal{N}$  (6.201467 , 0.001) & $6.201812^{+0.000009}_{-0.000009}$& $6.201813^{+0.000009}_{-0.000009}$\\
T0 [d] & $\mathcal{N}$  (7740.96111 , 0.01) & $7740.9588^{+0.0008}_{-0.0009}$& $7740.9586^{+0.0009}_{-0.0009}$\\
R [R${_\oplus}$] & $\mathcal{U}$ (0 , 3) & $1.89^{+0.16}_{-0.14}$& $1.83^{+0.17}_{-0.16}$\\
M [M${_\oplus}$] & $\mathcal{U}$ (0 , 10) & $3.02^{+0.58}_{-0.57}$& $3.17^{+0.61}_{-0.56}$\\
b & $\mathcal{U}$ (0 , 1) & $0.85^{+0.02}_{-0.03}$& $0.89^{+0.06}_{-0.05}$\\
$\sqrt{e}\cdot \cos{\omega}$ & $\mathcal{N}$ (0 , 0.03) & -- & $-0.19^{+0.20}_{-0.16}$\\
$\sqrt{e}\cdot \sin{\omega}$ & $\mathcal{N}$ (0 , 0.03) & -- & $0.22^{+0.19}_{-0.23}$\\
\\
Star \\
LD linear & $\mathcal{U}$  (0 , 1) & $0.38^{+0.23}_{-0.22}$& $0.42^{+0.25}_{-0.23}$\\
LD quadratic & $\mathcal{U}$  (0 , 1) & $0.52^{+0.29}_{-0.30}$& $0.55^{+0.29}_{-0.32}$\\
M [M${_\odot}$] & $\mathcal{N}$  (0.606 , 0.045) & $0.62^{+0.04}_{-0.04}$& $0.62^{+0.04}_{-0.04}$\\
R [R$_{\odot}$] & $\mathcal{N}$  (0.592 , 0.049) & $0.58^{+0.03}_{-0.03}$& $0.58^{+0.03}_{-0.03}$\\
\\
Model statistics \\
ln Z & & $40420.0689 \pm 0.4039$ & $40420.0682 \pm 0.3856$\\
rms K2 [ppm] & & 7.71 & 7.72 \\
rms TESS [ppm] & & 29.03 & 29.00  \\
rms ESPRESSO [ms$^{-1}$] & & 0.30 & 0.27 \\
rms HARPS [ms$^{-1}$] & & 0.83 & 0.82 \\
rms HARPS-N [ms$^{-1}$] & & 1.34 & 1.31 \\
rms HIRES [ms$^{-1}$] & & 1.40 & 1.42 \\

\noalign{\smallskip}
\end{longtable}
\newpage 

\clearpage
\twocolumn

\begin{table*}[htb]
\caption{Derived planetary parameters for GJ~9827 for the circular case.}
\label{tab:results_combined}
\centering 
\renewcommand{\arraystretch}{1.4}
\begin{tabular}{l ccc}
    \hline 
    \hline 
    \noalign{\smallskip}
    Parameter & combined & RV only$^a$ & Photometry only \\
\noalign{\smallskip}
\hline
\noalign{\smallskip}
Planet b \\
$R_p$ [R$_{\oplus}$] &	$1.44_{-0.07}^{+0.09}$ &	--	&	$1.43_{-0.07}^{+0.09}$	\\
$M_p$ [M$_{\oplus}$] &	$4.28_{-0.33}^{+0.35}$ &	$4.47_{-0.23}^{+0.25}$	&	--\\
$T0_p$ [d] &		$7738.8259_{-0.0005}^{+0.0005}$ &	$7738.8244_{-0.0088}^{+0.0091}$	&	$7738.8260_{-0.0005}^{+0.0005}$	\\
$P_p$ [d]	&		$1.208974 \pm 10^{-6}$ &	$1.208967_{-0.000025}^{+0.000025}$	&	$1.208974 \pm 10^{-6}$	\\
$b_p$	&		$0.30_{-0.16}^{+0.14}$ &	-- &	$0.27_{-0.16}^{+0.14}$	\\
$i_p$ [deg] &		$87.60_{-1.27}^{+1.31}$ &	-- &	$87.75_{-1.31}^{+1.35}$	\\
$a_p$ [AU] &		$0.018915_{-0.000410}^{+0.000419}$ &	$0.0189102 \pm 3 \cdot 10^{-7}$	&	-- \\
$K_p$ [ms$^{-1}$] &	$3.53_{-0.22}^{+0.22}$ &	$3.69_{-0.19}^{+0.21}$	&	-- \\
\\
Planet c \\
$R_p$ [R$_{\oplus}$] & 		$1.13_{-0.05}^{+0.07}$ &	--	&	$1.12_{-0.06}^{+0.07}$	\\
$M_p$ [M$_{\oplus}$] &		$1.86_{-0.39}^{+0.37}$ &	$1.65_{-0.35}^{+0.37}$	&	--	\\
$T0_p$ [d] &		$7742.2000_{-0.0014}^{+0.0014}$ &	$7742.1985_{-0.0093}^{+0.0095}$	&	$7742.2000_{-0.0015}^{+0.0015}$	\\
$P_p$ [d]	&		$3.64810_{-0.00001}^{+0.00001}$ &	$3.64845_{-0.00055}^{+0.00051}$	&	$3.64810_{-0.00001}^{+0.00001}$	\\
$b_p$	&		$0.23_{-0.15}^{+0.17}$ &	--&	$0.25_{-0.16}^{+0.16}$	\\
$i_p$ [deg] &		$89.09_{-0.68}^{+0.60}$ &	--&	$89.04_{-0.67}^{+0.62}$	\\
$a_p$ [AU] &		$0.039495_{-0.000857}^{+0.000876}$ &	$0.039488 \pm 4 \cdot 10^{-6}$	&	--\\
$K_p$ [ms$^{-1}$] &		$1.06_{-0.21}^{+0.21}$ &	$0.94_{-0.20}^{+0.21}$	&	--\\
\\
Planet d \\
$R_p$ [R$_{\oplus}$] &		$1.89_{-0.14}^{+0.16}$ &	--	&	$1.91_{-0.14}^{+0.16}$	\\
$M_p$ [M$_{\oplus}$] &		$3.02_{-0.57}^{+0.58}$ &	$3.19_{-0.43}^{+0.48}$	&	--\\
$T0_p$ [d] &		$7740.9588_{-0.0009}^{+0.0008}$ &	$7740.9601_{-0.0094}^{+0.0093}$	&	$7740.9588_{-0.0009}^{+0.0009}$	\\
$P_p$ [d]	&		$6.201812_{-0.000009}^{+0.000009}$ &	$6.201438_{-0.000734}^{+0.000694}$	&	$6.201813_{-0.000001}^{+0.000001}$	\\
$b_p$	&		$0.85_{-0.03}^{+0.02}$ &	-- &	$0.85_{-0.03}^{+0.02}$	\\
$i_p$ [deg] &		$87.66_{-0.16}^{+0.13}$ &	--&	$87.64_{-0.14}^{+0.13}$	\\
$a_p$ [AU] &		$0.056255_{-0.001220}^{+0.001247}$ &	$0.056239 \pm 4 \cdot 10^{-6}$	&	--\\
$K_p$ [ms$^{-1}$] &		$1.44_{-0.27}^{+0.27}$ &	$1.53_{-0.21}^{+0.23}$	&	-- \\
 
\noalign{\smallskip}
\hline

\end{tabular}
\tablefoot{
\tablefoottext{a}{Planet masses for 'RV only' are $M\sin{i}$.}
}
\end{table*}

\newpage
\clearpage

\begin{figure*}[!ht]
\centering
\includegraphics[width=0.5\textwidth]{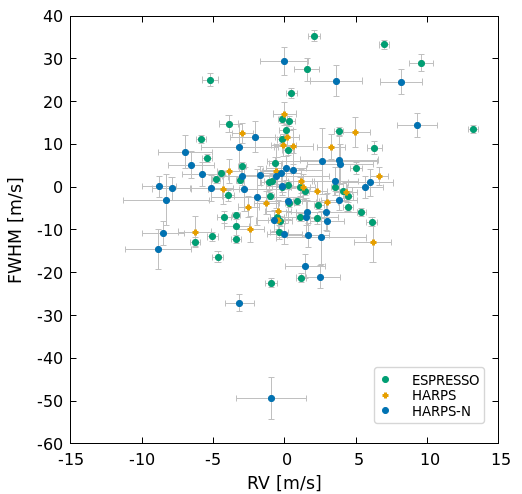}
\caption{Plotting RV vs FWHM for ESPRESSO, HARPS, and HARPS-N shows no correlation.}

\label{fig:RV-FWHM}
\end{figure*}

\begin{figure*}[ht]
\centering
\includegraphics[width=0.9\textwidth]{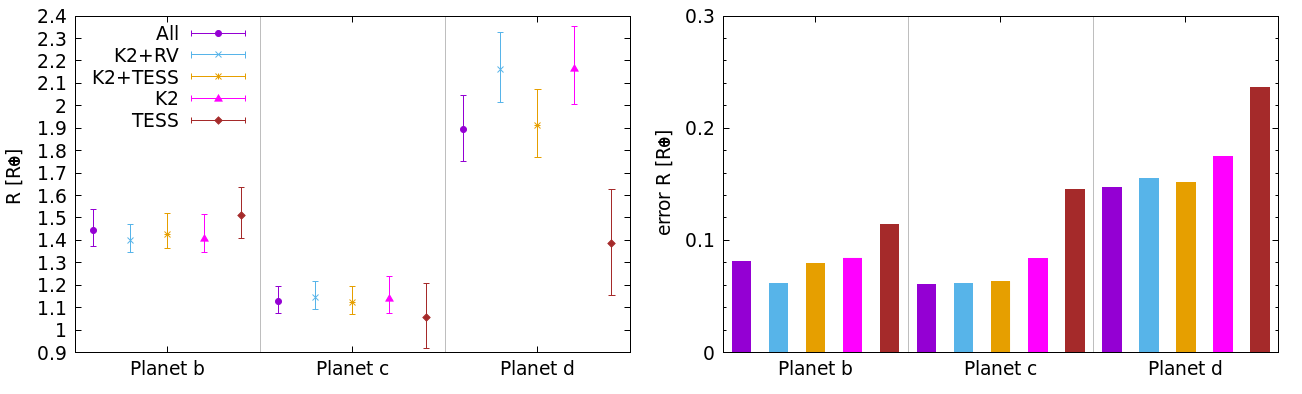}
\caption{Comparison of planetary radius from different GP runs using different combinations of datasets. $RV$ refers to ESPRESSO, HARPS, HARPS-N, and HIRES, and $All$ refers to $RV$, K2, and TESS combined.}
\label{fig:GP_radius}
\end{figure*}

\begin{figure*}[ht]
\centering
\includegraphics[width=0.9\textwidth]{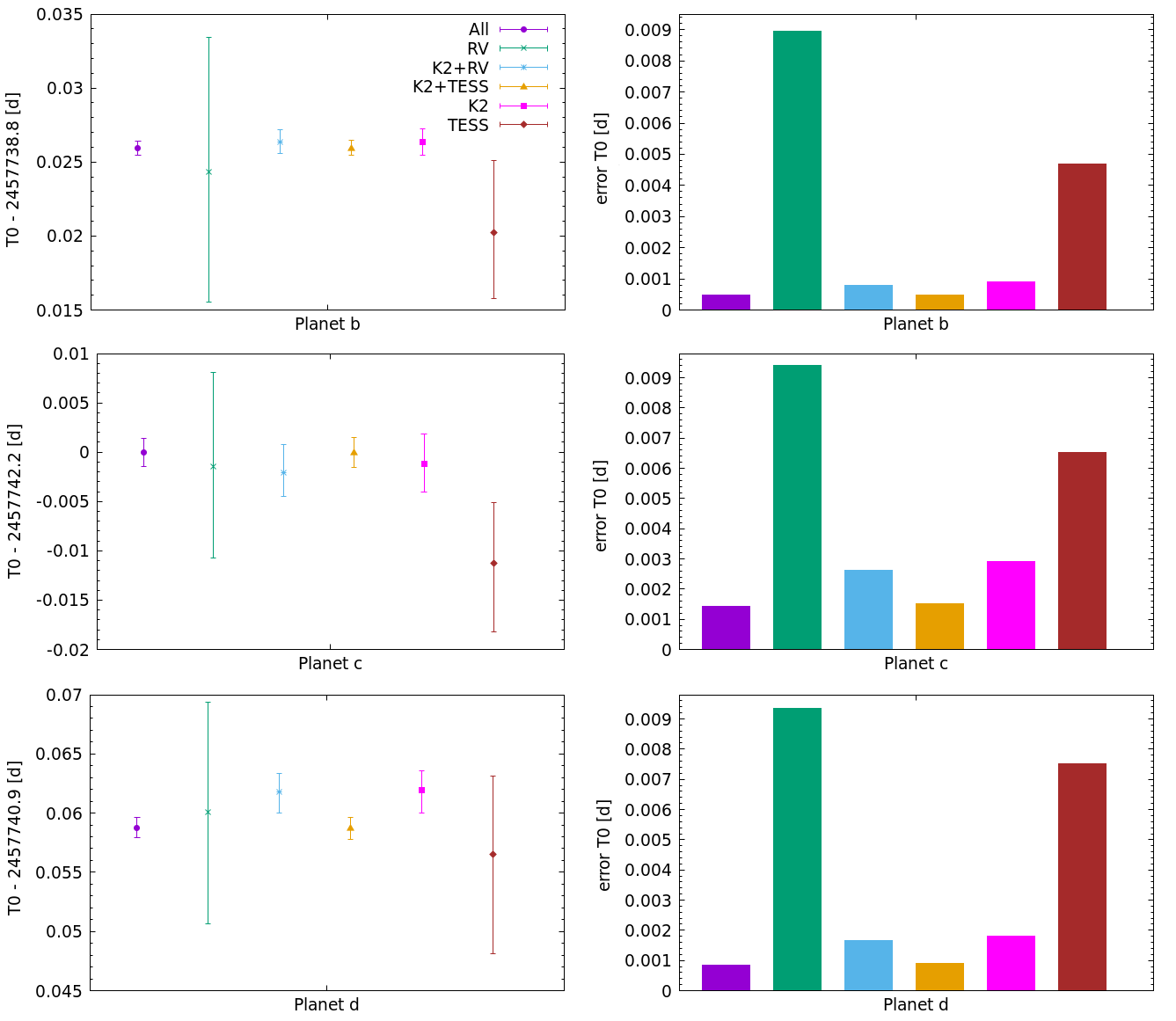}
\caption{Comparison of time of periastron from different GP runs using different combinations of datasets. $RV$ refers to ESPRESSO, HARPS, HARPS-N, and HIRES, and $All$ refers to $RV$, K2, and TESS combined.}
\label{fig:GP_T0}
\end{figure*}

\begin{figure*}[ht]
\centering
\includegraphics[width=0.95\textwidth]{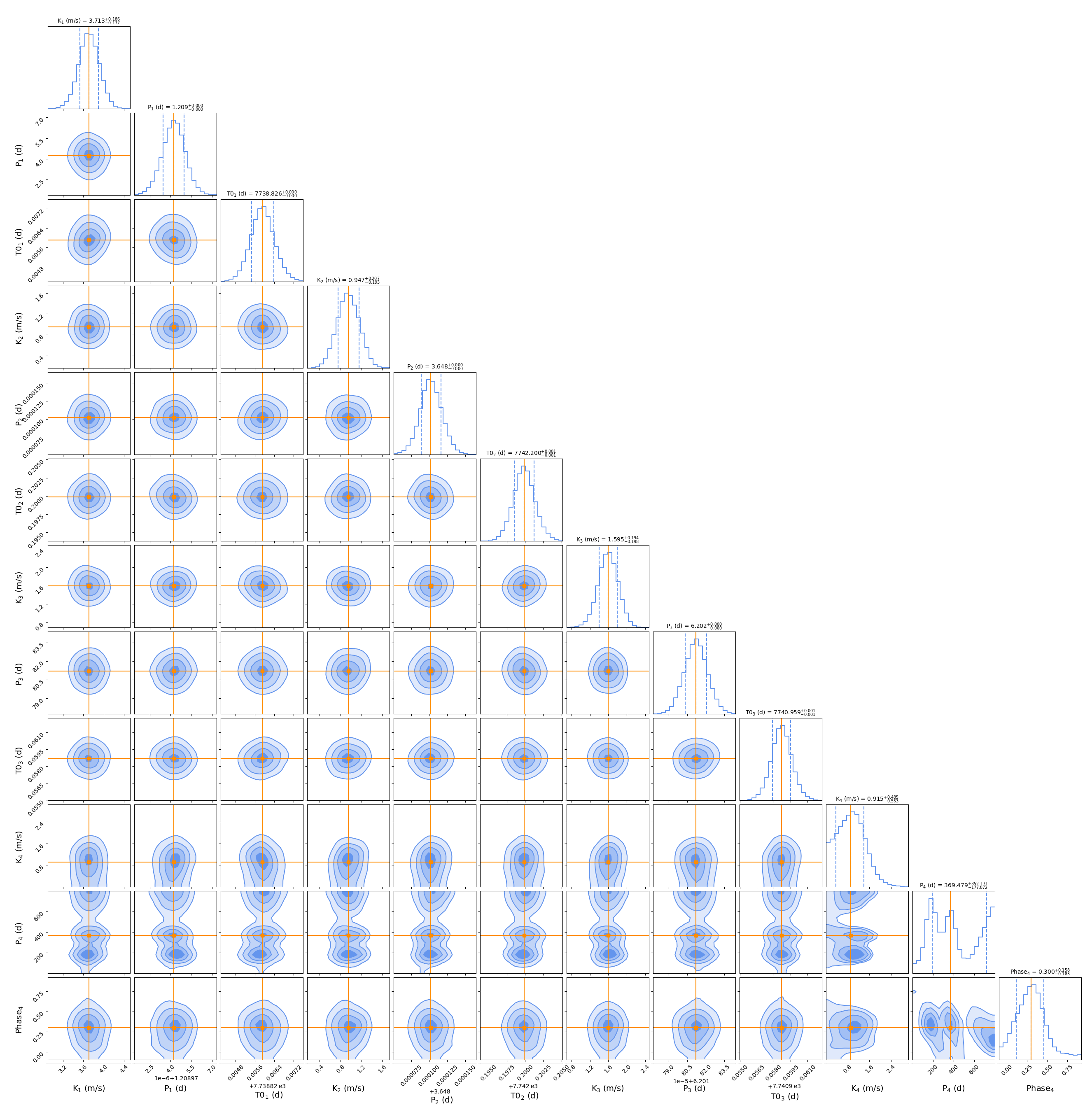}
\caption{Corner plot of the posterior distributions of Kepler amplitude (K), period (P), and T0/Phase for the three known planets and the potential fourth planet. }
\label{fig:planet4}
\end{figure*}

\end{appendix}

\end{document}